\documentclass{aa}  
\bibpunct{(}{)}{;}{a}{}{,} 
\usepackage{graphicx}

\usepackage{txfonts} 
\usepackage[dvipsnames]{xcolor} 
\usepackage{natbib} 
\usepackage{footmisc}

\usepackage[normalem]{ulem}  

\begin{document} 

   \title{Modelling of surface brightness fluctuation measurements}

   \subtitle{Methodology, uncertainty, and recommendations}

   \author{P. Rodríguez-Beltrán\inst{1,2}, M. Cerviño\inst{3}, A. Vazdekis\inst{1,2} \and M. A. Beasley\inst{1,2,4}
          }

   \institute{Instituto de Astrofísica de Canarias (IAC), E-38200 La Laguna, Tenerife, Spain.
   \and
   Departamento de Astrofísica, Universidad de La Laguna, E-38205 Tenerife, Spain
   \and
   Centro de Astrobiología (CSIC/INTA), ESAC Campus, Camino Bajo del Castillo s/n, E-28692 Villanueva de la Cañada, Spain.
   \and
   Centre for Astrophysics and Supercomputing, Swinburne University, John Street, Hawthorn VIC 3122, Australia  \\ \email{pablorb@iac.es}
            }

   \date{Received July 25, 2023; accepted Month Day, Year}

 
  \abstract 
   {}
   {The goal of this work is to scrutinise the surface brightness fluctuation (SBF) calculation methodology. 
   We analysed the SBF derivation procedure, 
   measured the accuracy of the fitted SBF under controlled conditions, 
   retrieved the uncertainty associated with the variability of a system that is inherently stochastic, and studied the SBF reliability under a wide range of conditions. 
   Additionally, we address the possibility of an SBF gradient detection. We also examine the problems related with biased measurements of the SBF and low luminosity sources. All of this information allows us to put forward guidelines to ensure a valid SBF retrieval.}
   {To perform all the experiments described above, we carried out Monte Carlo simulations of mock galaxies as an ideal laboratory. Knowing its underlying properties, we attempted to retrieve SBFs under different conditions. The uncertainty was evaluated through the accuracy, the precision, and the standard deviation of the fitting.}
   {We demonstrate how the usual mathematical approximations taken in the SBF theoretical derivation have a negligible impact on the results and how modelling the instrumental noise reduces the uncertainty.    
   We conducted various studies where we varied the size of the mask applied over the image, the surface and fluctuation brightness of the galaxy, its size and profile, its point spread function (PSF), and the sky background. It is worth highlighting that we find a strong correlation between having a high number of pixels within the studied mask and retrieving a low uncertainty result. We address how the standard deviation of the fitting underestimates the actual uncertainty of the measurement. Lastly, we find that, when studying SBF gradients, the result is a pixel-weighted average of all the SBFs present within the studied region. Retrieving an SBF gradient requires high-quality data and a sufficient difference in the fluctuation value through the different radii. We show how the SBF uncertainty can be obtained and we present a collection of qualitative recommendations for a safe SBF retrieval.} 
   {Our main findings are as follows. It is important to model the instrumental noise, rather than fitting it. The target galaxies must be observed under appropriate observational conditions. In a traditional SBF derivation, one should avoid pixels with fluxes lower than ten times the SBF estimate to prevent biased results. The uncertainty associated with the intrinsic variability of the system can be obtained using sets of Monte Carlo mock galaxy simulations. We offer our computational implementation in the form of a simple code designed to estimate the uncertainty of the SBF measurement. This code can be used to predict the quality of future observations or to evaluate the reliability of those already conducted.}

   
   \maketitle


\section{Introduction}
\label{sec:introduction}

The concept of surface brightness fluctuations (SBFs) was first introduced by \citet{tonry1988new} and \citet{tonry1990observations}. Since then, SBFs have been extensively studied as a powerful tool for understanding the properties of galaxies and their environments \citep[e.g.][]{jensen2003measuring}. Traditionally, SBFs are used to determine extragalactic distances \citep[e.g.][]{blakeslee2010surface,cantiello2018next}. However, besides distance indicators, SBFs have shown potential to constrain stellar populations in galaxies. Stellar population analysis is generally performed by comparing the \textit{mean}\footnote{We want to note that, throughout this work, when using \textit{mean} we are addressing the proper statistical meaning of the mean value of the stellar population luminosity distribution \citep{CL06, rodriguez2021surface}.} 'standard' luminosity of a given population with stellar population synthesis models. In this sense, SBFs arise as a complement to obtain stellar properties \citep{buzzoni1993statistical,worthey1994comprehensive,raimondo2004surface,raimondo2007tracing,cervino2013,vazdekis2020surface,rodriguez2021surface}, among others. 

Surface brightness fluctuations refer to the variation in the light across the surface of a galaxy, which arises from fluctuations in the distribution of stars among different pixels \citep{tonry1988new}. Surface brightness fluctuations are calculated by subtracting a mean reference image, which is the modelled surface brightness of the galaxy, correspondent with the average luminosity of the stellar population at each pixel, and then measuring the local variance of the light. From a theoretical point of view, SBFs are defined as the ratio of the variance and the \textit{mean} of the luminosity distribution of individual stars. It can be demonstrated that this ratio is independent of the number of stars when a stellar population is considered \citep{CL06, cervino2008surface}. In this sense, SBFs are the consequence of the pixel-to-pixel variations in the sampling of the luminosity distribution function of the stellar population, that is, pixels have a different luminosity even with a similar evolutionary status and number of stars. 

Surface brightness fluctuations have been used as a tool for studying the properties of galaxies across a wide range of masses and types. However, there are several limitations to their use that should be considered. 
One major limitation is the dependence of SBF measurements on factors such as the quality of the image data, the signal-to-noise ratio (S/N), the point spread function (PSF), or the brightness, the size of the object and its photometry, among others \citep{jensen2015measuring,moresco2022unveiling,cantiello2023surface}. 
Additionally, the presence of background galaxies, globular clusters (GCs), foreground stars, or other
sources of contamination must be masked, as they interfere with the quality of the SBF measurement.

In this context, it is necessary to establish a way of retrieving an uncertainty from the measured SBF. For instance, in \citet{jensen2015measuring}, the uncertainty of the SBF was measured from several sources: the standard deviation of the SBF fitting, the PSF adjustments, the background variability, and the subtracted mean galaxy model. Since \citet{jensen2015measuring} focussed on the use of SBFs to obtain distances, they also included other additional sources of uncertainty related to the colour and distance calibrations. The total statistical uncertainty given in \citet{jensen2015measuring} reaches 0.1 magnitudes. Among other methods, previous authors have also provided SBF uncertainties associated with the stochasticity in the residual signal from unmasked sources, an estimate through Monte Carlo simulations while slightly varying the galaxy mask and the fitted frequencies, applying bootstrap resampling, considering the PSF mismatch, or a combination of the above \citep{blakeslee2001stellar,cohen2018dragonfly, Carlsten2019}. 
To the best of our knowledge, most authors employ the deviation of the fitting to measure the uncertainty. Besides the uncertainties inherent to the observation, such as an unknown PSF or sky background, we aim to study another source of uncertainty: 
the variability of a system that is intrinsically stochastic. This is how the same SBF value could coincide with different pixel distributions of the light. In the current paper, we investigate the accuracy of the SBF measurement under controlled conditions and propose a way of evaluating the precision of the observations.

Among the applications of the SBF, multiple authors have obtained SBFs from dwarf or diffuse galaxies in recent years, for instance 
\citet{kim2021calibration, greco2021measuring, jerjen1998surface, jerjen2000testing} and  \citet{jerjen2004distances}. As low-mass systems, dwarf galaxies are thought to be the building blocks of larger galaxies \citep{grebel2001dwarf,tosi2003understanding}. Understanding their properties is crucial for constraining models of galaxy formation and evolution.  
On account of this, SBFs offer a powerful tool for studying such objects, as they are sensitive to variations in the underlying stellar population. 
Additionally, SBFs have been used to measure the distances of dwarf galaxies, which are notoriously difficult to determine using other methods. Nevertheless, sources with a low number of stars per pixel might not be a reliable representation of the SBF stellar population \citep{CL06, cervino2008surface, cervino2013}. In this work we discuss the limitations of using SBFs to study faint sources and dwarf galaxies.

Aside from studying the applicability of SBFs on dwarf galaxies, we are interested in evaluating the detection of SBF gradients in massive galaxies. 
Different formation histories can produce very different radial gradients in galaxies \citep{sanchez2007spatially, martin2018timing}. 
The spatial distribution of SBFs across a galaxy can provide additional information about  galaxy structure and evolution. Specifically, the gradient of SBFs changing as a function of the radius can reveal important information about the underlying stellar population, the presence of a substructure, and the history of galaxy interactions and mergers \citep{cantiello2007surface}. 
As was put forward in \citet{rodriguez2021surface}, a combination of \textit{mean} and SBF colours is able to constrain composite stellar populations and, so, predict galaxy formation models. 
Several authors have studied SBF gradients by applying annular masks over different regions of the galaxy, and they have been able to retrieve the SBF gradient; this includes \citet{cantiello2005detection, cantiello2007surface, sodemann1995gradient, sodemann1995variation} and \citet{jensen2015measuring}. However, those authors addressed the precariousness of the observations or did not conclusively detect SBF gradients \citep{jensen1996infrared}. Hence, investigating the possibility and limitations of the SBF gradient measurement is a necessary task.

Having presented the state of the art on the topic, we introduce the main goals of this work: 
(1) to analyse the SBF derivation methodology, addressing the approximations taken and proposing improvements for its estimation; 
(2) to provide a measure of the uncertainty of SBF estimations due to the variability of a system that is intrinsically stochastic;
(3) to evaluate the reliability of SBF retrieval under a wide range of conditions (varying parameters such as the brightness, the fluctuation, the mask applied, the PSF, the sky background, the size of the galaxy, etc.); 
(4) to analyse the possibility of SBF gradient detection and, if present, to address the influence when measuring the whole galaxy; 
and (5) to propose a recommended procedure to retrieve the SBF.
We used mock galaxies as an ideal laboratory in which  to perform such experiments, free from the inherent challenges associated with actual observations. 

This paper is organised as follows. In Sect.~\ref{sec:Methodology} we present the galaxy data we use as a reference, the modelling of our mock galaxy, the SBF derivation, and its consequent fitting. In Sect.~\ref{sec:Results} we display the results of retrieving the SBF while varying a wide range of parameters.
In Sect.~\ref{sec:discussion} we address the SBF derivation procedure, and we warn readers about SBF biased measurements due to low flux level pixels (as in dwarf galaxies) or other sources of offset and the possibility of tracing SBF gradients. 
In Sect.~\ref{sec:summary} we summarise the contents of this work. In Sect.~\ref{sec:codePresentation} we provide our computational implementation in the form of a straightforward code that estimates the uncertainty of the measured SBF. As a set of conclusions, in Sect.~\ref{sec:conclusion} we give recommendations and ideal conditions for a proper SBF retrieval. 
Finally, in Appendix~\ref{sec:appendixApplyMask} we give a more detailed description of the SBF derivation and in Appendix~\ref{sec:notation} we summarise the notation used in the paper.


\section{Methodology}
\label{sec:Methodology}


\subsection{Reference galaxy data}
\label{sec:galData} 

In order to study the reliability when obtaining SBFs, we have created a synthetic galaxy image using NGC 4649 (VCC 1978) as reference. The apparent $i-$band magnitude of the galaxy is $m_{\mathrm{i}} = 9.493\pm0.001$~mag~(AB) from Sloan Digital Survey \citep{abazajian2009seventh}, its effective radius is $R_{\mathrm{eff}}=82$~arcsec \citep{van1991velocity} and its velocity dispersion is $\sigma=330.5\pm4.6$ km~s$^{-1}$ \citep{davies1987spectroscopy}. The rest of data used to define the galaxy are derived from the work of \citet{cantiello2018next} and CFHT/MegaCam imaging data from the NGVS survey \citep{ferrarese2012next}. The MegaCam general specifications state the plate scale at the centre of the field is 0.187~arcsec/pixel, so the effective radius is $R_{\mathrm{eff}} \approx 438$ pixels. 
According to \citet{cantiello2018next} the SBF magnitude of the galaxy is $\bar{m_{\mathrm{i}}}= 30.64\pm0.07$~mag~(AB) and its distance, derived from the SBF magnitude, is $D=16.7 \pm0.6$~Mpc. From now on, we abandon the $i$-band notation, which is only taken as an initial reference value. The exposure time can be retrieved from \citet{ferrarese2012next} as a summed stacking of five exposures of 411 seconds each, leading to a total of $t_{\mathrm{exp}}=2055$ seconds. We obtain the sky background from the header of the stacked NGVS+3+0.I2 image\footnote{Found in the Canadian Astronomy Data Center website, belonging to the CFHTMEGAPIPE catalogue collection.}, where NGC 4649 is located. It is specified that the minimum and the maximum sky counts found among the 5 exposures are 1725 and 1938 counts, respectively. Therefore, we take an average value for the sky background of $N_{\mathrm{Sky}}=5\cdot1831 = 9155$~counts associated with each pixel, when all the images are added. 

Aside from the observational values we have gathered, we need other additional parameters to model a synthetic galaxy, such as a Sersic index and a PSF. The observed values of the Sersic index found in the literature for NGC 4649 range approximately from $n=3.4\pm0.5$ \citep{vika2013megamorph} to $n=5.36^{+0.38}_{-0.32}$ \citep{kormendy2009structure}. So, we choose an intermediate Sersic index value of $n=4$, typical of an elliptical galaxy \citep{de1953distribution, caon1993shape}. We show in Sect.~\ref{sec:ResultsRestOfParams} that the selected Sersic index does not drastically affect the SBF measurement. 
On the other hand, we generate a Gaussian PSF with a standard deviation such that $3\times\sigma_{\mathrm{PSF}}=4$~px ($\sigma_{\mathrm{PSF}}=1.33$~px), centred in a square frame of size $2\cdot 3\times \sigma_{\mathrm{PSF}}+1$~px.
In comparison, the full width at half maximum (FWHM) of the observed galaxy is $\mathrm{FWHM}\approx 0."55$, according to \citep{cantiello2018next}. With the MegaCam pixel scale, this corresponds to $\mathrm{FWHM}\approx2.94$~px. If we assume a Gaussian PSF for the NGVS observation, we find that $\sigma_{\mathrm{PSF}}\approx \mathrm{FWHM}/2.355\approx 1.24$~px. This is similar to our assumption, as $3\times\sigma_{\mathrm{PSF}}=3\cdot 1.24=3.72\approx 4$~px. Finally, the galaxy is centred in a squared $n_{\mathrm{pix}} \times n_{\mathrm{pix}}$ image, where $n_{\mathrm{pix}} = 1605$ pixel wide.

Using the total magnitude we calculate the number of counts in every pixel of the galaxy. In order to do so, first we integrate the total light of the galaxy as 

\begin{equation}
    L=2 \pi \int_0^{\infty} r \,I(r) \,\, \mathrm{d}r,
    \label{eq:totalLightIntegral}
\end{equation}

\noindent where we describe the light profile of the galaxy $I(r)$ as a Sersic profile: 

\begin{equation}
    I(r)=I_{\mathrm{eff}}\; e^{-b_n\left[ \left(\frac{r}{R_{\mathrm{eff}}}\right)^{1/n} - 1 \right]}.
    \label{eq:sersicProf}
\end{equation}

\noindent Here, $b_n\approx 2 n - 1/3$ \citep{ciotti1999analytical} and $I_{\mathrm{eff}}$ is the intensity per unit area at the effective radius. Then, integrating Eq.~(\ref{eq:totalLightIntegral}) returns:

\begin{equation}
    L= I_{\mathrm{eff}}\,2\pi \, n \, e^{b_n} b_n^{-2n } R_{\mathrm{eff}}^2 \Gamma (2n), 
    \label{eq:totalLightIntegrated}
\end{equation}

\noindent with $\Gamma$ being the mathematical gamma function.
If we apply the negative logarithm multiplied by 2.5 we find the enclosed magnitude profile as:

\begin{equation}
  m = \mu_{\mathrm{eff}} - 2.5 \log_{10}(R_{\mathrm{eff}}^2) -2.5 \log_{10}\left(2\pi \, n\, e^{b_n} b_n^{-  2n} \Gamma(2n)\right),  
\label{eq:enclosedMagProf}
\end{equation}

\noindent where $\mu_{\mathrm{eff}}$ is the surface brightness at the effective radius. Using our reference data, we obtain 
$\mu_{\mathrm{eff}} = 22.45$ mag/arcsec$^2$
or $26.09$~mag per pixel. 

Finally, we transform the surface brightness at the effective radius ($\mu_{\mathrm{eff}}$) to counts, $N_{\mathrm{Reff}}$, according to MegaCam general specifications\footnote{\url{https://www.cfht.hawaii.edu/Instruments/Imaging/Megacam/megaprimecalibration.html}}:

\begin{equation}
    N_{\mathrm{Reff}} = t_{\mathrm{exp}} 10^{(\mu_{\mathrm{eff}} - \mathrm{PHOT\_C0})/-2.5},
    \label{eq:muEFFtoCounts}
\end{equation}

\noindent where $t_{\mathrm{exp}}$ is the exposure time and $\mathrm{PHOT\_C0} = 25.743$ is the nominal camera zero point defined by ELIXIR-LSB software  used in the image reduction process. Then, according to Eq.~(\ref{eq:muEFFtoCounts}) we get $N_{\mathrm{Reff}} = 1494.96$~counts at the effective radius. Even though the count number returned is not an integer, we note that this value does not represent individual counts from the galaxy, but an estimation obtained from the magnitude\footnote{After applying the Poisson noise associated with the detector (see Sect.~\ref{sec:mockGalaxyCreation}), our resulting galaxy image consists of integer digits.}. 

Subsequently, the number of counts in each pixel of the galaxy is then obtained from the Sersic profile presented in Eq.~(\ref{eq:sersicProf}). The steepness of the Sersic profile might overestimate the number of counts in the innermost region of the galaxy, so the experiments performed in this work do not consider pixels in the centre.

On the other hand, the count number associated with the SBF, $\bar{N}$, is obtained similarly to that of $N_{\mathrm{Reff}}$, in order to keep a coherent procedure. We introduce $\bar{m_{\mathrm{i}}}$ in Eq.~(\ref{eq:muEFFtoCounts}) as:

\begin{equation}
    \bar{N} = t_{\mathrm{exp}} 10^{(\bar{m} - \mathrm{PHOT\_C0})/-2.5}.
    \label{eq:mSBFtoCounts}
\end{equation}

\noindent We find an SBF value of $\bar{N}_{\mathrm{ref}} = 22.59$~counts associated with each pixel. Here, we want to address a detail of the nomenclature applied throughout this document. $\bar{N}$ is a general way of addressing the number of counts associated with the SBF. If $\bar{N}$ has a subscript the connotations are different: 'input', refers to SBF values used for building a mock galaxy, it can be replaced either for 'real' (if the value is actually known, as for example, in a simulation) or for 'obs' (if the value is measured from an observation). Subscript 'ref' alludes to the reference value of 22.59 counts. Subscript 'fit' is used if the value is the result from the fitting.

The parameters presented in this section (shown in Table~\ref{table:refgal}) are the values chosen for our reference image in most of the experiments of this work, unless otherwise stated.

\begin{table}[h]
\centering
\caption{Parameters used for the creation of our mock galaxies and their respective values associated with the reference galaxy.}
\renewcommand{\arraystretch}{1.4}
\begin{tabular}{|l|l|}
\hline
$n_\mathrm{pix} \times n_\mathrm{pix}$ (image size) & $1605 \times 1605$~px \\ \hline
$\sigma_{\mathrm{PSF}}$ (standard deviation  & 1.33~px \\ 
\hfill of a 2-D Gaussian PSF) & \\ \hline
$R_\mathrm{eff}$ (effective radius) & 438~px \\ \hline
n (Sersic index) & 4 \\ \hline
$t_{\mathrm{exp}}$ (exposure time) & 2055 sec \\ \hline
$N_\mathrm{Reff}$ (counts at $R_\mathrm{eff}$) & 1494.96 counts \\ \hline
$N_{\mathrm{Sky}}$ (sky background) & 9155 counts \\ \hline
$\bar{N}_{\mathrm{ref}}$ (SBF value) & 22.59 counts \\ \hline
\end{tabular}
\tablefoot{These are some of the principal parameters that can be defined in our code, presented in Sect.~\ref{sec:codePresentation} (except the exposure time).}
\label{table:refgal}
\end{table}


\subsection{Mock galaxy creation}
\label{sec:mockGalaxyCreation}

Having specified the effective radius ($R_\mathrm{eff}$), the image size ($n_{\mathrm{pix}} \times n_\mathrm{pix}$), the number of counts at the effective radius ($N_{\mathrm{Reff}}$), the number of counts associated with the fluctuation ($\bar{N}$) in each pixel, the sky background counts ($N_{\mathrm{Sky}}$), the Sersic index ($n$) and the PSF model ($\sigma_{\mathrm{PSF}}$), consequently, the synthetic galaxy is created as follows: 

\begin{enumerate}

\item We create a two-dimensional Sersic image ($\mathrm{Gal_{mean}}(x,y)$) enclosed in a square of $n_{\mathrm{pix}} \times n_{\mathrm{pix}}$ size, assuming that the galaxy can be well described by a Sersic model (Eq.~(\ref{eq:sersicProf})), with index ($n$), an effective radius ($R_\mathrm{eff}$) in pixels and an intensity per unit area expressed in counts ($N_{\mathrm{Reff}}$). 
So, the count value for each pixel of the mean galaxy model is denoted as $N(\mathrm{Gal_{mean}}(x,y))$.
To simplify notation, from now on we do not write the $(x,y)$ dependence of the images. Additionally, in the notation used throughout this work we define the number of counts '$N$' of any magnitude '$\mathcal{X}$' as $N(\mathcal{X})$ or $N_{\mathcal{X}}$. The magnitude '$\mathcal{X}$' can refer to an image or a location within an image. For instance, $N(\mathrm{Gal_{mean}})$ is the number of counts of the mean model image ($\mathrm{Gal_{mean}}$) for each $(x,y)$ pixel, $N_{\mathrm{Reff}}$ is the number of counts at the effective radius or $N_{\mathrm{sky}}$ is the number of counts of the sky (which is the same for every pixel). 

\item We replicate the fluctuation of the stellar population luminosity of every pixel as a random Gaussian distribution ($\mathrm{Rand_{Gauss}}$)\footnote{Using \texttt{python} function \texttt{np.random.normal} \citep{harris2020array}.} with mean $N(\mathrm{Gal_{mean}})$ and variance\footnote{Obtained from the theoretical SBF definition itself, i.e. the variance divided by the mean $\bar{N}=\sigma^2/N(\mathrm{Gal}_{\mathrm{mean}})$.} $\sigma_{\mathrm{fluc}}^2=\bar{N}\cdot N(\mathrm{Gal_{mean}})$, according to each pixel count value. This step returns an image of the modelled galaxy with its fluctuations $\mathrm{Gal}_{\mathrm{mean,fluc}}$. 
Also, we can describe this fluctuation as the addition of a random Gaussian distribution ($\mathrm{Gal_{fluc}}$) with mean value in zero and the same variance\footnote{We remind that the normal distribution is invariant with respect to any scale translation, so the shape is independent of the selected mean value.} to the mean galaxy image $\mathrm{Gal}_{\mathrm{mean}}$:

\begin{equation}
\begin{split}
\mathrm{Gal}&_{\mathrm{mean,fluc}} = \mathrm{Rand_{Gauss}}\left[ \mathrm{Gal_{mean}}, \sigma_{\mathrm{fluc}}^{2} \right]\\
& = \mathrm{Gal_{mean}} + \mathrm{Rand_{Gauss}}\left[0, \sigma_{\mathrm{fluc}}^{2} \right] = \mathrm{Gal_{mean}} + \mathrm{Gal_{fluc}}.
\end{split}
\label{eq:GalMeanFluc}
\end{equation}

\noindent At this point, the presence of globular clusters or background sources could be added, although we will not consider them in our experiments. For instance, GCs are influenced by a combination of factors, including galaxy properties (e.g. brighter galaxies tending to host more GCs), the observational setup or factors such as the PSF of the image. We acknowledge that we vary parameters such as the galaxy brightness or the PSF in the current work (Sect.~\ref{sec:Results}), so the fluctuation contribution due to GCs may be relevant. However, in our analysis, we neglect the GCs impact, as our goal is to determine when the SBF fitting is reliable under ideal conditions. If the SBF retrieval is not trustworthy without the GCs and the background sources, it certainly will not be with them. 

\item We sum a flat image of size $n_{\mathrm{pix}} \times n_{\mathrm{pix}}$ with a value of $N_{\mathrm{Sky}}$ in every pixel as the sky background ($\mathrm{Sky} 
$),

\begin{equation}
\mathrm{Gal_{mean,fluc,sky}}=\mathrm{Gal_{mean,fluc}}+ \mathrm{Sky}.
    \label{eq:GalMeanFlucSky}
\end{equation}

\item We create a PSF as a two-dimensional Gaussian\footnote{Using \texttt{python} function \texttt{astropy.modeling.models.Gaussian2D}  \citep{astropy:2013, astropy:2018, astropy:2022}.} with a standard deviation of
$\sigma_{\mathrm{PSF}}$, centred in a square frame of size  $2\cdot 3\sigma_{\mathrm{PSF}}+1$~px. 
Then, $\mathrm{Gal_{mean,fluc,sky}}$ is convolved\footnote{We perform the PSF convolution in the Fourier space with the \texttt{python} function \texttt{astropy.convolution.convolve$\_$fft} \citep{astropy:2013, astropy:2018, astropy:2022}. In this case we should consider parameter boundary='wrap' for the conditions in the borders of the image.} with the PSF:

\begin{equation}
    \mathrm{Gal_{mean,fluc,sky,PSF}} = \mathrm{Gal_{mean,fluc,sky}}\otimes  \mathrm{PSF}
    \label{eq:GalMeanFlucSkyPSF}
.\end{equation}

\item Finally, we imitate the instrumental noise. 
To do so, we vary every pixel count number with a random Poisson distribution ($\mathrm{Rand_{Poi}}$)\footnote{Using \texttt{python} function \texttt{np.random.poisson}  \citep{harris2020array}.} centred around each pixel value, that is, $N(\mathrm{Gal_{mean,fluc,sky,PSF}})$. 
Thus, our mock galaxy is built as:

\begin{equation}
    \mathrm{Gal_{mock}} = \mathrm{Rand_{Poi}}\left[ \mathrm{Gal_{mean,fluc,sky,PSF}} \right].
\label{eq:mockGalPre}
\end{equation}

\noindent Moreover, we can express the addition of instrumental noise as $\mathrm{Gal_{mock}}=\mathrm{Gal_{mean,fluc,sky,PSF}}+R$. Here, $R$ depicts the instrumental noise, which would be calculated separately as: 

\begin{equation}
R = \mathrm{Rand_{Poi}}\left[ \mathrm{Gal_{mean,fluc,sky,PSF}} \right]-\mathrm{Gal_{mean,fluc,sky,PSF}}
\label{eq:Readout}
.\end{equation}

\end{enumerate}

\noindent In this work we make use of Poisson noise to represent any source of variance that is not convolved with the PSF and we call it generically 'instrumental noise'. 

In brief, the unfolded expression for the mock galaxy is:
\begin{equation}
    \mathrm{Gal_{mock}} = \mathrm{Rand_{Poi}}\left[ \left( \mathrm{Rand_{Gauss}}\left[ \mathrm{Gal_{mean}}, \sigma_{\mathrm{fluc}}^2 \right]+ \mathrm{Sky} 
    \right)\otimes  \mathrm{PSF} \right].
\label{eq:mockGal}
\end{equation}

\begin{figure*}
\centering
\includegraphics[width=\textwidth]{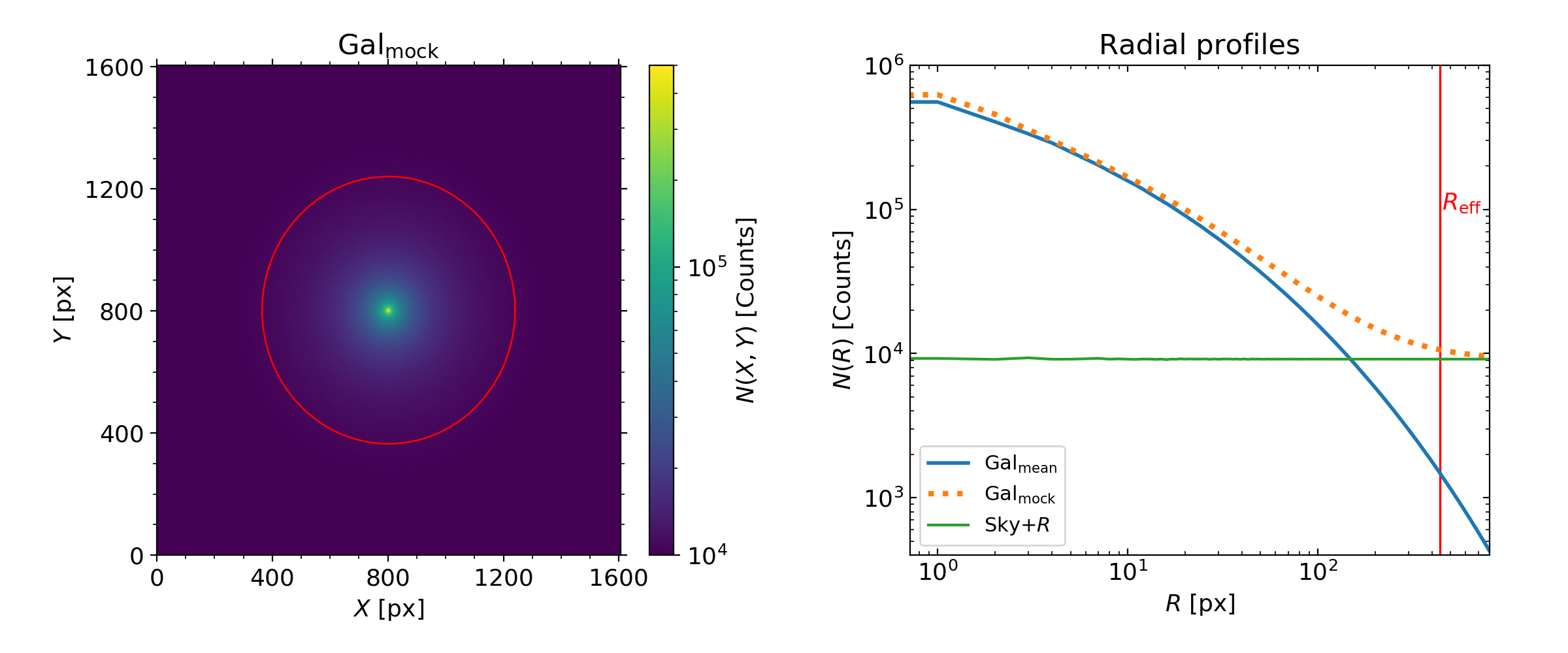}
\caption{Representation of our models computed with the data presented in Sect.~\ref{sec:galData}. Left: Image of the mock galaxy model ($\mathrm{Gal_{mock}}$). Right: Radial profiles of the mean model ($\mathrm{Gal_{mean}}$) as a blue solid line, the mock model ($\mathrm{Gal_{mock}}$) as an orange dotted line, and the sky background with the instrumental noise as a solid green line. The effective radius $R_{\mathrm{eff}}$ is shown as a red circumference (left) and a vertical line (right).}
\label{fig:mockGal}
\end{figure*}

\noindent As an example, in the left panel of Fig.~\ref{fig:mockGal} we show the image of a mock galaxy ($\mathrm{Gal_{mock}}$), where the input data are taken from Sect.~\ref{sec:galData}.
In the right panel of the same figure, we show its associated radial profiles for the mean model ($\mathrm{Gal_{mean}}$), the final synthetic galaxy ($\mathrm{Gal_{mock}}$) and the background sky with the instrumental noise ($\mathrm{Sky}+R$).

\subsection{SBF derivation}
\label{sec:SBFdef}

For measuring the SBF amplitude in the mock galaxy
we start by rewriting Eq.~(\ref{eq:mockGal}) as:

\begin{equation}
    \mathrm{Gal_{mock}} = \left( \mathrm{Gal_{mean}}+\mathrm{Gal_{fluc}}+ \mathrm{Sky} 
    \right)\otimes  \mathrm{PSF} + R.
\label{eq:obsGal}
\end{equation}

\noindent Again, $\mathrm{Gal_{fluc}}$ represents the Gaussian fluctuation around the mean value ($\mathrm{Gal_{mean}}$) due to the stellar population luminosity variation, as introduced in Eq.~(\ref{eq:GalMeanFluc}). 
The term $R$ is the instrumental (Poisson) noise introduced in Eqs.~(\ref{eq:mockGalPre}) and (\ref{eq:Readout}). 
In this case, the PSF, which is contained in a square of side $2\cdot 3\sigma_{\mathrm{PSF}}+1$~px, is now re-inscribed in the centre of a blank-template of the same size as the rest of images. This is done by convolving the PSF with a $n_{\mathrm{pix}}\times n_{\mathrm{pix}}$ image of zero values for all the pixels except the central one, which is assigned a value of one. Also, we note that $\mathrm{Sky} 
\otimes \mathrm{PSF} = \mathrm{Sky}$, since $\mathrm{Sky}$ is a constant image. 
We emphasise that the experiments of this study are conducted under ideal conditions, and we treat the sky as known and flat. In observations, however, the sky value has its own uncertainty and is not necessarily spatially or temporally invariant.

In order to arrive to the SBF term, first we subtract the sky background and the mean model (smoothed with the PSF) to the observed galaxy:

\begin{equation}
    \mathrm{Gal_{mock}} - \mathrm{Sky} 
    - \mathrm{Gal_{mean}}\otimes  \mathrm{PSF} = \mathrm{Gal_{fluc}}\otimes  \mathrm{PSF} + R.
    \label{eq:flucVSrest}
\end{equation}

\noindent Next, we normalise by the square root of the mean model convolved with the PSF. This step is necessary to arrive at the SBF definition, that is, the stellar luminosity distribution variance divided by its \textit{mean} (it will be squared later on, in Eq.~(\ref{eq:PS_obsSBFraw})).

\begin{equation}
    \frac{\mathrm{Gal_{mock}} - \mathrm{Sky} 
    - \mathrm{Gal_{mean}}\otimes  \mathrm{PSF}}{\sqrt{\mathrm{Gal_{mean}}\otimes  \mathrm{PSF}}} = \frac{\mathrm{Gal_{fluc}}\otimes  \mathrm{PSF} + R}{\sqrt{\mathrm{Gal_{mean}}\otimes  \mathrm{PSF}}}.
    \label{eq:sqrtSBF}
\end{equation}

\noindent We denote as $\mathrm{Gal_{mock\;fluc}}$ the left side of Eq.~(\ref{eq:sqrtSBF}). 
The right-hand side of the equation contains two noise components with a null mean value: the first one is the population fluctuation (the SBF), which is convolved with the PSF; the second one is the instrumental noise, both varying from pixel to pixel. 
The fluctuation contribution $\mathrm{Gal_{fluc}}$ and the PSF can be disentangled in the Fourier space. We do so applying the power spectrum to Eq.~(\ref{eq:sqrtSBF}), this is $PS(f)=|\mathfrak{F}(f)|^2 = \mathfrak{F}(f) \cdot \mathfrak{F}(f)^{\dag}$, where $\dag$ denotes the complex conjugate.

\begin{equation}
    PS(\mathrm{Gal_{mock\;fluc}}) = PS\left( \frac{\mathrm{Gal_{fluc}}\otimes  \mathrm{PSF} +R}{\sqrt{\mathrm{Gal_{mean}}\otimes  \mathrm{PSF}}}\right).
    \label{eq:PS_obsSBFraw}
\end{equation}

\noindent As the power spectrum involves a squared Fourier transform, the summed terms of Eq.~(\ref{eq:PS_obsSBFraw}) are developed as the square of complex numbers\footnote{Taking into account that the result of $\mathfrak{F}(f)$ is a complex number (let its result be denoted as $z$), and its square uses the identity $|z|^2 = z z^{\dag}$, then, if we denote $\mathfrak{F}(f)=z_1$ and $\mathfrak{F}(g)=z_2$, the squared sum of complex numbers is derived as: $|z_1+z_2|^2=(z_1+z_2)(z_1+z_2)^{\dag}=(z_1+z_2)(z_1^{\dag}+z_2^{\dag})=z_1z_1^{\dag}+z_2^{\dag}z_2+z_1^{\dag}z_2+z_1z_2^{\dag}=|z_1|^2+|z_2|^2+z_1^{\dag}z_2+z_1z_2^{\dag}$.} ($PS(f+g)=|\mathfrak{F}(f)+\mathfrak{F}(g)|^2=|\mathfrak{F}(f)|^2+\mathfrak{F}(f)^{\dag}\cdot\mathfrak{F}(g)+\mathfrak{F}(f)\cdot\mathfrak{F}(g)^{\dag}+|\mathfrak{F}(g)|^2=PS(f)+\mathfrak{F}(f)^{\dag}\cdot\mathfrak{F}(g)+\mathfrak{F}(f)\cdot\mathfrak{F}(g)^{\dag}+PS(g)$). Then, by applying the convolution theorem, $\mathfrak{F}(f\otimes  g)=\mathfrak{F}(f)\cdot \mathfrak{F}(g)$, we separate $\mathrm{Gal_{fluc}}$ and the PSF contribution.

\begin{equation}
    \begin{split}
    & PS(\mathrm{Gal_{mock\;fluc}}) = PS\left( \frac{\mathrm{Gal_{fluc}}}{\sqrt{\mathrm{Gal_{mean}}\otimes  \mathrm{PSF}}}\right) 
    \cdot PS(\mathrm{PSF}) \\
    &+ PS\left(\frac{R}{\sqrt{\mathrm{Gal_{mean}}\otimes  \mathrm{PSF}}}\right) \\
    & + \mathfrak{F}\left(\frac{\mathrm{Gal_{fluc}}}{\sqrt{\mathrm{Gal_{mean}}\otimes  \mathrm{PSF}}} \right)^\dag \cdot \mathfrak{F}\left( \mathrm{PSF} \right)^\dag \cdot \mathfrak{F}\left( \frac{R }{\sqrt{\mathrm{Gal_{mean}}\otimes  \mathrm{PSF}}} \right) \\
    & + \mathfrak{F}\left(\frac{\mathrm{Gal_{fluc}}}{\sqrt{\mathrm{Gal_{mean}}\otimes  \mathrm{PSF}}}\right) \cdot \mathfrak{F}\left( \mathrm{PSF} \right)\cdot \mathfrak{F}\left( \frac{R }{\sqrt{\mathrm{Gal_{mean}}\otimes  \mathrm{PSF}}} \right)^\dag.
    \end{split}
    \label{eq:PS_obsSBFlong}
\end{equation}

\noindent Finally, we apply an azimuthal average to these power spectra, which in the following we denote with the subindex '$\mathrm{r}$' to indicate a radial profile. 
Applying an azimuthal average reduces the dimension of the 2D Fourier space images ($n_{\mathrm{pix}} \times n_{\mathrm{pix}}$) of Eq.~(\ref{eq:PS_obsSBFlong}) into 1D radial profiles, each with length $n_{\mathrm{pix}}/2$, dependent on the frequency ($k$) in $\mathrm{px}^{-1}$ units. 
Note that, as far as only sums and scalar multiplications are involved, we can consider the total azimuthal average as a sum of the azimuthal averages of the different components. 
Thus, the SBF value to recover ($\bar{N}$) appears as a constant value corresponding to the $PS\left( \mathrm{Gal_{fluc}}/\sqrt{\mathrm{Gal_{mean}}\otimes  \mathrm{PSF}}\right)_{\mathrm{r}}$ term: 

\begin{equation}
\begin{split}
PS(\mathrm{Gal_{mock\;fluc}})_{\mathrm{r}} = \bar{N}\cdot PS(\mathrm{PSF})_{\mathrm{r}} + PS\left(\frac{R}{\sqrt{\mathrm{Gal_{mean}}\otimes  \mathrm{PSF}}}\right)_{\mathrm{r}} \\
+ \sqrt{\bar{N}^\dag} \cdot \left(\mathfrak{F}\left( \mathrm{PSF} \right)^\dag \cdot \mathfrak{F}\left( \frac{R }{\sqrt{\mathrm{Gal_{mean}}\otimes  \mathrm{PSF}}} \right) \right)_\mathrm{r\;} \\
+ \sqrt{\bar{N}} \cdot \left(\mathfrak{F}\left( \mathrm{PSF} \right)\cdot \mathfrak{F} \left(\frac{R }{\sqrt{\mathrm{Gal_{mean}}\otimes  \mathrm{PSF}}} \right)^\dag\right)_\mathrm{r\;}.
\end{split}
\label{eq:PS_obsSBFlong_rad}
\end{equation}

\begin{figure}
\centering
\includegraphics[width=0.5\textwidth]{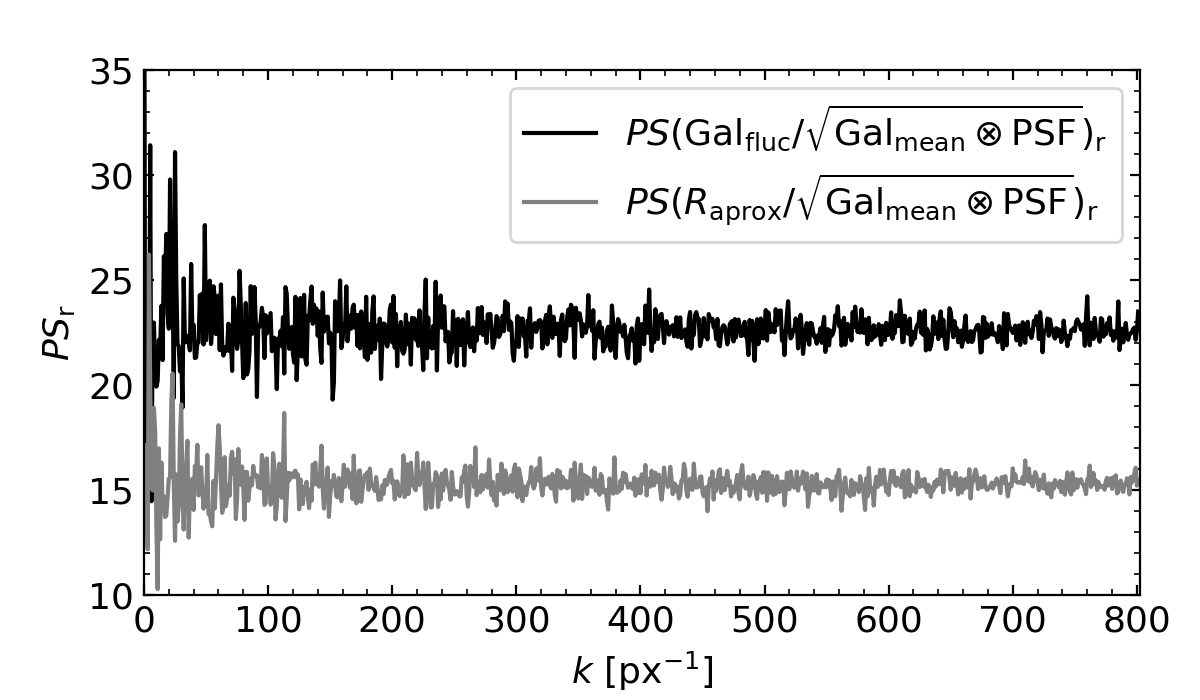}
\caption{Azimuthally averaged power spectrum of both fluctuation contributions: the stellar population luminosity variability ($PS(\mathrm{Gal_{fluc}}/\sqrt{\mathrm{Gal_{mean}}\otimes \mathrm{PSF}})_{\mathrm{r}}$) is in black and the instrumental noise ($PS(R_{\mathrm{approx}}/\sqrt{\mathrm{Gal_{mean}}\otimes \mathrm{PSF}})_{\mathrm{r}}$) is in grey. In this case we consider a mask, introduced as in Eq.~(\ref{eq:PS_obsSBFshort_rad_Mask}), with the same data used in Fig.~\ref{fig:PS}. Note that the input SBF value ($\bar{N}_{\mathrm{ref}}$=22.59~counts) coincides with the average value of the fluctuation term.}
\label{fig:PsGalFluc_R_AreCte}
\end{figure}

\noindent The two fluctuation terms considered here, in essence, the stellar population luminosity variability ($PS(\mathrm{Gal_{fluc}}/\sqrt{\mathrm{Gal_{mean}}\otimes  \mathrm{PSF}})_{\mathrm{r}}$) and the instrumental noise ($PS(R/\sqrt{\mathrm{Gal_{mean}}\otimes \mathrm{PSF}})_{\mathrm{r}}$), are constant on average with respect to the frequency; as the example of Fig.~\ref{fig:PsGalFluc_R_AreCte} shows, using the data of our reference galaxy. In this figure, both radial profiles are flat and, therefore, neither of their associated images have any structure. Thus, the fluctuation term ($PS\left( \mathrm{Gal_{fluc}}/\sqrt{\mathrm{Gal_{mean}}\otimes  \mathrm{PSF}}\right)_{\mathrm{r}}$) can be represented by a constant value, which in this case matches the input SBF ($\bar{N}_{\mathrm{ref}}$).  

\begin{figure}
\centering
\includegraphics[width=0.5\textwidth]{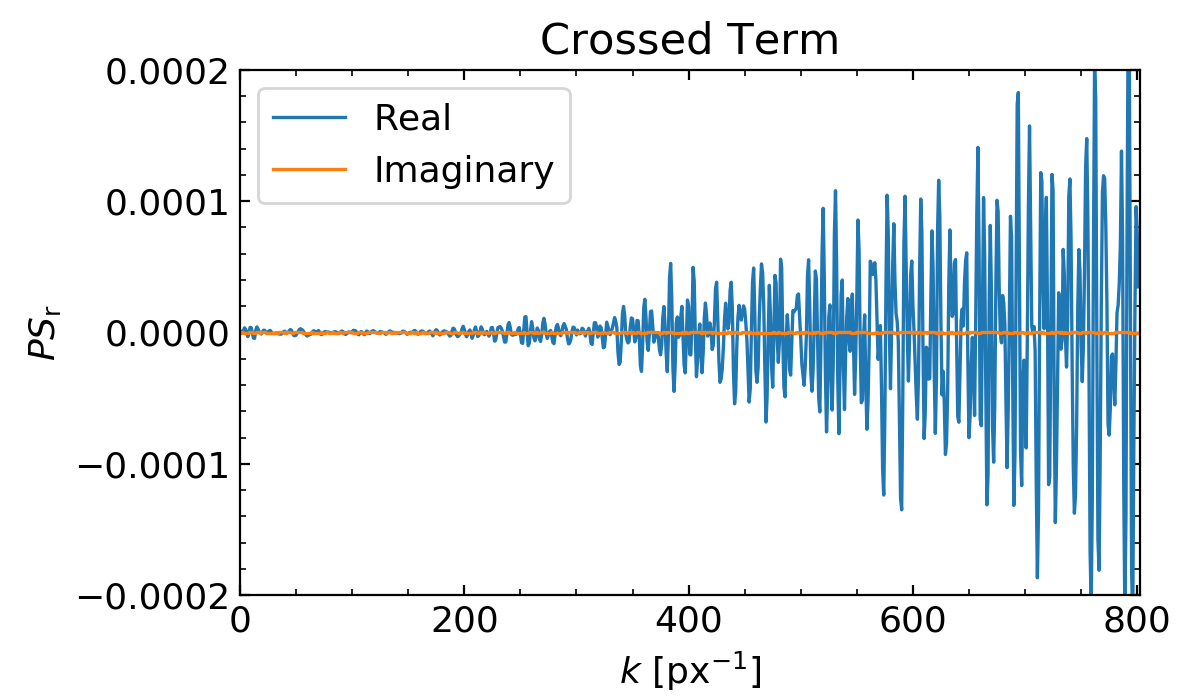}
\caption{Real and imaginary components of the crossed term presented in Eq.~(\ref{eq:PS_obsSBFlong_rad}). Using the same data as in Fig.~\ref{fig:PS}.} 
\label{fig:CT_components}
\end{figure}

At this point, most of the literature assumes neglecting the crossed term that appears due to the square modulus of the power spectrum. In Fig.~\ref{fig:CT_components} (again, using the data of our reference galaxy) we demonstrate that the power spectrum of the imaginary component is null and the real component is three orders of magnitude lower than the rest of the terms. Additionally, numerical calculations show that the influence of the real part of the crossed term is negligible (see Sect.~\ref{sec:RelianceSBF}, Table~\ref{table:relianceSBF_Can}).
Consequently, Eq.~(\ref{eq:PS_obsSBFlong_rad}) is rewritten and fitted as:

\begin{equation}
    PS(\mathrm{Gal_{mock\;fluc}})_{\mathrm{r}} \approx \bar{N}\cdot PS(\mathrm{PSF})_{\mathrm{r}} + PS\left(\frac{R}{\sqrt{\mathrm{Gal_{mean}}\otimes  \mathrm{PSF}}}\right)_{\mathrm{r}}.
    \label{eq:PS_obsSBFshort_rad}
\end{equation}

\noindent In Appendix~\ref{sec:appendixApplyMask} we show this mathematical development with more detail, also including the presence of a mask. 

\subsection{Applying a mask}
\label{sec:appMask}

Commonly, during the SBF measurement, masks are applied in order to cover areas of the image that could hinder obtaining the SBF. 
For instance, in the centre of the galaxy either the number of counts could saturate the detector or the profile obtained from a theoretical model could overestimate the light, as in our case.  
Moreover, the mean model ($\mathrm{Gal_{mean}}$) subtracted from the galaxy image is not representative in the pixels with the highest count value (this corresponds to the central pixels in our work), as the mean will always return lower values than the maximum. Thus, in this work all the experiments with mock masked galaxies are performed, at least, 4 pixels away from the centre. 
On the other hand, in the external regions of the galaxy there might be not enough light to obtain the SBF, either because the observed galaxy is too faint or because the theoretical model undervalues the profile. 
Additionally, there might be other saturated or intrusive elements that should be hidden, such as foreground stars, globular clusters, other galaxies, cosmic rays, etc.

These masks are applied by multiplying zero-one (False-True) $n_{\mathrm{pix}} \times n_{\mathrm{pix}}$ images to our synthetic galaxy image ($\mathrm{Gal_{mock}}\cdot \mathrm{Mask} = \mathrm{Gal_{mock\;mask}}$). 
In so doing, Eq.~(\ref{eq:obsGal}) would look:

\begin{equation}
    \mathrm{Gal_{mock}}\cdot \mathrm{Mask} = \left( \left( \mathrm{Gal_{mean}}+\mathrm{Gal_{fluc}}+ \mathrm{Sky} 
    \right)\otimes  \mathrm{PSF} + R \right)\cdot \mathrm{Mask}.
\label{eq:obsGalMask}
\end{equation}

\noindent Applying a similar procedure as in Sect.~\ref{sec:SBFdef}, we find that Eq.~(\ref{eq:PS_obsSBFshort_rad}) with a mask is\footnote{We perform the convolution between the mask and the PSF in the Fourier space with the \texttt{python} function \texttt{scipy.signal.fftconvolve} \citep{2020SciPy-NMeth}. In this case the boundary conditions should consider parameter mode='same' in the borders of the image.}:

\begin{equation}
\begin{split}
PS(\mathrm{Gal_{mock\;fluc\;mask}})_{\mathrm{r}}\approx \\
\bar{N}\cdot \left(PS(\mathrm{PSF})\otimes  PS(\mathrm{Mask})\right)_{\mathrm{r}} +PS\left(\frac{R\cdot \mathrm{Mask}}{\sqrt{\mathrm{Gal_{mean}}\otimes  \mathrm{PSF}}} \right)_{\mathrm{r}}.
 \end{split}
 \label{eq:PS_obsSBFshort_rad_Mask}
\end{equation}

\noindent This expression is commonly found in the literature as $P(k) = P_0\cdot  E(k) + P_1$ \citep[e.g.][]{tonry1990observations}. Here, $P(k)$ corresponds to the fluctuation frame term, $P_0$ is the average flux from the fluctuations to fit and $E(k)$ is the expectation power spectrum (this is, the convolution of the PSF power spectrum and the mask power spectrum). $P_1$ is the constant instrumental noise component (or any other sources of variance that are not convolved with the PSF).

The derivation of the power spectrum is mathematically not completely rigorous, for example, by neglecting the crossed term or altering the operational order of the fluctuation, the PSF and the mask. Nonetheless, the effect of these approximations on the power spectrum is minimal \citep{jensen1998measuring}, as demonstrated in Sect.~\ref{sec:RelianceSBF} and in the Appendix~\ref{sec:appendixApplyMask}.


\subsection{Measuring SBFs from images}

Our goal consists in fitting the fluctuation value from Eq.~(\ref{eq:PS_obsSBFshort_rad_Mask}) and comparing the measured $\bar{N}_{\mathrm{fit}}$ with the known input value $\bar{N}_{\mathrm{real}}$. In the procedure of deriving an SBF from an observation there are some considerations to take into account: on the one hand, for calculating $\mathrm{Gal_{mock\;fluc\;mask}}$ it is necessary to use a model for $\mathrm{Gal_{mean}}$, which serves as a reference to obtain the fluctuations. 
When using real observations, such a mean model is obtained by smoothing the image, by an isophote fitting of the galaxy, with Sersic profiles or other methods (e.g. \citealt{pahre1999detection,cantiello2018next,Carlsten2019}). The sky value is often obtained from an empty region of the observed image and the PSF can be derived from the profile of isolated stars, always considering the particularities of each observation. However, for our experiments, we already know the mean model behind our mock galaxy, the sky count number and the shape of the PSF. So, we can subtract them directly and recover an accurate version of the fluctuation image.
Here, we reiterate that we do not account for fluctuations coming from other sources than the galaxy and the instrumental noise. That is, in our ideal image we neglect globular clusters, background galaxies, foreground stars, etc.  


\subsubsection{Modelling the instrumental noise}

In our work, we already know the non-correlated noise added to the image ($R$), as shown in Eq.~(\ref{eq:Readout}). 
However, for each fitting we assume as unknown any fluctuation present on the image, that is, neither the fluctuation of the stellar population luminosity nor the instrumental noise. 
Therefore, $R$ needs to be calculated differently, so, we propose modelling the instrumental noise with a slight difference with respect to Eq.~(\ref{eq:Readout}), this is, without the term $\mathrm{Gal_{fluc}}$:

\begin{equation}
R_{\mathrm{approx}}=\mathrm{Rand_{Poi}}\left[\mathrm{Gal_{mean}}\otimes  \mathrm{PSF}+\mathrm{Sky} 
\right]-(\mathrm{Gal_{mean}}\otimes  \mathrm{PSF}+\mathrm{Sky}).
\label{eq:ReadoutComp}
\end{equation}

\noindent With this equation we derive a map for $R_{\mathrm{approx}}(x,y)$. This approach could be used both for mock galaxies and for some observations, as far as the mean galaxy, PSF and sky background are known. We note that this work considers Poisson noise, but this approach should be adapted to the nuances of each observation.
 
Modelling independently the instrumental noise leaves Eq.~(\ref{eq:PS_obsSBFshort_rad_Mask}) with only the SBF left to fit. This reduces the uncertainty, as we demonstrate in Sect.~\ref{sec:RelianceSBF} (see Table~\ref{table:relianceSBF_Can}). This approximation is reliable if the $\mathrm{Gal_{fluc}}$ contribution is small enough compared to the rest of the terms. In the current work this is backed up by the third criterion presented in Eq.~(\ref{eq:criteria}), in which we require the contribution of galaxy counts to be 10 times larger than the SBF counts.


\subsubsection{SBF fitting example}
\label{sec:SBFfitting}

In this section we create mock galaxy images using Eq.~(\ref{eq:mockGal}) based on the data from Sect.~\ref{sec:galData}. Then we attempt to recover its SBF value following Eq.~(\ref{eq:PS_obsSBFshort_rad_Mask}), where we already know $\mathrm{Gal_{mean}}$, the PSF, the $\mathrm{Sky}$ and $R_{\mathrm{approx}}$ images. We fit\footnote{Using \texttt{python} function \texttt{scipy.optimize.curve\_fit} \citep{2020SciPy-NMeth} \label{fn:scipy.curve_fit}.} the fluctuation value ($\bar{N}_{\mathrm{fit}}$) and we compare it with the actual input value ($\bar{N}_{\mathrm{real}}$) of $\mathrm{Gal_{mock}}$.

As an example, we take the reference galaxy created in Fig.~\ref{fig:mockGal} (with size of $n_\mathrm{pix}\approx1605$~px) and attempt to fit its SBF value imitating the mask used in \citet{cantiello2018next}. The results are shown in Fig.~\ref{fig:PS}. In the inset panel we show the resulting image of calculating $\mathrm{Gal_{mock\;fluc\;mask}}$ with a centred annular mask of radii $r_1\approx80$~px (with $N(\mathrm{Gal_{mean}}(r_1))\approx21132$~counts) and $r_2\approx332$~px (with $N(\mathrm{Gal_{mean}}(r_2))\approx2492$~counts). We choose these values based on figure 1 and table 2 ($\langle Rad\rangle$ column) of \citet{cantiello2018next}, with an inner radius of $r_1=15$~arcsec and an external radius of $r_2=r_1+\langle Rad\rangle=15+47.2$~arcsec. As our work considers an ideal laboratory for the mock galaxies, no contaminants are present, so we only assume an annular mask. In a real observation, any other light source needs to be taken into account and covered, the $\mathrm{Mask}$ term must be a combination of all of these contributions.

In the main panel of Fig.~\ref{fig:PS} 
we show the logarithm of the radial power spectrum obtained for the different components of Eq.~(\ref{eq:PS_obsSBFshort_rad_Mask}): we show with a cyan line the power spectrum of the normalised (by $\sqrt{\mathrm{Gal_{mean}}\otimes \mathrm{PSF}})_{\mathrm{r}}$) and masked instrumental noise ($PS(R_{\mathrm{approx}}^{\mathrm{norm}})_{\mathrm{r}}=PS(R_{\mathrm{approx}}\cdot \mathrm{Mask}/\sqrt{\mathrm{Gal_{mean}}\otimes \mathrm{PSF}})_{\mathrm{r}}$); with a blue line we show the power spectrum of the ('observed') mock galaxy fluctuation ($PS(\mathrm{Gal_{mock\;fluc\;mask}})_{\mathrm{r}}$); with a red solid line and a green dashed line we show the right part of Eq.~(\ref{eq:PS_obsSBFshort_rad_Mask}) for the real input value of the SBF ($\bar{N}_{\mathrm{real}}$) and for the fitted fluctuation ($\bar{N}_{\mathrm{fit}}$), respectively. As commonly done in the literature, each one of these power spectra has been rescaled by multiplying with the ratio of the PSF loss due to the mask, that is, $PS(\mathrm{PSF})_{\mathrm{r}}/\left(PS(\mathrm{PSF})\otimes  PS(\mathrm{Mask})\right)_{\mathrm{r}}$. In such way, the y-axis of different SBF figures can be compared independently of the mask used. 

The selected range of frequencies where the fitting is performed (between $k_{\mathrm{fit,i}}=75$~px$^{-1}$ to $k_{\mathrm{fit,f}}=400$~px$^{-1}$) is marked with a pale-yellow vertical region. All the examples presented in this work fit the fluctuation between these frequencies (exceptions are mentioned when necessary). 
After numerous tests, in this range we can ensure a proper following of $\mathrm{Gal_{mock\;fluc}}$ shape, without entering too much into the noisy or flat, non-informative power spectrum frequency intervals (at low and high frequencies, respectively). These $k_{\mathrm{fit}}$ values are selected for an image size of $n_{\mathrm{pix}}=1605$~px and a given PSF of $3\times\sigma_{\mathrm{PSF}}=4$~px, when these parameters are changed, it is necessary to adjust the interval of fitting.

Our modelled masked instrumental noise term ($PS(R_{\mathrm{approx}}\cdot \mathrm{Mask}/\sqrt{\mathrm{Gal_{mean}}\otimes \mathrm{PSF}})_{\mathrm{r}}=PS(R_{\mathrm{approx}}^{\mathrm{norm}})_{\mathrm{r}}$, or commonly $P_1$ in the literature) is displayed to show the contrast between it and the mock masked fluctuation ($PS(\mathrm{Gal_{mock\;fluc\;mask}})_{\mathrm{r}}$, as a blue line). In reality, $PS(\mathrm{Gal_{mock\;fluc\;mask}})_{\mathrm{r}}$ would correspond to the observed fluctuation. A significant contrast between the two is key for a reliable measurement.
Then, $PS(\mathrm{Gal_{mock\;fluc\;mask}})_{\mathrm{r}}$ is meant to be compared with the fitted fluctuation (the right part of Eq.~(\ref{eq:PS_obsSBFshort_rad_Mask}), with $\bar{N}=\bar{N}_{\mathrm{fit}}$, as a green dashed line). This would correspond to comparing an observed fluctuation (commonly $P(k)$ in the literature) and its fitting ($P_0\cdot E(k)$), respectively. We also plot input 'real' introduced fluctuation (the right part of Eq.~(\ref{eq:PS_obsSBFshort_rad_Mask}), with $\bar{N}=\bar{N}_{\mathrm{real}}$, as a red line), which is meant to be compared to the green line associated with $\bar{N}_{\mathrm{fit}}$. Even if both are almost identical, we show them because they represent our methodology for evaluating the input against the output values (see next section).

Although not shown in Fig.~\ref{fig:PS}, the PSF is responsible for the shape of the PS. If the PSF is narrower in the physical plane the shape of $PS(\mathrm{Gal_{mock\;fluc\;mask}})_{\mathrm{r}}$ increases its frequency width. It is worth mentioning other features found when changing parameters in Eq.~(\ref{eq:PS_obsSBFshort_rad_Mask}), such as the size of the image, the mask or the sky background. 
The number of pixels ($n_{\mathrm{pix}}$) employed for the SBF measurement fixes the final frequency $k=n_{\mathrm{pix}}/2$, constraining the width of the PS and, therefore, the range of frequencies where the fitting is worth. For instance, an image with a larger $n_{\mathrm{pix}} \times n_{\mathrm{pix}}$ presents a larger range of frequencies where the fitting can be performed. For a given mask and brightness, an image with a larger $n_{\mathrm{pix}} \times n_{\mathrm{pix}}$ value provides a lower power spectrum.
In a different sense, applying a mask reduces the value of the power spectrum of the SBF and the uncorrelated noise; but the contrast between both remains constant. However, the larger the number of masked pixels the less information available for the measurement. And finally, the sky background is responsible for the difference between the value of $PS(R_{\mathrm{approx}}^{\mathrm{norm}})_{\mathrm{r}}$ and the expected value of $PS(\mathrm{Gal_{mock\;fluc\;mask}})_{\mathrm{r}}$ at $k=0$ 
(corresponding to the $\bar{N}$ value). A high sky count value enlarges the effect of $R$, making such contrast lower and increasing the noise effect when fitting.

In Fig.~\ref{fig:PS}, our known input fluctuation was $\bar{N}_{\mathrm{real}}=\bar{N}_{\mathrm{ref}}=22.59$~counts or $\bar{m}_{\mathrm{ref}}=30.64$~mag, while the fitted result is $\bar{N}_{\mathrm{fit}}=22.47$~counts or $\bar{m}_{\mathrm{fit}}=30.646$~mag. Additionally, we find clear similarities in the shape of the power spectrum when comparing the results of Fig.~\ref{fig:PS} with the results of figure 1 in \citet{cantiello2018next} for the galaxy selected in Sect.~\ref{sec:galData}.


\subsubsection{Reliability of the SBF measurement}
\label{sec:SBFreliance}

We evaluate the reliability of the SBF estimate with two parameters, the relative error and the relative standard deviation of the fitting:

\begin{equation}
\varepsilon_{\mathrm{rel}}=\frac{|\bar{N}_{\mathrm{real}}-\bar{N}_{\mathrm{fit}}|}{\bar{N}_{\mathrm{real}}},
\end{equation}
\begin{equation}
\sigma_{\mathrm{fit}}=\frac{\sigma_{\mathrm{cov}}}{\bar{N}_{\mathrm{fit}}},
\end{equation}

\noindent where $\sigma_{\mathrm{cov}}$ is the standard deviation in counts returned from the computational fitting\footref{fn:scipy.curve_fit}.

\begin{figure*}
\centering
\includegraphics[width=\textwidth]{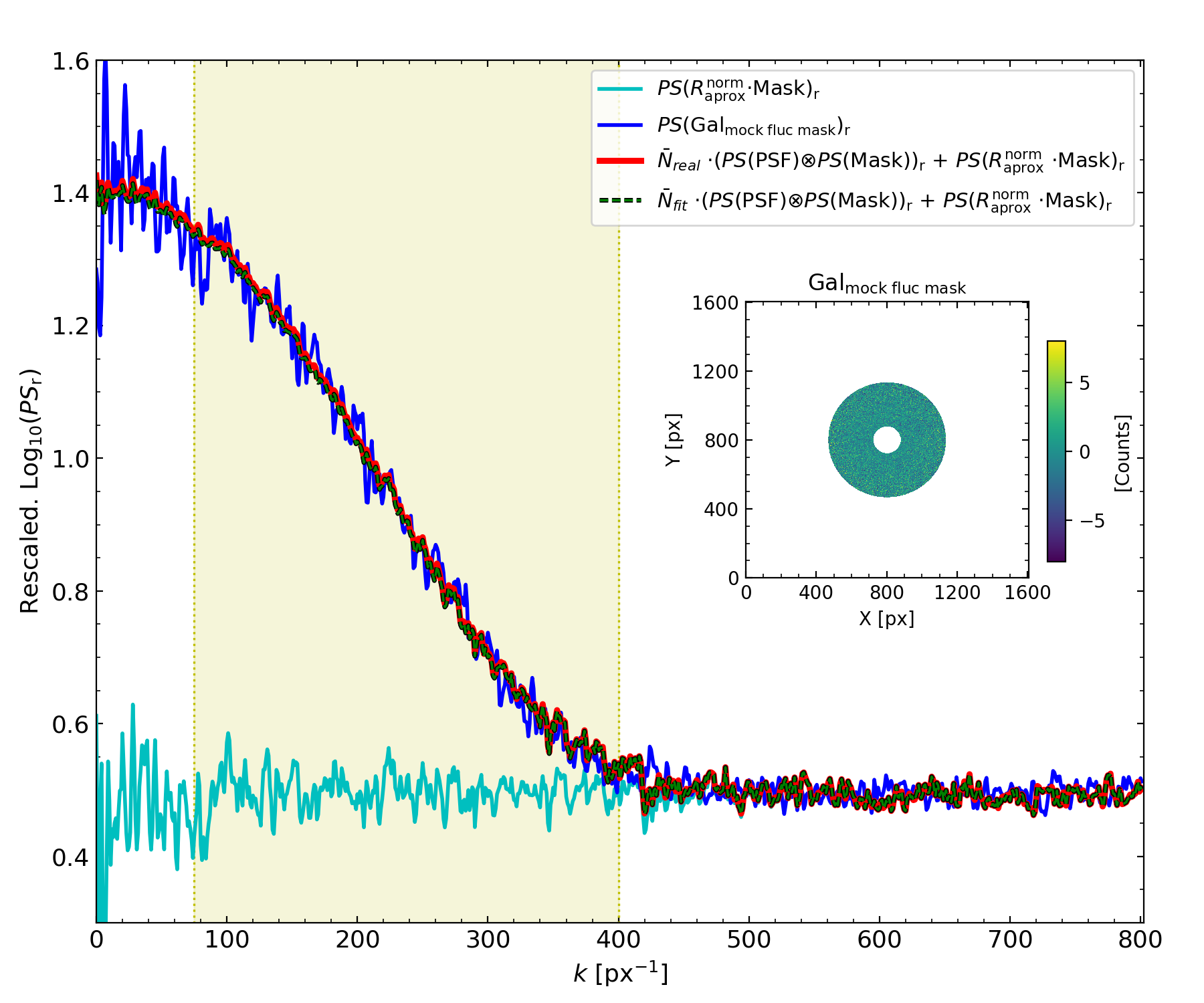}
\caption{Rescaled (multiplied by $PS(\mathrm{PSF})_{\mathrm{r}}/\left(PS(\mathrm{PSF})\otimes  PS(\mathrm{Mask})\right)_{\mathrm{r}}$) logarithm of the radial power spectrum profiles: the masked instrumental noise term ($PS(R_{\mathrm{approx}}\cdot \mathrm{Mask}/\sqrt{\mathrm{Gal_{mean}}\otimes \mathrm{PSF}})_{\mathrm{r}}=PS(R_{\mathrm{approx}}^{\mathrm{norm}})_{\mathrm{r}}$) is the cyan line; the observed masked fluctuation ($PS(\mathrm{Gal_{mock\;fluc\;mask}})_{\mathrm{r}}$) is the blue line; the right part of Eq.~(\ref{eq:PS_obsSBFshort_rad_Mask}) with the real input value of the SBF ($\bar{N}_{\mathrm{real}}$) is the red line, and with the fitted fluctuation ($\bar{N}_{\mathrm{fit}}$) is the dashed green line. The selected range of frequencies where the fitting is performed happens between $k_{\mathrm{fit,i}}=75$~px$^{-1}$ and  $k_{\mathrm{fit,f}}=400$~px$^{-1}$, marked with a pale yellow vertical region.
In this example, the input SBF value is $\bar{N}_{\mathrm{real}}=\bar{N}_{\mathrm{ref}}=22.59$ and the fitted SBF is $\bar{N}_{\mathrm{fit}}=22.47$. The embed image is the observed fluctuation $\mathrm{Gal_{mock\;fluc\;mask}}$ with a circular mask of $r_1\approx80$~px and $r_2\approx332$~px applied.}
\label{fig:PS}
\end{figure*}

The relative standard deviation ($\sigma_{\mathrm{fit}}$) estimates the quality of the least squares fitting, while the relative error ($\varepsilon_{\mathrm{rel}}$) refers to the accuracy of the $\bar{N}_{\mathrm{fit}}$ result. Thus, the experiment performed in Fig.~\ref{fig:PS} returns a relative error of $\varepsilon_{\mathrm{rel}}=0.53$\% and a relative standard deviation of $\sigma_{\mathrm{fit}}=0.18$\%.

\subsection{Criteria}
\label{sec:Criteria}

For this work we take as a valid SBF estimation those measurements with a 
$\sigma_{\mathrm{fit}}$ and a $\varepsilon_{\mathrm{rel}}$ lower than a 10\%.
In addition, we only consider non-masked pixels of the galaxy ($\mathrm{Gal_{mock\;mask}}$) with values larger than 10 times the input SBF in counts. We check if every pixel fulfils this criterion (stated in the third line of the equation below), otherwise the measurement is not performed. 
This condition assures a Gaussian probability distribution of the integrated light among the pixels, as approximated from the galaxy modelling of \citep{tonry1988new, cervino2008surface}. For lower count values, the Gaussian condition is not assured \citep{CL06}, so the traditional SBF modelling is not necessarily physically correct. This is discussed further in Sect.~\ref{sec:biasedSBF}. 

In summary, our adopted criteria are:

\begin{equation} \mathrm{Criteria}
    \begin{cases} 
      & \varepsilon_{\mathrm{rel}}<10\%, \\ 
      & \sigma_{\mathrm{fit}}<10\%, \\
      & N(\mathrm{Gal_{mock\;mask}}(x,y)) - N(\mathrm{Sky_{mask}}(x,y))\ge 10\cdot \bar{N}.
\end{cases}
\label{eq:criteria}
\end{equation}

\noindent Throughout this work, we take as known the luminosity distribution in each pixel, on the contrary, in real observations the SBF value is unknown \textit{a priori}. 
Therefore, in observations we can calculate $\sigma_{\mathrm{fit}}$, but we are not able to calculate $\varepsilon_{\mathrm{rel}}$, as $\bar{N}_{\mathrm{real}}$ is unknown. 
This is one reason why the modelling presented in this work is a very useful tool: we select an input value of $\bar{N}_{\mathrm{real}}$ with which we can foresee if an observation will return a reliable result. Moreover, the relative standard deviation of the fitting $\sigma_{\mathrm{fit}}$ is not representative of the accuracy of the returned SBF, but only the quality of the fit. Thus, for the purposes of this work $\varepsilon_{\mathrm{rel}}$ serves as a guide for obtaining a trustworthy SBF under different conditions. 

In observations, other consequence of being oblivious to the real SBF value is that the third criterion cannot be guaranteed before performing the measurement. However, we suggest checking the condition $N(\mathrm{Gal_{mock\;mask}}) - N(\mathrm{Sky_{mask}})\ge 10\cdot \bar{N}_{\mathrm{obs}}$ on every non-masked pixel after the fitting, to assure the reliability of the observed SBF. We are aware that the parameters we vary are not necessarily independent (e.g. they depend on distance or stellar population properties), but we consider here that these parameters are unrelated and selected \textit{ad hoc}, as a way to explore several scenarios. 

Note that for the relative error and the relative standard deviation of the fitting, the sky is implicitly embedded. As we show in Sect.~\ref{sec:ResultsRestOfParams}, the uncertainty increases for large sky background values, due to a smaller contrast between $P_0$ and $P_1$. The third condition of Eq.~\ref{eq:criteria} already includes the sky contribution to the image. The latter criterion should only consider the brightness of the galaxy stellar population itself, without any other contribution.

Besides the criteria of Eq.~(\ref{eq:criteria}), in real observations the sky itself has its own uncertainty and an offset in its estimation can distort the SBF measurement. Similarly, a bad modelling of the mean reference galaxy or the PSF lead to an offset in the SBF. 
These cases are quoted in more detail in Sect.~\ref{sec:otherBias}. 


\section{Results}
\label{sec:Results}

In order to evaluate the SBF estimation procedure we create mock galaxies for a wide range of conditions. We compare the $\bar{N}_{\mathrm{fit}}$ obtained from the mock galaxies with our known input $\bar{N}_{\mathrm{real}}$ and we study the uncertainty of the fitting, as explained in Sects.~\ref{sec:SBFfitting} and \ref{sec:SBFreliance}. In this way we can explore the parameter space looking for those galaxies where the retrieved SBF is reliable based on the criteria shown in Sect.~\ref{sec:Criteria}. 
The galaxy parameters we choose to vary are: the magnitude of the galaxy, the magnitude of the fluctuation, the effective radius, the PSF, the Sersic index and the exposure time. For these variations of the galaxy parameters we study the fitting results when applying masks of different sizes. As in an observational setup, for each iteration the image remains unchanged when different masks are applied. We emphasise that the mean galaxy model, the sky and the PSF are assumed to be known in the fitting process. Therefore, the uncertainties addressed in this section are representative of the variability of the system due to the intrinsic stochasticity of the stellar population luminosity distribution and the instrumental noise. 

With these experiments, we aim to evaluate the uncertainty of the SBF via Monte Carlo simulations. To do so, we create $n_{\mathrm{sim}}=50$ mock galaxies each time a parameter is varied\footnote{Note that the mock galaxies require two probability distributions, the population SBF (modelled as a Gaussian) and the instrumental noise (modelled as a Poissonian).} except the mask. For each mock galaxy we compute $\varepsilon_{\mathrm{rel}}$ and $\sigma_{\mathrm{fit}}$. From the resulting distribution of those parameters we choose the higher 90\% percentile ($\varepsilon_{\mathrm{rel},90\%},\sigma_{\mathrm{fit},90\%}$), while also
satisfying the criteria of Eq.~(\ref{eq:criteria}). In a distribution of 50 simulations, this corresponds to the 45$^{\mathrm{th}}$ highest value. This provides a conservative estimate of how unfavourable the obtained SBF would be. Additionally, we calculate the relative 90\% width of the $\bar{N}_{\mathrm{fit}}$ distribution results. This is calculated as the subtraction of the 95\% and 5\% percentile values, divided by the mean value of the $\bar{N}_{\mathrm{fit}}$ distribution $\langle \bar{N}_{\mathrm{fit}} \rangle$. In our case those 5\% and 95\% percentile values are the 2$^{\mathrm{nd}}$ lowest and the 48$^{\mathrm{th}}$ highest $\bar{N}_{\mathrm{fit}}$ values found after sorting the 50 simulations results. Thus, $\Delta_{90\%}$ represents the width covered by the 90\% of the distribution of $\bar{N}_{\mathrm{fit}}$ values, giving a measurement of the precision of the fitting results: 

\begin{equation}
\Delta_{90\%}=\frac{\bar{N}_{\mathrm{fit,95th\%}}-\bar{N}_{\mathrm{fit,5th\%}}}{\langle\bar{N}_{\mathrm{fit}}\rangle}.
    \label{eq:Delta90}
\end{equation}

In summary, the key parameters for the analysis of this work are: $\varepsilon_{\mathrm{rel},90\%}$,  $\sigma_{\mathrm{fit},90\%}$ and $\Delta_{90\%}$, as introduced above. All the following figures are calculated with the previous procedure, making use of the distribution of results obtained from the mock galaxy simulations, except for those in Sect.~\ref{sec:ResultsRestOfParams}, where further explanations are provided. Figure~\ref{fig:flowchart} is an illustrative flowchart of the procedure explained in this section.


\subsection{Parameter space: Masks}
\label{sec:ParamSpaceMasks}

The calculation of SBFs is performed using masks of different sizes. Having masked point sources or bad pixels, the SBF is then commonly measured within an annulus (of properly selected width and eccentricity). This annular mask permits hiding regions of the galaxy that negatively affect the SBF measurement, as explained in Sect.~\ref{sec:appMask}. Also, in order to study the possible detection of SBF gradients we require annuli of different width and radius.
Thus, we study the SBF fitting for all the varying combinations of annular rings defined by an internal and an external radius ($r_1,r_2$), with circular shape. We move from 4 pixels (avoiding the overestimation of the centre) to $R_{\mathrm{eff}}$ (as representative of a region of the galaxy with proper signal-to-noise), with a step of $\Delta r=20$~px. The rest of parameters are taken as described in Sect.~\ref{sec:galData}. We perform 50 Monte Carlo simulations of our mock galaxy, then, each one of these realisations is analysed for every ($r_1,r_2$) pair.

\begin{figure*}
\centering
\includegraphics[width=\textwidth]{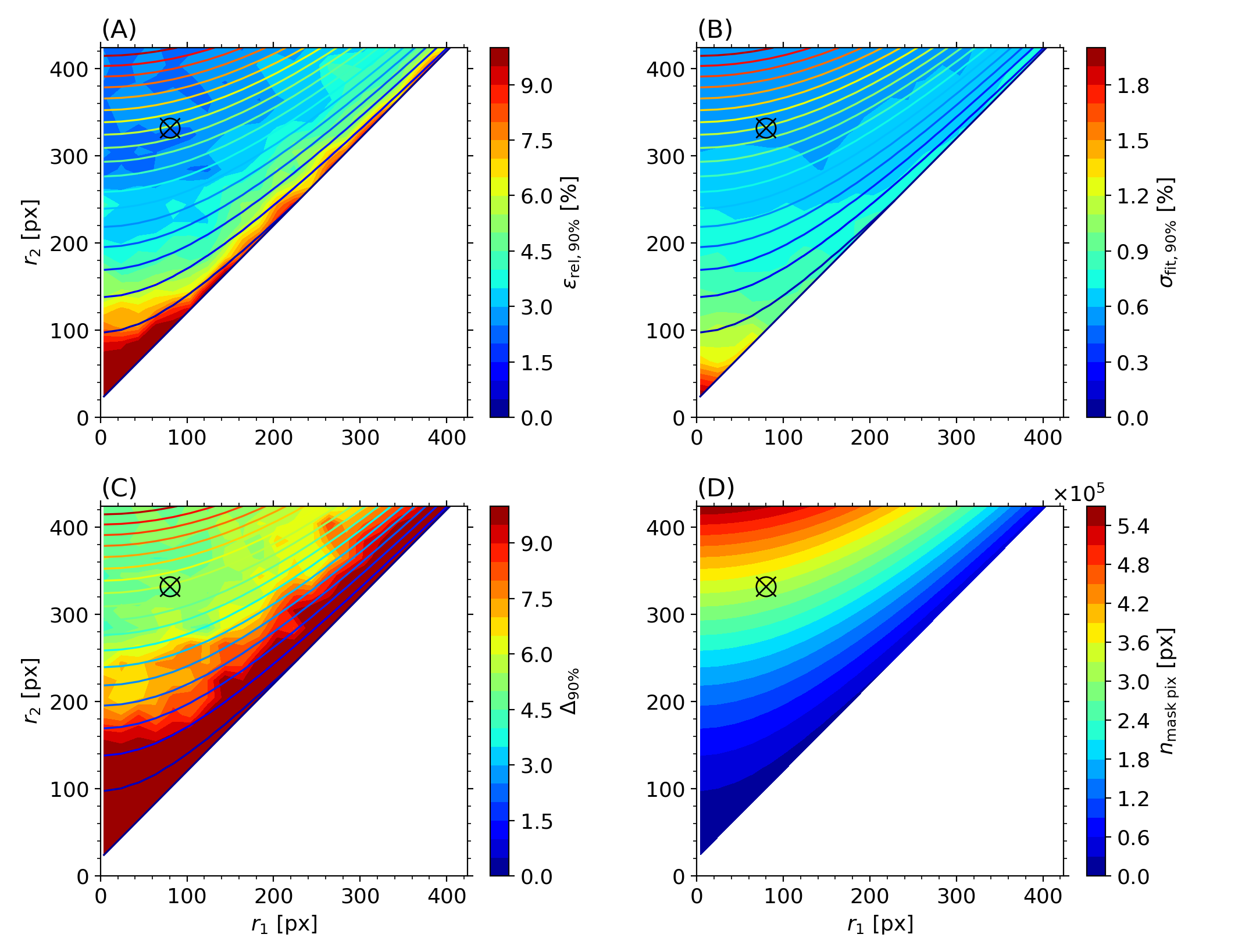}
\caption{Colour maps for the reference galaxy when applying different masks ($r_1,r_2$): the 90\% percentile of the relative error in panel (A), up to $\varepsilon_{\mathrm{rel},90\%}\ge10\%$; the 90\% percentile of the relative standard deviation in panel (B), up to $\sigma_{\mathrm{fit},90\%}\le2\%$; the relative 90\% width of the $\bar{N}_{\mathrm{fit}}$ distribution in panel (C), up to $\Delta_{90\%}\le10\%$; and the number of pixels within each one of the masks ($n_{\mathrm{mask\;pix}}$) in panel (D). The contour lines found over the map in panels (A), (B) and (C) correspond to the number of pixels shown in panel (D). The black circled-cross signals the reference galaxy presented in Sect.~\ref{sec:galData} with the mask applied in its source reference (\citet{cantiello2018next}; $r_1\approx80$~px, $r_2\approx332$~px).}
\label{fig:maps_r1_r2_opt}
\end{figure*}

In Fig.~\ref{fig:maps_r1_r2_opt} we show the colour maps of the 90\% percentile values of the relative error ($\varepsilon_{\mathrm{rel},90\%}$) in panel (A), 90\% percentile values of the relative standard deviation ($\sigma_{\mathrm{fit},90\%}$) in panel (B), the relative 90\% width of the $\bar{N}_{\mathrm{fit}}$ distribution  ($\Delta_{90\%}$) in panel (C) and the number of pixels within a mask ($n_{\mathrm{mask\;pix}}$) in panel (D), all of them dependent on the masks size determined by ($r_1,r_2$). The contour lines found over the map in panels (A, B, C) correspond to the number of pixels shown in panel (D). We mark as a black circled-cross the reference galaxy presented in Sect.~\ref{sec:galData} with the mask applied in its source reference ($r_1\approx80$~px, $r_2\approx332$~px; \citet{cantiello2018next}).

Comparing panels (A, B, C) with (D) we find a relation between the number of pixels of the mask and $\varepsilon_{\mathrm{rel},90\%}$, $\sigma_{\mathrm{fit},90\%}$ and $\Delta_{90\%}$. This behaviour is especially clear for the relative error. Comparing those panels with (D) there are larger errors in regions of low numbers of pixels, that is, in very thin annuli or in small masks. For instance, any mask with 60000 pixels or fewer will most likely produce an error higher than $\varepsilon_{\mathrm{rel},90\%}>9\;\%$. On the other hand, panel (B) shows that the relative standard deviation of the fitting fulfils the criterion for every mask, with values lower than $\sigma_{\mathrm{fit},90\%}<2\;\%$. In panel (C) we find  $\Delta_{90\%}<10\%$ for masks with a number of pixels approximately larger than 120000. Among these three parameters (panels A, B, C), we find that $\varepsilon_{\mathrm{rel},90\%}$ is about a factor 10 larger than $\sigma_{\mathrm{fit},90\%}$, then, $\Delta_{90\%}$ is about 2 to 4 times larger than $\varepsilon_{\mathrm{rel},90\%}$. 

From the results of this section we can draw some general notions:

\begin{itemize}

    \item First, the relative error ($\varepsilon_{\mathrm{rel},90\%}$), the relative standard deviation of the fitting ($\sigma_{\mathrm{fit},90\%}$) and the relative 90\% width of the $\bar{N}_{\mathrm{fit}}$ distribution ($\Delta_{90\%}$) are tightly related to the number of pixels within the mask ($n_{\mathrm{mask\;pix}}$).
    \item Second, the relative error ($\varepsilon_{\mathrm{rel},90\%}$) estimates the accuracy of the result and is a more restrictive constraint than the relative standard deviation ($\sigma_{\mathrm{fit},90\%}$). The relative standard deviation just measures the quality of the least squares fit, but is not representative of how close the fitted SBF ($\bar{N}_{\mathrm{fit}}$) is to the real SBF value ($\bar{N}_{\mathrm{real}}$).
    \item Third, the relative 90\% width of the $\bar{N}_{\mathrm{fit}}$ distribution ($\Delta_{90\%}$) is the most restrictive parameter. Although, it does not provide information about the accuracy for finding a $\bar{N}_{\mathrm{fit}}$ value similar to $\bar{N}_{\mathrm{real}}$, it provides a much more conservative uncertainty estimate than $\varepsilon_{\mathrm{rel},90\%}$ and $\sigma_{\mathrm{fit},90\%}$.
    
\end{itemize}

Since we are interested in the accuracy in the SBF estimate, henceforth we use as a proxy the number of pixels ($n_{\mathrm{mask\;pix}}$) against the relative error ($\varepsilon_{\mathrm{rel},90\%}$) in our experiments. Moreover, studying  $\varepsilon_{\mathrm{rel},90\%}$ takes advantage of creating mock galaxies, as in this work. The other parameters, $\sigma_{\mathrm{fit},90\%}$ and $\Delta_{90\%}$, are presented only in certain cases of interest. Additionally, we would like to propose that in real observations it is always possible to build mock galaxies as in Sect.~\ref{sec:mockGalaxyCreation} using the retrieved (observationally fitted) SBF, then calculate $\sigma_{\mathrm{fit},90\%}$ and $\Delta_{90\%}$ with respect to our initially fitted $\bar{N}_{\mathrm{obs}}$.

\subsection{Parameter space: Galaxy brightness and SBF magnitude}

Once our reference galaxy has been studied through different mask sizes, we study the parameter space when varying the brightness of the galaxy and its fluctuation contribution. We use as a reference the galaxy described in Sect.~\ref{sec:galData} and we vary both its apparent magnitude from $m = 3$~mag to $m=14$~mag, in intervals of $\Delta(m)=0.5$~mag, and the SBF magnitude from $\bar{m}=25$~mag to $\bar{m}=37$~mag, with $\Delta(\bar{m})=0.5$~mag. We recognise that some of these values might be unrealistic, but we keep such ranges for illustrative purposes and exploring the parameter space. 
As justified in the previous section, for these experiments we study the behaviour of the relative error only.

\begin{figure*}
\centering
\includegraphics[width=\textwidth]{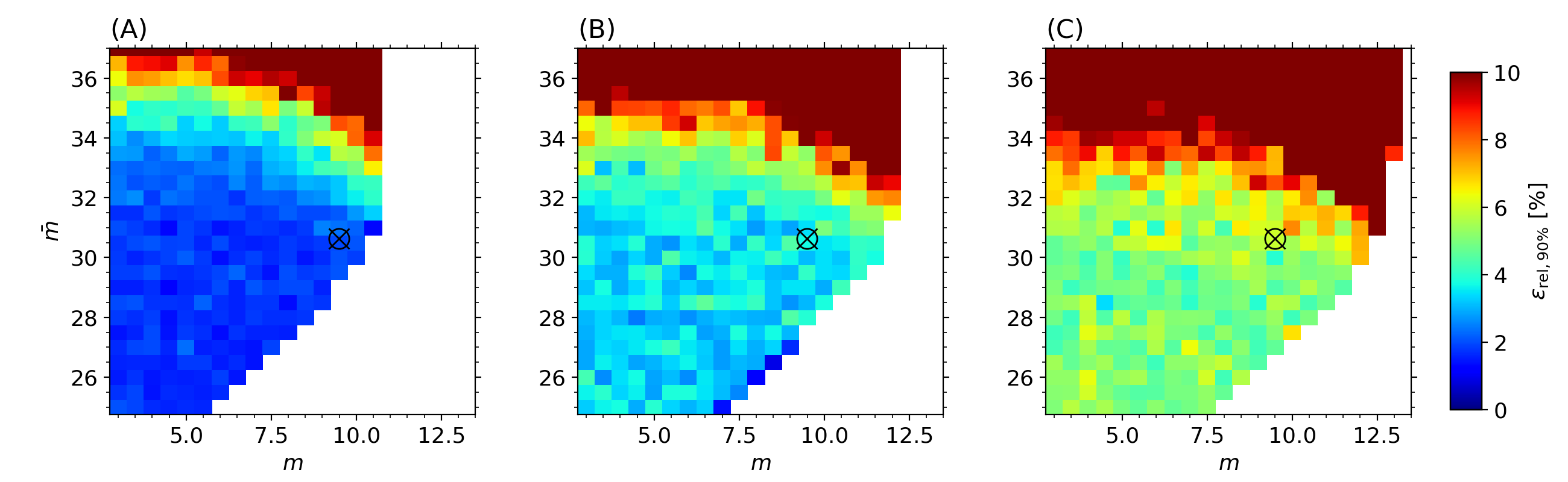}
\caption{Colour map of the relative error ($\varepsilon_{\mathrm{rel},90\%}$) obtained after fitting the SBF for mock galaxies of different brightness and SBF magnitudes ($m,\bar{m}$). The fitting was performed for three centred annular masks, with fixed $r_1=4$~px, then $r_2=R_{\mathrm{eff}}$ for panel (A), $r_2=R_{\mathrm{eff}}/2$ (B), and $r_2=R_{\mathrm{eff}}/3$ (C). The black circled cross signals the reference galaxy presented in Sect.~\ref{sec:galData} ($m=9.493,\bar{m}=30.64$).}
\label{fig:SearchingOpt_varMask}
\end{figure*}

Again, we perform 50 galaxy simulations for each case and we obtain the 90\% percentile of the distribution of $\varepsilon_{\mathrm{rel},90\%}$.
In Fig.~\ref{fig:SearchingOpt_varMask} we show the relative error ($\varepsilon_{\mathrm{rel},90\%}$) for each ($m,\bar{m}$) combination. The SBF estimate is obtained for three centred annular masks: we fix $r_1=4$~px, then we select $r_2=R_{\mathrm{eff}}$ for panel (A), $r_2=R_{\mathrm{eff}}/2$ (B) and $r_2=R_{\mathrm{eff}}/3$ (C). Also, in Fig.~\ref{fig:SearchingOpt_varMask} we mark with a black circled-cross the reference galaxy based on Sect.~\ref{sec:galData} data ($m=9.493,\bar{m}=30.64$). 

All cases where the error is higher than the criterion, $\varepsilon_{\mathrm{rel},90\%}\ge10$, are shown with the same red colour as $\varepsilon_{\mathrm{rel},90\%}=10$. In general, high brightness and low fluctuation magnitudes, that is, increasing the count number of both, returns lower relative errors. 
When increasing $\bar{m}$ the relative error ascends up to $\varepsilon_{\mathrm{rel},90\%}=10$. We find errors higher than our criterion for $\bar{m}\approx34-37$~mag, depending on the mask applied. When the fluctuation luminosity is larger than the galaxy luminosity itself, our mock galaxy is no longer physically realistic. This happens for lower fluctuation magnitudes than $\bar{m}\approx24$. 

The empty region found in the right and bottom-right corner represents galaxies where the third criterion of Eq.~\ref{eq:criteria}, $N(\mathrm{Gal_{mock\;mask}}) - N(\mathrm{Sky_{mask}})\ge 10\cdot \bar{N}$, is not fulfilled, therefore the SBF is not obtained. In our scenario, total galaxy magnitudes higher than $m\approx10.5-13.5$ (depending on the mask) do not fulfil the condition and neither do cases where $\bar{m}-m\gtrsim19$. Such a region is larger for larger masks, as we are considering regions of the galaxy with lower brightness. 
We observe that using smaller masks increases the relative error. As we have shown in Sect.~\ref{sec:ParamSpaceMasks}, the larger the number of pixels within the mask, the lower the relative error. 

It is worth highlighting the differences when studying a galaxy through the integrated surface luminosity, or when studying it through surface brightness fluctuations. 
The first case depends on the luminosity profile, so a few pixels with high luminous stars can dominate the flux. In the second case, the SBF strongly depends on the number of pixels with a given fluctuation, since the SBF is defined as the luminosity normalised by variance.   
Additionally, the SBF range of possible values is more restricted, unlike the large variation of the integrated surface luminosity. 
The integrated surface luminosity and the SBF complement each other and a combination of both provides more constrained information about the galaxy \citep{rodriguez2021surface}.

\subsection{Varying other parameters}
\label{sec:ResultsRestOfParams}

After studying general SBF measurements related to the brightness of the galaxy and the size of the masks applied (their number of pixels), we proceed to analyse other parameters such as: the PSF size ($\sigma_{\mathrm{PSF}}$), the exposure time ($t_{\mathrm{exp}}$), the sky number of counts ($N_{\mathrm{Sky}}$), the Sersic index ($n$) and the effective radius ($R_{\mathrm{eff}}$). Again, we use as a reference the galaxy described in Sect.~\ref{sec:galData}, then, we vary a certain parameter while fixing the rest. In the figures of this subsection we show the reference case as black dashed line. 
For every case we show the relative error ($\varepsilon_{\mathrm{rel},90\%}$) against the number of pixels within a mask ($n_{\mathrm{mask\;pix}}$). In Sect.~\ref{sec:ParamSpaceMasks} we demonstrated how the uncertainty is similar in masks with the same number of pixels.
Therefore, we make use $n_{\mathrm{mask\;pix}}$ as a proxy for our analysis, with the procedure we explain below.

First, we select which parameters are held constant based on our reference galaxy and which one is varied. Second, for this chosen configuration we perform simulations for every mask combination ($r_1, r_2$), as we did in Fig.~\ref{fig:maps_r1_r2_opt}. Third, the results are grouped into \textit{subsets} with similar number of pixels. 
The \textit{subset} size is determined by the maximum number of pixels, which corresponds to a mask with $r_1=4$ and $r_2=R_{\mathrm{eff}}$, divided in 10 partitions. 
These \textit{subsets} are analogous to each one of the regions displayed with different colours in panel (D) of Fig.~\ref{fig:maps_r1_r2_opt}  (although this figure is divided in 20 \textit{subsets}, instead of 10).  
Fourth, we repeat the process of the previous two points (second and third) using 50 Monte Carlo simulations. Fifth, from this group of simulations we combine every \textit{subset} with similar number of pixels. This is, every \textit{subset} is now fifty times larger. Sixth, from each one of these \textit{subsets} we calculate the 90th percentile of the relative error. With this method we ensure that the percentile is derived from masks with a similar number of pixels, and that it is applied simultaneously to the 50 simulations.

We note that, as expected, in all our experiments the relative error generally decreases when the number of pixels increases. The curves do not descend smoothly due to the finite number of simulations that we have performed and their random nature. However, even our limited number of simulations is sufficient to address the general behaviour shown in the plots.

\subsubsection{PSF size}

\begin{figure}
\centering
\includegraphics[width=0.5\textwidth]{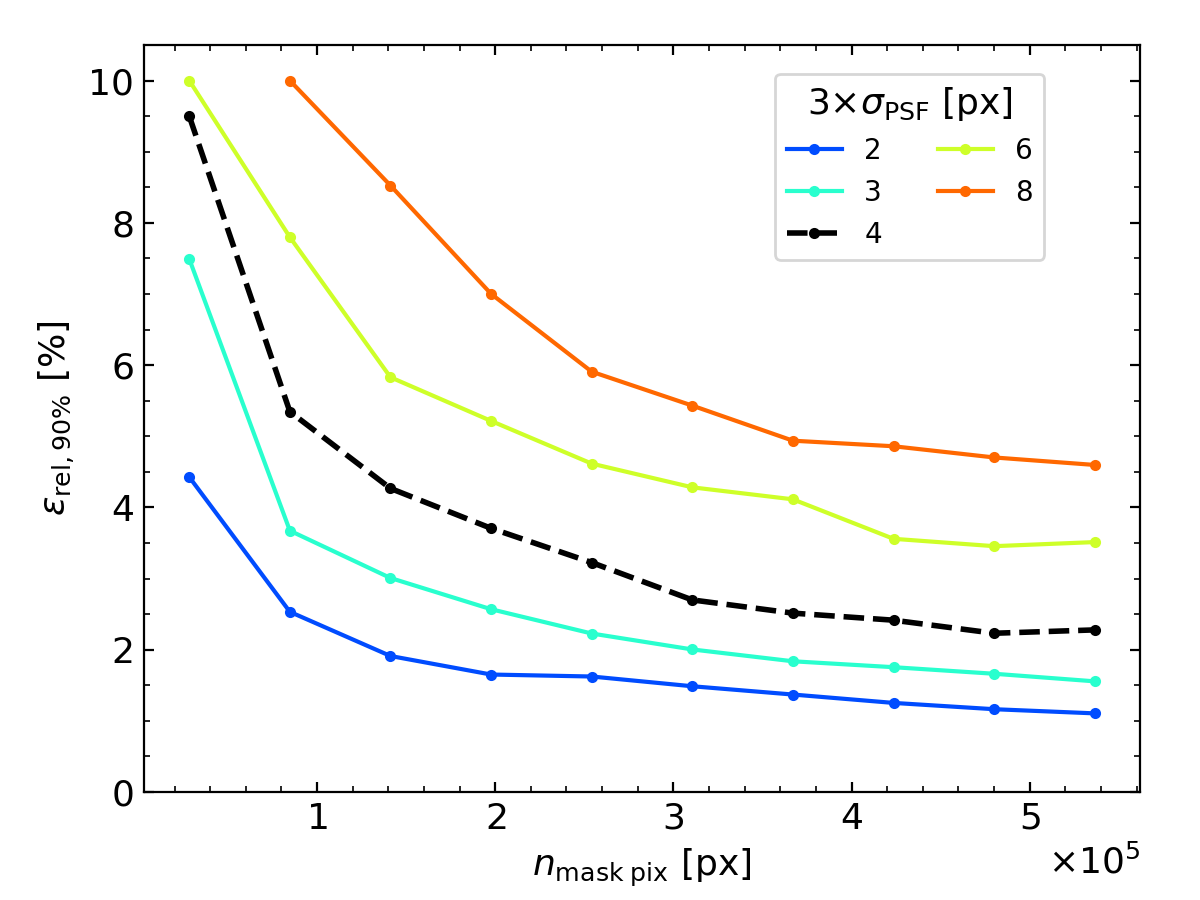}
\caption{Upper 90\% relative error against the number of pixels within the mask when varying the PSF pixel size: $3\times\sigma_{\mathrm{PSF}}=2$ is in blue, $3\times\sigma_{\mathrm{PSF}}=3$ in cyan, $3\times\sigma_{\mathrm{PSF}}=4$ in dashed black, $3\times\sigma_{\mathrm{PSF}}=6$ in lime yellow, and $3\times\sigma_{\mathrm{PSF}}=8$ in orange.}
\label{fig:PSF_ErrVSnpix}
\end{figure}

In Fig.~\ref{fig:PSF_ErrVSnpix} we show how the relative error changes when varying the PSF width, using $3\times\sigma_{\mathrm{PSF}}=2, 3, 4, 6$ and 8~px. The power spectrum shrinks in the frequency domain  ($k-\mathrm{axis}$) when the PSF is wider in the physical domain ($\mathrm{px}-\mathrm{axis}$). In order for the comparison to be fair, we adjust the range of frequencies used for the fitting with respect to the point spread function of reference ($\sigma_{\mathrm{PSF,ref}}$) from Sect.~\ref{sec:galData} and the fitting frequencies of reference ($k_{\mathrm{fit,ref}}$) from Sect.~\ref{sec:SBFfitting}. The new fitting frequencies are calculated as $k_{\mathrm{fit}}=k_{\mathrm{fit,ref}}/(\sigma_{\mathrm{PSF}}/\sigma_{\mathrm{PSF,ref}})$.

The $\varepsilon_{\mathrm{rel},90\%}$ worsens as the physical width of the PSF increases: the power spectrum is narrower, the contribution of the noise is enlarged and there is less relevant information to fit. Hence, in the physical plane the smaller the PSF the better, nonetheless, we remind that the presence of the PSF is required to disentangle the instrumental noise from the spatially correlated SBF signal.

\subsubsection{Exposure time}

\begin{figure}
\centering
\includegraphics[width=0.5\textwidth]{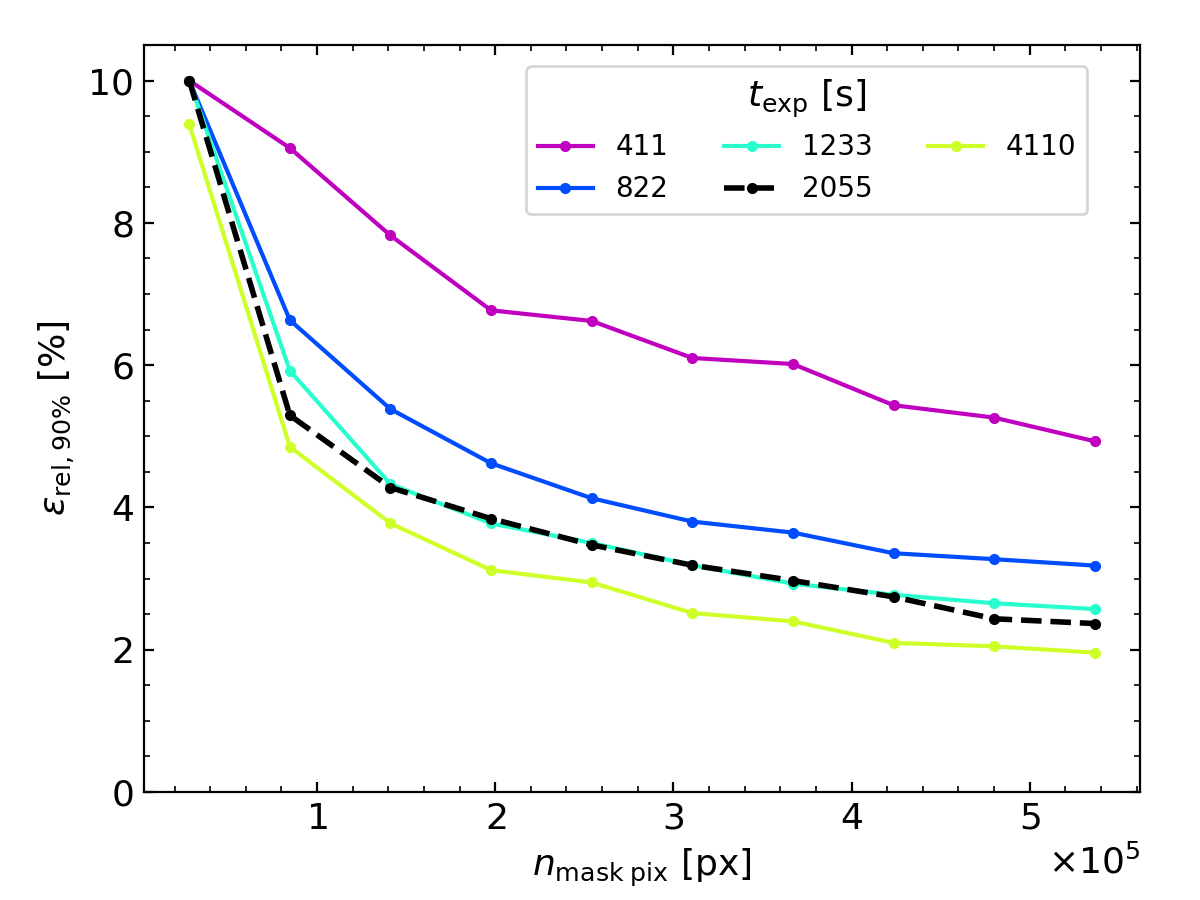}
\caption{Upper 90\% relative error against the number of pixels within the mask when varying the exposure time (in seconds) of the mock galaxy: $t_{\mathrm{exp}}=411$ is in magenta, $t_{\mathrm{exp}}=822$ in blue, $t_{\mathrm{exp}}=1233$ in cyan, 
$t_{\mathrm{exp}}=2055$ in dashed black, and $t_{\mathrm{exp}}=4110$ in lime yellow.}
\label{fig:TempExp_ErrVSnpix}
\end{figure}

In Fig.~\ref{fig:TempExp_ErrVSnpix} we show $\varepsilon_{\mathrm{rel},90\%}$ versus $n_{\mathrm{mask\;pix}}$ while changing the exposure time $t_{\mathrm{exp}}=411, 822, 1233, 2055$ and 4110 seconds.
This is equivalent to changing the number of counts of $N_{\mathrm{Reff}}$, $\bar{N}$ and $N_{\mathrm{Sky}}$. 
We find that very short exposure times lead to larger relative errors, as the number of counts of the image is not high enough compared to the instrumental noise source. 

\subsubsection{Sky background}
\label{sec:ResultsSky}

\begin{figure}
\centering
\includegraphics[width=0.5\textwidth]{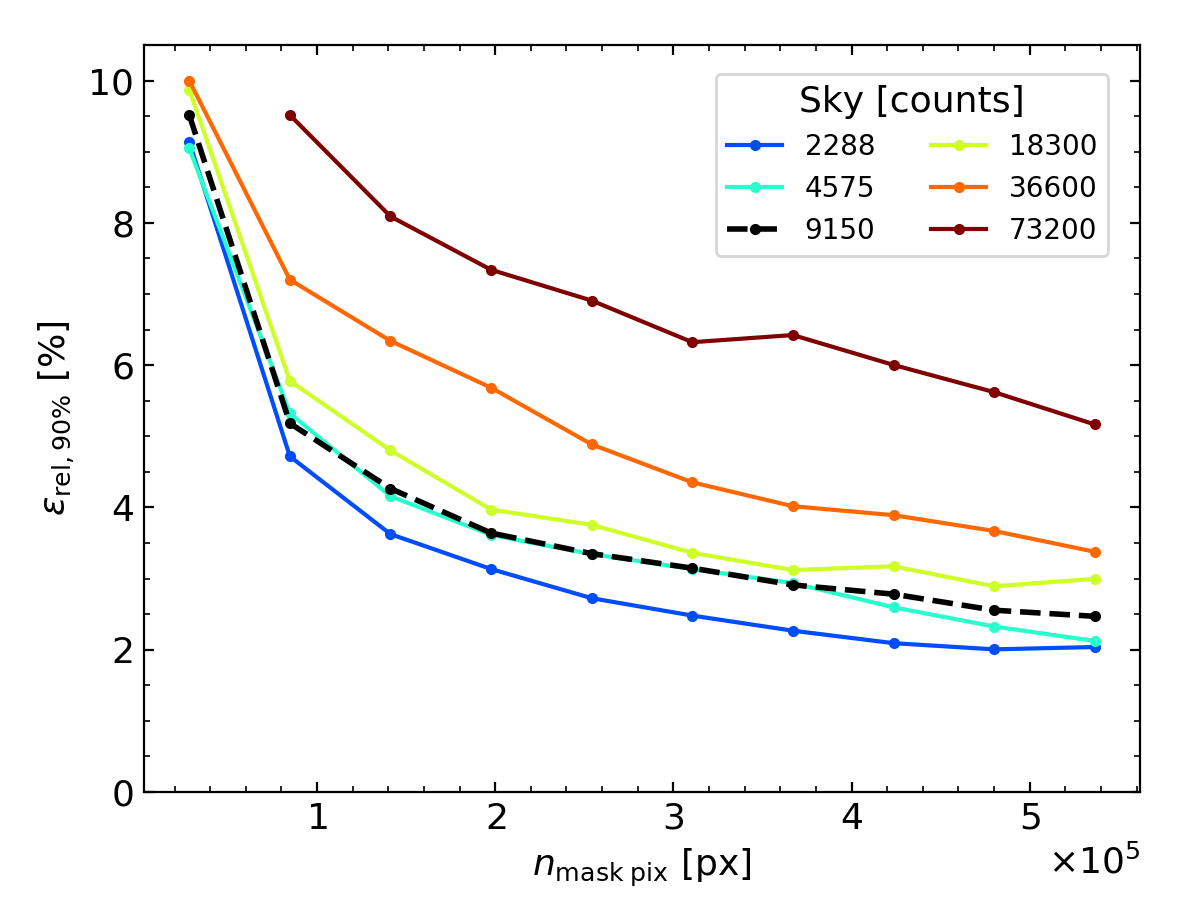}
\caption{Upper 90\% relative error against the number of pixels within the mask when varying the sky background count number: $N_{\mathrm{Sky}}=2288$ is in blue, 4575 in cyan, 9150 in dashed black, 18300 in lime yellow, 36600 in orange, and 73200 in brown.}
\label{fig:Sky_ErrVSnpix}
\end{figure}

In Fig.~\ref{fig:Sky_ErrVSnpix} we show $\varepsilon_{\mathrm{rel},90\%}$ versus $n_{\mathrm{mask\;pix}}$ while changing the sky background of the image ($N_{\mathrm{Sky}}$), which is similar to varying the S/N of the image. 
We show the results for a lower and higher sky count with respect to Sect.~\ref{sec:galData}: Sky=2288, 4575, 9150, 18300, 36600 and 73200~counts.

For a given galaxy flux ($\mathrm{Gal}_\mathrm{mean,fluc}$), we find that decreasing the sky background reduces the relative error, while increasing the sky worsens the SBF retrieval, up to a limit where the criteria of this work are not fulfilled (around $\mathrm{Sky}\approx 120000$~counts). These results are similar to those found in Fig.~\ref{fig:SearchingOpt_varMask}, where instead of varying the sky, we vary the luminosity of the galaxy (both $m$ and $\bar{m}$, for a fixed sky value).  
The instrumental noise increases with higher sky values. If there is not enough contrast between the correlated noise (the SBF with the PSF or, traditionally, $P_0$) and the uncorrelated noise (instrumental noise or $P_1$) the fitting will worsen. This is different from a bad evaluation of the observed sky, introducing an offset, which also leads to high relative errors (see Sect.~\ref{sec:otherBias}).

\subsubsection{Sersic index}

\begin{figure}
\centering
\includegraphics[width=0.5\textwidth]{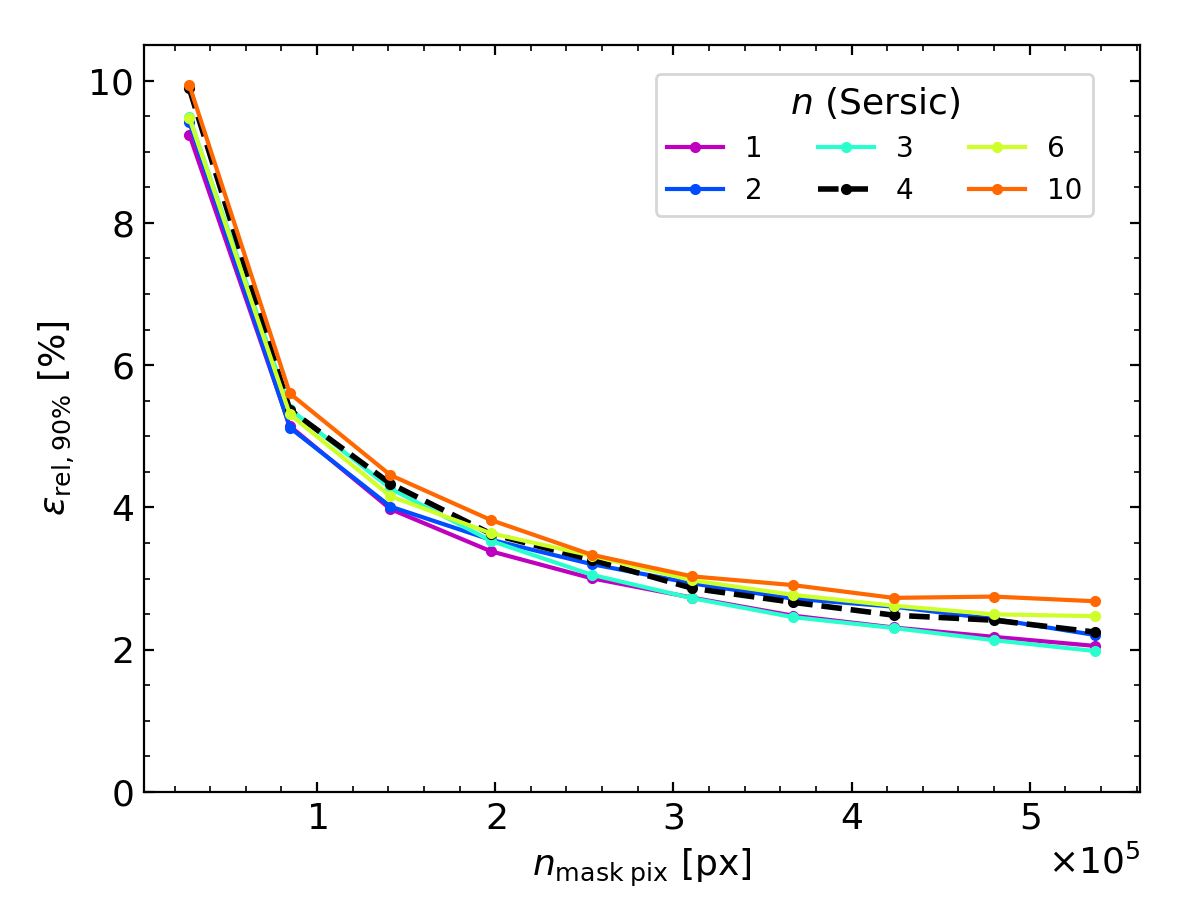}
\caption{Upper 90\% relative error against the number of pixels within the mask when varying the Sersic profile of the mock galaxy: $n=1$ is in magenta, $n=2$ in blue, $n=3$ in cyan, $n=4$ in dashed black, $n=6$ in lime yellow, and $n=10$ in orange.}
\label{fig:Nsersic_ErrVSnpix}
\end{figure}

In Fig.~\ref{fig:Nsersic_ErrVSnpix} we test changing the Sersic index $n=1,2,3,4,6$ and 10, which is equivalent to varying the steepness of the light profile. We find that, for our reference galaxy at least, the Sersic index is not a dominant factor when evaluating the SBF fitting. 

\subsubsection{Effective radius}

\begin{figure}
\centering
\includegraphics[width=0.5\textwidth]{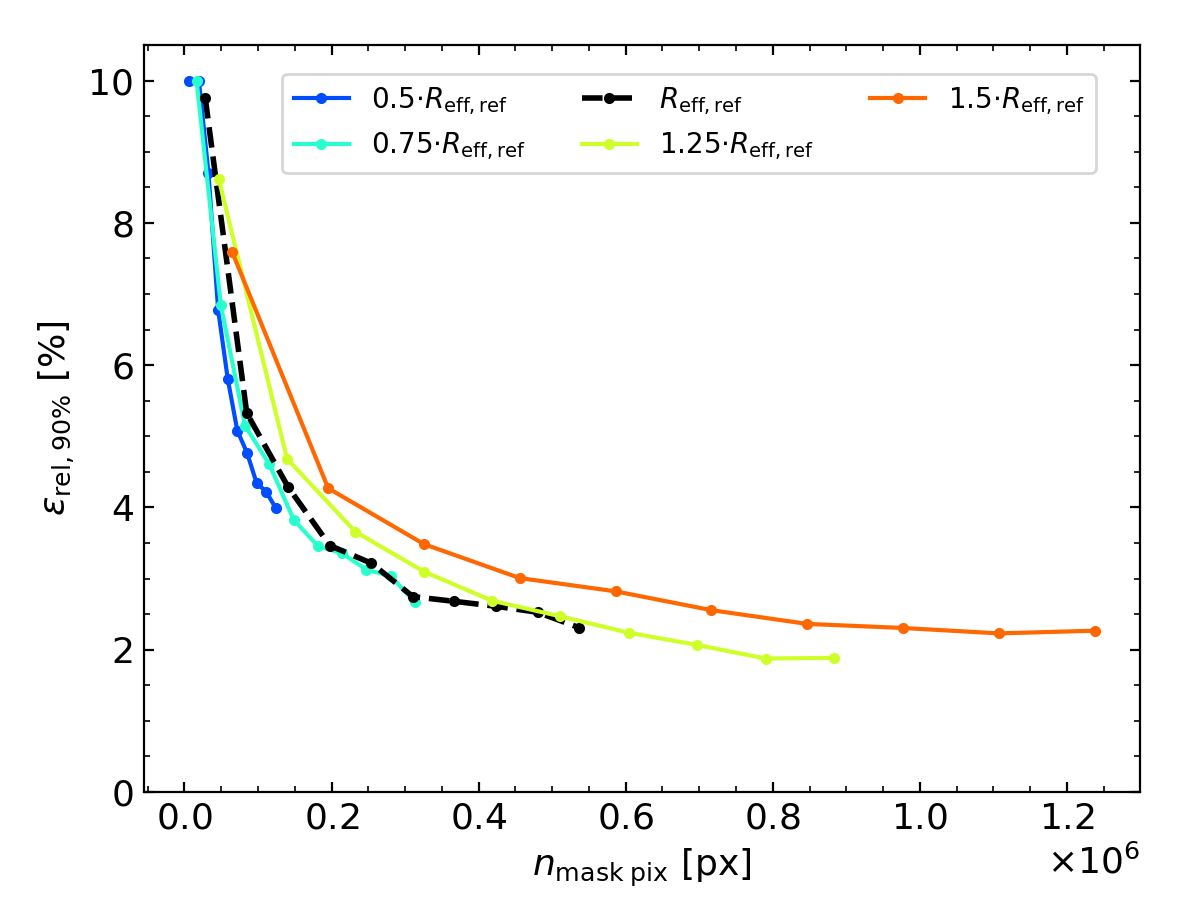}
\caption{Upper 90\% relative error against the number of pixels within the mask when varying the effective radius of the mock galaxy: $0.5\cdot R_{\mathrm{eff,ref}}$ is in blue, $0.75\cdot R_{\mathrm{eff,ref}}$ in cyan, $R_{\mathrm{eff,ref}}$ in dashed black, $1.25\cdot R_{\mathrm{eff,ref}}$ in lime yellow, and $1.5\cdot R_{\mathrm{eff,ref}}$ in orange.}
\label{fig:Reff_ErrVSnpix}
\end{figure}

In Fig.~\ref{fig:Reff_ErrVSnpix} we vary the size of our mock galaxy by changing its effective radius $R_\mathrm{eff}$. We take the effective radius presented in Sect.~\ref{sec:galData} ($R_{\mathrm{eff,ref}}$) and present the fitting for galaxies with a size proportional to this radius ($0.5R_{\mathrm{eff,ref}},\;0.75R_{\mathrm{eff,ref}},\;R_{\mathrm{eff,ref}},\;1.25R_{\mathrm{eff,ref}},\;1.5R_{\mathrm{eff,ref}}$). 
The range of each line is limited by the maximum number of pixels of each effective radius, as we only study masks from $r_1=4$~px to $r_2=R_{\mathrm{eff}}$. 

As expected, the number of pixels limits the reliability of the fitting, for instance the case of $0.5R_{\mathrm{eff,ref}}$ does not reach $\varepsilon_{\mathrm{rel},90\%}\lesssim
4$\%, while $R_{\mathrm{eff,ref}}$ and $1.5R_{\mathrm{eff,ref}}$ do. 
Aside from limiting the region to study, larger effective radii only increase slightly the relative error when sharing the same number of pixels.


    \hfill
    \hfill
    \hfill
    \hfill


\section{Discussion}
\label{sec:discussion}

The results presented in Sect.~\ref{sec:Results} together with the diverse literature previously discussed in Sect.~\ref{sec:introduction} are aimed at addressing the reliability and applicability of the SBF retrieval. In this section we discuss the limitations of the SBF computations, such as some of the approximations taken in the procedure, calculating biased measurements and the possibility of measuring robust SBF gradients.  


\subsection{Reliability of SBF derivation}
\label{sec:RelianceSBF}

In Sects.~\ref{sec:SBFdef}, \ref{sec:appMask} and Appendix~\ref{sec:appendixApplyMask} we describe the mathematical development for deriving the SBF. Some approximations must be taken to derive the final expression, Eq.~(\ref{eq:PS_obsSBFshort_rad_Mask}). Hence, in this section we analyse the influence of the crossed term, the azimuthal average, the operational order when applying the convolution theorem and the fitting of the instrumental noise.

First, after applying the power spectrum (Eq.~(\ref{eq:PS_obsSBFraw}) or Eq.~(\ref{eq:PS_obsSBFraw_Mask}) with a mask), a crossed term appears (Eq.~(\ref{eq:PS_obsSBFlong}) or Eq.~(\ref{eq:PS_obsSBFlong_Mask})) which, to our knowledge, is not discussed in the literature. This crossed term appears to be negligible. In Fig.~\ref{fig:CT_components} we test this by calculating the contribution of the crossed term. We found that the crossed term has a null imaginary component and a real component three orders of magnitude lower than the radial power spectrum of the galaxy. We presently do not have an interpretation for the non-zero real component of the crossed term. We demonstrate this contribution is insignificant by fitting the SBF of the galaxy presented in Fig.~\ref{fig:mockGal} considering the crossed term as in Eq.~(\ref{eq:PS_conTerminoCruzado_radMask}) (the galaxy is masked as in Fig.~\ref{fig:PS}). The results are shown in Table~\ref{table:relianceSBF_Can}, where we present the upper 90\% percentile relative error, the 90\% percentile of the relative standard deviation and the relative 90\% width of the $\bar{N}_{\mathrm{fit}}$ distribution, all from 50 simulations of the galaxy. In column 1 (named '1 Param.') we present the results for the fitting obtained as in Fig.~\ref{fig:PS} and the rest of the work, that is, using Eq.~(\ref{eq:PS_obsSBFshort_rad_Mask}). In column 2 (named 'C.T.') we present the results for the fitting when considering the crossed term. Both columns are equal, finding again that the crossed term does not change the results of the fitting of $\bar{N}_{\mathrm{fit}}$ (at least for this case, up to the fifth decimal digit). 

\begin{figure}
\centering
\includegraphics[width=0.5\textwidth]{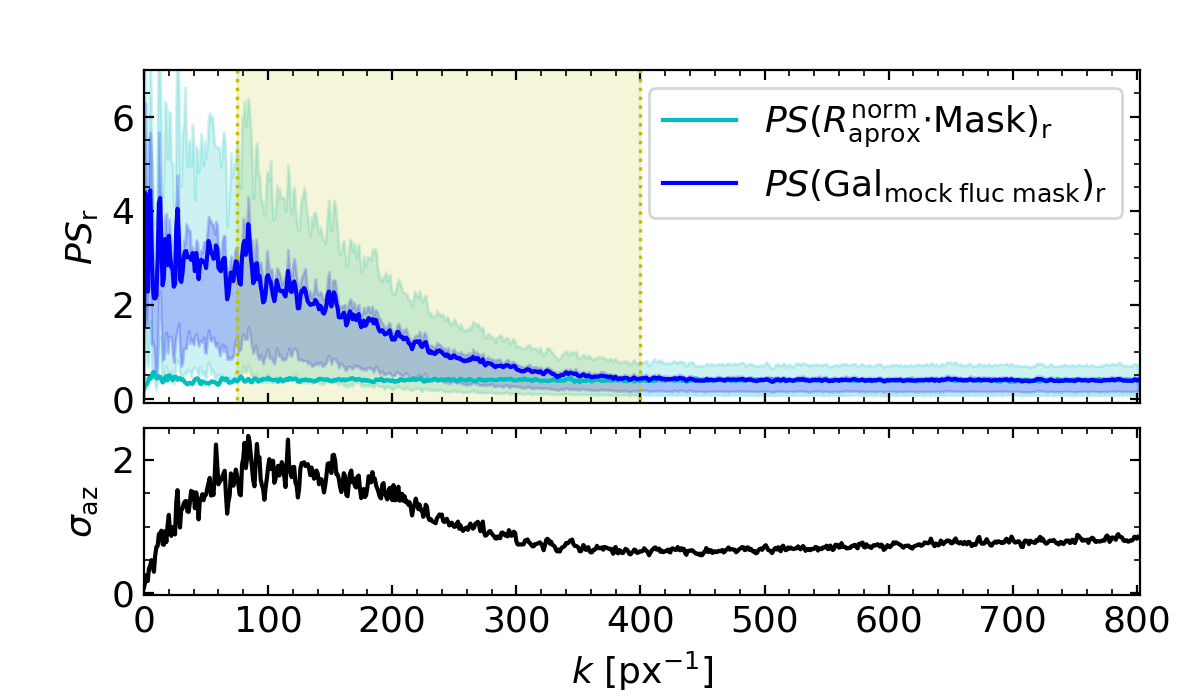}
\caption{Power spectrum (not rescaled) similar to the one of Fig.~\ref{fig:PS}, using the same data and mask. Top panel: blue line as the radial power spectrum of the observed fluctuation, $PS(\mathrm{Gal_{mock\;fluc\;mask}})_{\mathrm{r}}$, with a scatter associated with the distribution of values found at each radius (lower 16\% and higher 84\% percentiles in pale-blue and lower 32\% and higher 68\% in pale-cyan); cyan line as the radial power spectrum of the masked instrumental noise term, $PS(R_{\mathrm{approx}}\cdot \mathrm{Mask}/\sqrt{\mathrm{Gal_{mean}}\otimes \mathrm{PSF}}_{\mathrm{r}}=PS(R_{\mathrm{approx}}^{\mathrm{norm}})_{\mathrm{r}}$; the range of frequencies where the fitting is performed ($k_{\mathrm{fit,i}}=75$~px$^{-1}$, $k_{\mathrm{fit,f}}=400$~px$^{-1}$) is marked with a pale-yellow vertical region. 
Bottom panel: standard deviation associated with the azimuthal average ($\sigma_\mathrm{az}$).}
\label{fig:PsGalMockWithMargins}
\end{figure}

Second, another source of uncertainty appears in Eq.~(\ref{eq:PS_obsSBFlong_rad}) (or Eq.~(\ref{eq:PS_conTerminoCruzadoLong_radMask}) when masked), where an azimuthal average is applied to the power spectrum of the images. 
Each point in the profile found after performing this average has a scatter, since this image is noisy and does not necessarily present radial symmetry.  
Therefore, the resulting distribution of values for a fixed frequency $k$ will follow, in general, a non-Gaussian asymmetric distribution. 
This is shown in the top panel of Fig.~\ref{fig:PsGalMockWithMargins} where, for each $k$ value, we display in pairs, the lower 16\% and higher 84\% percentiles and the lower 32\% and higher 68\% percentiles of the scatter associated with the distribution. Note that the azimuthal average is close to the 68\% percentile instead of the 50\% percentile, as it should happen in a symmetrical distribution. 
This means that, if the SBF extraction procedure makes use of an azimuthal mode or median instead of the azimuthal average, it could lead to distorted SBF results. 
In addition, we have an estimate of the standard deviation of the averaged value at each $k$, $\sigma_\mathrm{az}(k)$, which is shown in the bottom panel of Fig.~\ref{fig:PsGalMockWithMargins}. 
In this manner, we can perform a weighted fitting\footnote{While using \texttt{python} function \texttt{scipy.optimize.curve\_fit} \citep{2020SciPy-NMeth} we introduce the $\sigma_\mathrm{az}(k)$ weights in the parameter 'sigma' and activate the argument \texttt{absolute\_sigma = 'True'}.} by considering this $\sigma_\mathrm{az}(k)$. 
The results obtained when performing the weighted fitting are found in column 3 (named $\sigma_\mathrm{az}$) of Table~\ref{table:relianceSBF_Can}, showing an increase in both the relative error and the relative standard deviation with respect to our standard way of fitting. This indicates that including the $\sigma_\mathrm{az}(k)$ margins of the radial profile is a more conservative way of fitting. On the other hand, the relative 90\% width of the $\bar{N}_{\mathrm{fit}}$ distribution ($\Delta_{90\%}$) is slightly lower than the original fitting. This shows less dispersion (better precision) in the fitting results when applying the $\sigma_\mathrm{az}(k)$ margins. 

Third, when tackling the fitting procedure, we contemplate different options. The results presented in this work consider a single fitting of the SBF as an unknown parameter, because we model the Poisson noise as in Eq.~(\ref{eq:ReadoutComp}). Most authors fit both the SBF and the instrumental noise term simultaneously, this is, with the traditional nomenclature, fitting together $P_0$ and $P_1$ (e.g. \citealt{pahre1999detection,mitzkus2017spectroscopic}). Other authors have already studied how to model the noise and the sky (e.g. for optimised drizzling algorithms,  \citealt{mei2005acs}), although it was not directly applied in the SBF fitting.
In column 4 (named '2 Param.') of Table~\ref{table:relianceSBF_Can} we present the results of the dual fitting of the SBF and $PS(R^{\mathrm{norm}})_{\mathrm{r}}$. We find an increase in the 90\% percentile of the relative error, the 90\% percentile of the relative standard deviation and the relative 90\% width of the $\bar{N}_{\mathrm{fit}}$ distribution, with respect to our fitting. This result shows how modelling the instrumental noise helps in measuring a more precise SBF, instead of fitting both parameters.

\begin{table}[ht]
\centering
\caption{Uncertainties ($\varepsilon_{\mathrm{rel},90\%}$,$\sigma_{\mathrm{fit},90\%}$ and $\Delta_{90\%}$) found after fitting the SBF of the galaxy presented in Fig.~\ref{fig:PS} under different considerations.}
\begin{tabular}{|c|c|c|c|l|}
\hline
 Fit: & 1 Param. & C.T. & $\sigma_\mathrm{az}$ & 2 Param. \\
 & This work. $\bar{N}$ & & & $\bar{N};PS(R^{\mathrm{norm}})_{\mathrm{r}}$ \\ \hline
$\varepsilon_{\mathrm{rel},90\%}$~[\%] & 1.40       & 1.40     & 1.49   & 1.96; 5.43 \\ \hline
$\sigma_{\mathrm{fit},90\%}$~[\%]      & 0.51      & 0.51    & 8.22   & 0.76; 2.52 \\ \hline
$\Delta_{90\%}$~[\%]      & 5.76      & 5.76    & 4.87   & 8.57; 12.2 \\ \hline
\end{tabular}
\tablefoot{Column~1 (1 Param.): fitting of Eq.~(\ref{eq:PS_obsSBFshort_rad_Mask}) considering the SBF is the only unknown parameter, i.e. as done in the rest of the work. Column~2 (C.T.): fitting of Eq.~(\ref{eq:PS_conTerminoCruzado_radMask}), i.e. the SBF is the only unknown parameter and we consider the crossed term (it returns the same results as Col.~1). Column~3 ($\sigma_{\mathrm{az}}$): weighted fitting of Eq.~(\ref{eq:PS_obsSBFshort_rad_Mask}) considering the uncertainties when performing the azimuthal average of $PS(\mathrm{Gal_{mock\;fluc\;mask}})$. Column~4 (2 Param.): fitting of Eq.~(\ref{eq:PS_obsSBFshort_rad_Mask}) considering both the SBF and the power spectrum of the instrumental noise as unknown parameters (uncertainties of both are shown in order).}
\label{table:relianceSBF_Can}
\end{table} 

Fourth, we review a certain step in the SBF derivation procedure that, to the best of our knowledge, was not previously assessed. \citet{tonry1990observations} defines the expectation power spectrum ($E(k)$) as the convolution of $PS(\mathrm{PSF})\otimes PS(\mathrm{Mask})$, which is scaled by the SBF value. Using the expressions of this work, the calculation is performed in the following order: $ \left|\mathfrak{F}\left( \frac{\mathrm{Gal_{fluc}}}{\sqrt{\mathrm{Gal_{mean}}\otimes  \mathrm{PSF}}} \right)\right|^2 \cdot \left(\left|\mathfrak{F}\left( \mathrm{PSF} \right)\right|^2  \otimes  \left|\mathfrak{F}\left( \mathrm{Mask} \right)\right|^2\right)$. 
Instead, the rigorous order according to Eq.~(\ref{eq:applyingTheConvTh}), consists of first multiplying the fluctuation term by the PSF and, then, convolving the result with the mask, this is, $\left( \left|\mathfrak{F}\left( \frac{\mathrm{Gal_{fluc}}}{\sqrt{\mathrm{Gal_{mean}}\otimes  \mathrm{PSF}}} \right)\right|^2 \cdot \left|\mathfrak{F}\left( \mathrm{PSF} \right)\right|^2 \right) \otimes  \left|\mathfrak{F}\left( \mathrm{Mask} \right)\right|^2$. 
We note that the order followed conventionally by the literature is correct only if the fluctuation term, $PS(\mathrm{Gal_{fluc}}/\sqrt{\mathrm{Gal_{mean}}\otimes \mathrm{PSF}})$, is constant. And so it appears to be, at least for the experiments of this work, as we demonstrated in Sect.~\ref{sec:SBFdef}, with Fig.~\ref{fig:PsGalFluc_R_AreCte}.
In a galaxy with an SBF gradient (as presented in Sect.~\ref{sec:SBFgradient}) we find the same constant behaviour. To this extent, and given our ideal experiments, the above approximation is valid. 

In summary, the fitting approach presented in this work appears to be a proper estimation for the SBF measurement. We note that considering the uncertainties when performing the azimuthal average shows a more conservative estimation for $\varepsilon_{\mathrm{rel},90\%}$) and $\sigma_{\mathrm{fit},90\%}$, but improves the precision of the fitting. Modelling the instrumental noise ($R$), instead of fitting it, reduces the uncertainty of the calculation (see Table~\ref{table:relianceSBF_Can}). We suggest modelling the instrumental noise as in Eq.~(\ref{eq:ReadoutComp}), which requires knowledge of the mean galaxy value, the PSF and the sky background, as well as considering  Poisson noise. In this regard we encourage adapting, if necessary, the equation for each observation or studying other procedures, such as the one presented in \citep{mei2005acs} for correlated noise. The rest of the approximations taken during the SBF derivation are negligible (i.e. the crossed term and the order after applying the convolution theorem). 


\subsection{Biased SBF measurements due to low flux levels}
\label{sec:biasedSBF}

This section is intended to point out an important caution to consider:  examining a low flux source using the traditional SBF extraction methodology could potentially introduce a bias into the result. Here, we explore how this bias appears, how to avoid or mitigate it, as well as the conditions necessary for applying confidently the standard SBF derivation.

To begin with, we analyse how this bias can appear in the procedure. A reference 'mean image' ($\mathrm{Gal}_\mathrm{mean}(x,y)$) always can be obtained by different methods, such as applying a mean filter, smoothing the image, by an isophote fitting, with Sersic profiles or other approaches. 
Usually, this image is used to obtain an SBF measurement by subtracting it to the original image, dividing the result by its square root and obtaining the power spectrum of the resulting image. 
However, having an adequate $\mathrm{Gal}_\mathrm{mean}(x,y)$ model image of the mean brightness profile is not a sufficient condition: it is required that this is a proper representation of the \textit{mean} of the stellar population luminosity distribution\footnote{This is, the distribution of the possible luminosities of a system with a given evolutionary condition and a given total number of stars.} in each pixel ($\left<{\cal{L}}(x,y)\right>$)\footnote{Note that our galaxy model assumes that $\mathrm{Gal}_\mathrm{mean,fluc}(x,y)$ is determined by $\left<{\cal{L}}(x,y)\right>$ and $\sigma_\mathrm{fluc}^2(x,y)$, which are implicitly taken as known.} and, consequently, obtaining the variance ($\sigma_\mathrm{fluc}^2(x,y)$) of such a distribution. 
We recall that the \textit{mean} and variance of the population luminosity distribution scale linearly with the number of stars in each pixel. This number of stars must cancel out, as a way for all the pixels to be equivalent when measuring the fluctuation \citep{tonry1988new,CL06,cervino2008surface,cervino2013}. Otherwise, it cannot be used neither for stellar population analysis nor for distance calculations.

To illustrate this, let us consider a scenario in which we have made a biased estimate of the \textit{mean} of the population luminosity distribution along the galaxy pixels, $\left< {\cal{L}}(x,y) \right>_{\mathrm{real}}$, by an amount of $\pm L_\mathrm{offset}(x,y)$, which could vary depending on the ($x,y$) positions (see \citet{cervino2008surface}). That is:

\begin{equation}
    \left< {\cal{L}}(x,y)\right>_{\mathrm{biased}}= \left< {\cal{L}}(x,y) \right>_{\mathrm{real}} \pm L_\mathrm{offset}(x,y).
\end{equation}

\noindent A simple calculation\footnote{The variance $\sigma_{\mathrm{fluc}}^2$ is the average of $({\cal{L}} - \left< {\cal{L}} \right>_{\mathrm{real}})^2$. If we use $\left< {\cal{L}} \right>_{\mathrm{biased}} = \left< {\cal{L}} \right>_{\mathrm{real}} \pm L_\mathrm{offset}$ instead of $\left< {\cal{L}} \right>_{\mathrm{real}}$, we will have the average of $({\cal{L}} - \left< {\cal{L}} \right>_{\mathrm{biased}})^2=({\cal{L}} - \left< {\cal{L}} \right>_{\mathrm{real}})^2 + L_\mathrm{offset}^2 \mp 2\,L_\mathrm{offset}\,({\cal{L}} - \left< {\cal{L}}\right>_{\mathrm{real}})$. Then, since the average of $({\cal{L}} - \left< {\cal{L}}\right>_{\mathrm{real}})^2=\sigma_{\mathrm{fluc}}^2$ and the average of $({\cal{L}} - \left< {\cal{L}}\right>_{\mathrm{real}})$ equals zero, we reach the result of Eq.~(\ref{eq:sigbias}).} 
shows that, if we use $\left< {\cal{L}}(x,y) \right>_{\mathrm{biased}}$ as the reference value at a position $(x,y)$, its associated variance is:

\begin{equation}
\sigma^2_\mathrm{fluc,biased}(x,y)=\sigma^2_\mathrm{fluc}(x,y) +  L_\mathrm{offset}(x,y)^2,
\label{eq:sigbias}
\end{equation}

\noindent and, therefore, we obtain a biased fluctuation at that position:

\begin{equation}
\mathrm{SBF}_\mathrm{biased}(x,y) = \frac{\sigma^2_\mathrm{fluc,biased}(x,y)}{\left< {\cal{L}}(x,y) \right>_{\mathrm{biased}}}=\frac{ \sigma^2_\mathrm{fluc}(x,y) +  L_\mathrm{offset}(x,y)^2}{\left< {\cal{L}}(x,y) \right>_{\mathrm{real}} \pm L_\mathrm{offset}(x,y)},
\label{eq:sbfbias}
\end{equation}

\noindent which deviates from the actual definition of the SBF (i.e. $\mathrm{SBF}_\mathrm{real}(x,y)=\sigma^2_{\mathrm{fluc}}(x,y)/\left< {\cal{L}} \right>_{\mathrm{real}}(x,y)$), with no straightforward method of eliminating this offset by factorisation. The situation is worse when several pixels are taken into consideration, since the fluctuations measured along the pixels cannot be compared in a common framework, and it is not assured that there is the required independence of the SBF with the number of stars per pixel.

To assure a
correct \textit{mean} value of the population luminosity distribution it is required that all possible evolutionary phases are well-sampled along the pixels. This is achieved if there is a large enough number of stars distributed in these pixels, about a total of $\sim10^7$ stars, depending on the band \citep{cervino2004physical,CL06}.
In practice, the galaxy model $\mathrm{Gal}_\mathrm{mean}(x,y)$ is commonly obtained from an observed image by considering the flux of nearby pixels and making some kind of average for each \textit{collection}\footnote{For the sake of explaining this section, we address this number of nearby pixels as a '\textit{collection}'.} of them. 
Therefore, it is mandatory that every \textit{collection} has enough stars to represent robustly the whole stellar population. Additionally, the pixels of each \textit{collection} should have similar characteristics, as number of stars and stellar populations properties.

In this context, we could analyse two opposite cases:
(a) if the pixels of the observation gather low number of stars per pixel, each \textit{collection} needs to assemble a large enough number of pixels to achieve this statistically meaningful estimate of the \textit{mean}; 
or, (b) if every pixel in the observation has a sufficient number of stars, we do not need a necessarily large number of pixels per \textit{collection}, as in a traditional SBF derivation.
Both cases are fully understood when knowing the shape of the population luminosity distribution function, which has been studied in detail in \citet{CL06} and \citet{cervino2008surface}.
These works show that the distribution of the integrated luminosity in the \textit{collection} of pixels follows an L-shape in the extreme case (a) with a single star per pixel, and a Gaussian shape in case (b) with infinite stars per pixel. There is a continuous transition between both cases, dependent on the number of star per pixel.

Remember that in most studies the methodology applied considers a situation as case~(b), typical of elliptical galaxies, where a traditional SBF derivation is considered. The mean value of each pixel is obtained with a reasonable small group of neighbouring pixels, each with sufficient number of stars.
Here, we warn about the risks of studying an observation of case~(a) applying the traditional procedure made for case~(b). In doing so, the mean is estimated from neighbouring pixels that might be dominated by a few luminous stars. If those pixels are considered, they lead to an overestimation, while, if absent, they lead to an underestimation (e.g. if pixels with extremely bright stars are masked due to its luminosity excess). In both situations we introduce an offset, such as illustrated in Eq.~(\ref{eq:sbfbias}). 

To tackle this situation it is necessary to stablish a threshold in which we are confident to be working in scenario (b). In this sense, \citet{cervino2004physical} states how many stars are required for a statistically meaningful distribution to be found.
A Gaussian-like regime can be reached with $\sim10^6$ stars per pixel for the optical bands (even a larger number for the infrared bands), assuming a simple stellar population and a standard Initial Mass Function \citep{CL06}. 
This requirement is similar to the assumption of having about 20 giant stars per pixel in old stellar populations, quoted by \citet{tonry1988new}, in the visible and infrared bands. In the case of an old stellar population, the SBF flux value is similar to the flux of a single giant star \citep{tonry1988new}. Hence, we find the requirement that the SBF should be obtained from pixels with fluxes larger than $\sim20$ times the SBF flux. Such a criterion can be extended to more complex stellar populations, as inferred from \citet{cervino2004physical} and \citet{CL06}. In Sect.~\ref{sec:Criteria} we relaxed this criterion to 10 times the SBF flux ($N(\mathrm{Gal_{mock\;mask}}) - N(\mathrm{Sky_{mask}})\ge 10\cdot \bar{N}$, in count numbers). 
This is a quick and easy test to minimise possible effects of non-Gaussianity in the stellar population distribution\footnote{It is important to note that when the luminosity distributions become Gaussian-like, the mean, the median and the mode coincide, and therefore any of them can be used to obtain the reference galaxy image.}. 
Finally, aside of having enough stars per pixel, it is desirable to avoid too steep light profiles or abrupt flux changes, as shown in \citet{cervino2008surface}.

In conclusion, in this section we highlight the problems derived when obtaining the SBF following the standard procedure, which assumes the condition of high number of star per pixel (case~(b)), when the observed source has a low number of stars per pixel (case~(a)). The caveats discussed here are particularly relevant when studying external parts of galaxies, faint sources, diffuse or dwarf galaxies (see references of Sect.~\ref{sec:introduction}). For instance, we found this scenario in Fig.~\ref{fig:SearchingOpt_varMask}, for low surface brightness and high fluctuation counts. In that figure, there is a blank region where the criterion of 'studying pixels with, at leats, fluxes of 10 times the SBF' is not fulfilled. For these cases, even if the estimated errors were low, the mean reference value is potentially incorrect.
Then, the chance of obtaining a biased SBF increases in a non-trivial manner, so it is less reliable for the calibration of distances or stellar populations. 
Therefore, when working with an observation it is advisable to verify whether all the pixels utilised in the measurement meet the criterion of Eq.~(\ref{eq:criteria}), once the SBF has been estimated.


\subsection{Other sources of bias}
\label{sec:otherBias}

Beyond the experimental conditions of an SBF measurement, such as an adequate number of counts associated with each pixel or achieving a favourable S/N, three essential parameters must be accurately estimated: the mean galaxy image, the PSF, and the sky background. These terms are responsible for obtaining the fluctuation image ($\mathrm{Gal_{mock\;fluc}}$) and modelling the instrumental noise ($R_{\mathrm{approx}}$). Additionally, the mean image and the PSF normalise Eq.~(\ref{eq:sqrtSBF}) by $\sqrt{\mathrm{Gal_{mean}}\otimes \mathrm{PSF}}$, following the SBF definition. Currently, the SBF measurement procedure is already well established by using several PSF templates, running experiments on the sky background and the galaxy profile until is well determined. Nevertheless, in this section we want to highlight how the modelling of these parameters should not be treated lightly. 

On the one hand, even with enough counts in each pixel (as discussed in the previous section), when $\mathrm{Gal}_{\mathrm{mean}}$ is not properly calculated, an offset is introduced in the SBF definition, as addressed in Eq.~(\ref{eq:sbfbias}). On the other hand, if a uniform sky is assumed in an image with a non-uniform real sky, the estimated value of $N_{\mathrm{Sky}}$ would present an offset with respect to the real one, at least in certain areas of the image. Even if the observed sky is flat, it carries its inherent uncertainty. Finally, the PSF is commonly obtained from one or more foreground stars using different techniques, introducing variability in the PSF characterisation. 
In addition, although not considered in our following examples, in a real observation, a wrong sky estimation implicitly could lead to an erroneous mean galaxy model ($\mathrm{Gal}_{\mathrm{mean}}$). 

To assess how an offset in any of these terms affects the SBF measurement, we propose some example experiments where we perform 50 simulations underestimating or overestimating the values of either $\mathrm{Gal}_\mathrm{mean}$, $N_{\mathrm{Sky}}$ or $\sigma_\mathrm{PSF}$ of our reference galaxy from Fig.~\ref{fig:PS}, modifying one parameter while leaving the other two fixed. 
We choose to introduce offsets of $\chi_{\mathrm{offset}}=\pm3\%$ and $\pm9\%$, applied as, for instance, $\mathrm{Gal}_{\mathrm{mean,offset}}=\mathrm{Gal}_{\mathrm{mean}} \pm \chi_{\mathrm{offset}}\times\mathrm{Gal}_{\mathrm{mean}}$. 
In the case of the PSF, it represents an offset of $\pm0.04$ and $\pm0.12$~px with respect to the reference value of $\sigma_{\mathrm{PSF}}=1.33$~px. 
For the sky, it represents an offset of $\pm275$ and $\pm824$~counts with respect to the original value of $N_{\mathrm{Sky}}=9155$~counts. 
The results for the mean fitted SBF $\langle \bar{N}_{\mathrm{fit}} \rangle$ and the 90\% percentile of the relative error ($\varepsilon_{\mathrm{rel},90\%}$) are presented in Table~\ref{table:offset}. 
They should be compared to those of the reference case~(Fig.~\ref{fig:PS}), which has an input SBF of $\bar{N}_{\mathrm{ref}}=22.59$ and, after performing 50 simulations, $\langle \bar{N}_{\mathrm{fit}} \rangle=22.53$~counts and $\varepsilon_{\mathrm{rel},90\%}=2.88$ \%.
In Table~\ref{table:offset} we do not show neither the relative standard deviation nor the precision, as they do not change significantly compared to the results found for Fig.~\ref{fig:PS}.

\begin{table}[]
\centering
\caption{Mean fitted SBF $\langle \bar{N}_{\mathrm{fit}} \rangle$ and upper 90\% percentile of the relative error ($\varepsilon_{\mathrm{rel},90\%}$), applying different offsets ($\pm3\%$ and $\pm9\%$) to $\mathrm{Gal}_{\mathrm{mean}}$, $N_{\mathrm{Sky}}$, and $\sigma_{\mathrm{PSF}}$ with respect to the reference case of Fig.~\ref{fig:PS}.}
\renewcommand{\arraystretch}{1.4}

\begin{tabular}{ccc}
    \begin{tabular}{|c|}
    
    \multicolumn{1}{c}{} \\
    \hline
    $\chi_{\mathrm{offset}}$ \\
    \hline
    0\%  \\
    \hline
    -3\%  \\
    \hline
    +3\%  \\  
    \hline
    -9\%  \\
    \hline
    +9\%  \\ 
    \hline
    \end{tabular}

    \begin{tabular}{|p{9mm}|p{7mm}|p{7mm}|}
    \hline
    \multicolumn{3}{|c|}{$\langle \bar{N}_{\mathrm{fit}} \rangle$ [counts]} \\
    \hline
    $\mathrm{Gal}_{\mathrm{mean}}$ & $N_{\mathrm{Sky}}$ & $\sigma_{\mathrm{PSF}}$ \\
    \hline
    \multicolumn{3}{|c|}{22.53} \\
    \hline
    \centering 24.98 & 29.87 & 21.95 \\
    \hline
    \centering 23.5 & 29.66 & 23.63 \\  
    \hline
    \centering 40.72 & 87.23 & 20.52 \\
    \hline
    \centering 33.81 & 86.33 & 24.87 \\ 
    \hline
    \end{tabular}

    \begin{tabular}{|p{9mm}|p{7mm}|p{7mm}|}
    \hline
    \multicolumn{3}{|c|}{$\varepsilon_{\mathrm{rel},90\%}$ [\%]} \\
    \hline
    $\mathrm{Gal}_{\mathrm{mean}}$ & $N_{\mathrm{Sky}}$ & $\sigma_{\mathrm{PSF}}$ \\
    \hline
    \multicolumn{3}{|c|}{2.88} \\
    \hline
    \centering 12.9 & 35.3 & 4.8 \\
    \hline
    \centering 5.9 & 33.9 & 7.2 \\  
    \hline
    \centering 82.9 & 291.4 & 11.1 \\
    \hline
    \centering 51.8 & 286.4 & 12.4 \\ 
    \hline
    \end{tabular}
\end{tabular}
\tablefoot{The reference case has an input SBF of $\bar{N}_{\mathrm{ref}}=22.59$ and, after performing 50 simulations, its fitting returned: $\langle \bar{N}_{\mathrm{fit}} \rangle=22.53$~counts and $\varepsilon_{\mathrm{rel},90\%}=2.88$ \%.}
\label{table:offset}
\end{table}

The results presented in Sect.~\ref{sec:Results} have a variety of different relative errors depending on the circumstances, but their mean value of the fitted SBF are found surrounding $\bar{N}_{\mathrm{real}}$. However, when an offset is introduced, the results of $\langle \bar{N}_{\mathrm{fit}} \rangle$ show a systematic bias with respect to the reference case of Fig.~\ref{fig:PS}, as it is shown in Table~\ref{table:offset}.
We note that for all the cases presented in Table~\ref{table:offset} the accuracy worsens with the offset. In this example the most sensitive parameter is the sky, followed by the mean galaxy model and the PSF.
Underestimating $\mathrm{Gal}_{\mathrm{mean}}$ returns worse accuracies than overestimating it, due to the reduced contrast between the sky background and the galaxy light.

We note that the impact of a biased sky depends on the luminosity of the galaxy ($\mathrm{Gal}_{\mathrm{mean,fluc}}$). A brighter galaxy will not be as highly distorted by an offset in the sky. Similarly, when applying a mask, the relative error will be lower when studying the inner (and more luminous) regions than the outer parts of the galaxy. For instance, the $\pm$275 counts of sky offset introduced in the previous example quantify differently depending on the number of counts at each part of the galaxy: at the effective radius, with $N_{\mathrm{Reff}}=1494.96$~counts, it represents an $\pm$18\% offset; at the outer radius of the mask ($r_2$), with $N_{\mathrm{Reff}}=2492$~counts, the offset is $\pm$11\%; at the inner radius of the mask ($r_1$) with $N_{\mathrm{Reff}}=21132$~counts it represents a $\pm$1\% offset. As already shown in Sect.~\ref{sec:Results}, studying bright galaxy pixels with respect to the sky count value facilitates retrieving a reliable SBF.

To summarise, these results show how an offset due to a non-properly estimated mean model, PSF or sky background, could severely increase the relative error. In particular, the accuracy drastically worsens when the sky is not properly measured.
Meanwhile, the relative standard deviation and the precision are not drastically affected. In a real observation, we are not able to obtain the accuracy, but the relative standard deviation and the precision can be used as measurements of the uncertainty. Therefore, it is worth emphasising that achieving a favourable relative standard deviation or precision values might lead into misinterpretations of how close our fitted SBF is from the real fluctuation value.
Even if this section aimed at pointing out how the results vary due to a biased modelling, generally such issues can be detected when checking the power spectrum plot used for the fitting (as the one shown in Fig.~\ref{fig:PS}), where any inconsistency should be apparent and serves as a warning that an offset is being introduced.


\subsection{SBF gradient detection}
\label{sec:SBFgradient}

As mentioned in the introduction, some authors have investigated SBF gradients in galaxies. Several studies have found such SBF profiles \citep{cantiello2005detection, cantiello2007surface, sodemann1995gradient, sodemann1995variation, jensen2015measuring}, but often show the radial outline to be relatively small, uncertain, or not conclusively detect the variability under given observational conditions \citep{jensen1996infrared}. Thus, we study the possibility of SBF gradient detection in a controlled environment, with the intention of establishing under what conditions such gradients (if present) would be found, and what is the optimal procedure to obtain them.

First, we need to determine what is the actual SBF value
when the fluctuations are not homogeneous in all regions of a studied mask. We have found that the value recovered after the fitting is a weighted average of all the SBFs present in the masked region. This is, the sum of every SBF value ($\bar{N}_1,\bar{N}_2,...,\bar{N}_f$) times the number of pixels where it appears ($n_{\mathrm{pix},1},n_{\mathrm{pix},2},...,n_{\mathrm{pix},f}$), divided by the total number of pixels within the mask ($n_{\mathrm{mask\;pix}}$): 

\begin{equation}
    \bar{N}_{\langle\mathrm{SBF} \rangle} = \frac{1}{n_{\mathrm{mask\;pix}}} \sum_{i=1}^{f} n_{\mathrm{pix},i}\; \bar{N}_i.
    \label{eq:SBFweightedAverage}
\end{equation}

\noindent Note that the SBF of an individual pixel is not measurable in observations, as more than one pixel is required for measuring a variance, but in practical terms we do know at each point of our mock galaxy the integrated luminosity distribution associated with the stellar population in that point and, hence, its associated SBF value \citep[see][]{CL06}.

We study the presence of a variable SBF using the reference galaxy presented in Sect.~\ref{sec:galData} and some complementary literature information. We built a gradient based on the work of \citet{martin2018timing}. From that work, in their figure 5 the average stellar population gradient versus the radius is analysed. In particular, we focus on the metallicity profile for galaxies with a velocity dispersion over 300~km s$^{-1}$, typical of massive galaxies. We obtain a metallicity of $[\mathrm{M/H}] \approx 0.26$ for 
$\log_{10}(r/R_{\mathrm{eff}})=-1.1$ and $[\mathrm{M/H}]\approx0.06$ for $\log_{10}(r/R_{\mathrm{eff}})=-0.3$. Additionally, the age is approximately constant with  radius, with $t\sim12.5$ Gyr. We can relate these values to SBF magnitudes using the models presented in \citet{vazdekis2020surface}, their figure~9. There, it is shown how a metallicity of $[\mathrm{M/H}]\approx0.26$ corresponds to $\bar{M}_i=0.1$~mag, while $[\mathrm{M/H}]\approx0.06$ corresponds to $\bar{M}_i=-0.25$~mag, both for a 13 Gyr. Then, we calculate the apparent SBF magnitude for our reference galaxy, using the distance presented in Sect.~\ref{sec:galData}, $D=16.7$~Mpc. We find $\bar{m}_i=31.21$~mag for $\log_{10}(r/R_{\mathrm{eff}})=-1.1$ and $\bar{m}_i=30.86$~mag for $\log_{10}(r/R_{\mathrm{eff}})=-0.3$. 
Finally, if we follow the gradient shown in figure~5 from \citet{martin2018timing} for the metallicity, we can approximate a linear profile for $\bar{m}_i$ versus $\log_{10}(r/R_{\mathrm{eff}})$. Thus, we can calculate $\bar{m}_i(r)$ for any given radius as a linear function and then transform it into counts using Eq.~(\ref{eq:mSBFtoCounts}). Note that the fluctuation count increases with the radius.

\begin{figure}
\centering
\includegraphics[width=0.5\textwidth]{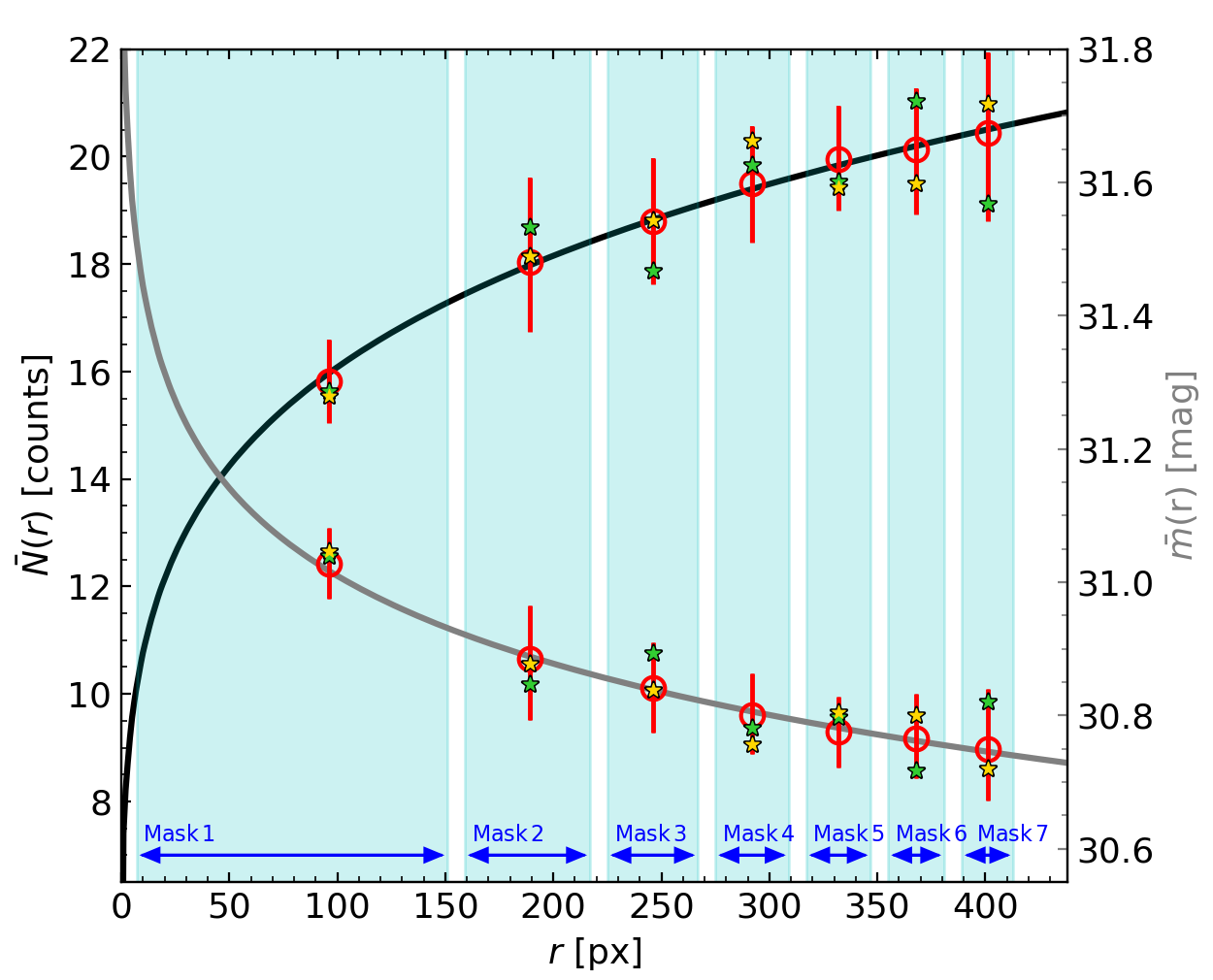}
\caption{Variable SBF radial profile shown in counts as a black line and in magnitudes as a grey line. Annular masks are indicated as cyan vertical regions. For each mask, the SBF mean fitting results over 50 simulations are marked as red dots, as well as its error bar. Results at each mask of two different example galaxies (retrieved from the 50 simulations) are shown with green and yellow star markers.}
\label{fig:DoubleAxisGradient75000px}
\end{figure}

Once a galaxy with an SBF gradient is created, we test the possibility of an actual SBF gradient detection. In Fig.~\ref{fig:DoubleAxisGradient75000px} we show the radial profile of the SBF gradient in counts as a black line and its respective profile in magnitudes as a grey line. We mark with cyan vertical regions each of the annular masks used to fit the SBF in different annuli, from 4 pixels up to the effective radius. Each one of these masks share approximately the same number of pixels ($n_{\mathrm{mask\;pix}}\approx75000$~px), this way we assure that the uncertainties are comparable. 
Within an annular mask, the weighted average SBF, obtained with Eq.~(\ref{eq:SBFweightedAverage}), is associated with a radius that depends on the SBF profile. 
In Fig.~\ref{fig:DoubleAxisGradient75000px}, we chose the radius ($r_{\langle\mathrm{SBF} \rangle}$) such that the value of the SBF at that distance from the centre matches the weighted average SBF of the studied annulus, this is $\mathrm{SBF}(r_{\langle\mathrm{SBF} \rangle}) = \langle\mathrm{SBF} \rangle$. 
We perform the SBF estimation for 50 galaxy simulations. Then, we present the mean SBF fitting value as red dots, with an error interval defined with the 5th\% minimum and the 95th\% maximum percentile values (as explained in Eq.~(\ref{eq:Delta90})) among the 50 iterations of the mock galaxy. 
We show the results at each mask for two different example mock galaxies retrieved from the 50 simulations, using green and yellow star-markers. 

In Table~\ref{table:tableGradient} we show in more detail the results for each one of the masks. Column 1 presents the radii ($r_1,r_2$) that define each annular mask, column 2 shows the exact number of pixels within the mask ($n_{\mathrm{mask\;pix}}$), column 3 marks the radius associated with the weighted average SBF of each mask ($r\,_{\langle\mathrm{SBF} \rangle}$), column 4 shows the weighted average of the SBF in counts ($N_{ \langle\mathrm{SBF}\rangle}$), column 5 corresponds to the measured SBF in counts 
($\langle \bar{N}_{\mathrm{fit}} \rangle$, with a lower and upper limit calculated as $\langle \bar{N}_{\mathrm{fit}} \rangle-\bar{N}_{\mathrm{fit,5th\%}}$ and $\bar{N}_{\mathrm{fit,95th\%}}-\langle \bar{N}_{\mathrm{fit}} \rangle$, respectively), 
column 6 and 7 show the higher 90\% percentile of the relative error and the relative standard deviation ($\varepsilon_{\mathrm{rel},90\%}$ and $\sigma_{\mathrm{fit},90\%}$) respectively, column 8 presents the relative 90\% width of the $\bar{N}_{\mathrm{fit}}$ distribution ($\Delta_{90\%}$), column 9 is the magnitude of the weighted average SBF and column 10 is its corresponding SBF measured magnitude ($m_{\left<\mathrm{SBF} \right> }$ and $\bar{m}_{\mathrm{fit}}$ magnitudes are retrieved from columns 4 and 5 making use of Eq.~(\ref{eq:mSBFtoCounts})). 

\begin{table*}[ht]
\centering
\caption{Results of tracing an SBF radial profile over different annular masks, as in Fig.~\ref{fig:DoubleAxisGradient75000px}.}
\renewcommand{\arraystretch}{1.4}
\begin{tabular}{|c|c|c|c|c|c|c|c|c|c|}
\hline
Mask ($r_1;r_2$) & $n_{\mathrm{mask\;pix}}$~[px] & $r\,_{\langle\mathrm{SBF} \rangle}$~[px] & $N_{ \langle\mathrm{SBF}\rangle}$ & $\langle \bar{N}_{\mathrm{fit}} \rangle$ & $\varepsilon_{\mathrm{rel},90\%}$~[\%] & $\sigma_{\mathrm{fit},90\%}$~[\%] & $\Delta_{90\%}$~[\%] & $m_{\left<\mathrm{SBF} \right> }$ & $\bar{m}_{\mathrm{fit}}$ \\\hline
  4; 154 & 74432 &  96 & 15.95 & 15.8$\pm$0.8          & 5.3 & 0.9 & 9.8  & 31.02 & 31.03$\pm$0.05 \\\hline
156; 220 & 75616 & 189 & 17.98 & 18.0$\pm^{1.6}_{1.3}$ & 7.5 & 0.8 & 15.9 & 30.89 & 30.89$\pm^{0.08}_{0.09}$ \\\hline
222; 270 & 74240 & 246 & 18.83 & 18.8$\pm$1.2          & 6.1 & 0.8 & 12.5 & 30.84 & 30.84$\pm$0.07 \\\hline
272; 312 & 73404 & 292 & 19.40 & 19.5$\pm$1.1          & 5.8 & 0.7 & 11.2 & 30.81 & 30.80$\pm$0.06 \\\hline
314; 350 & 75068 & 332 & 19.84 & 20.0$\pm$1            & 5.5 & 0.7 & 9.8  & 30.78 & 30.78$\pm$0.05 \\\hline
352; 384 & 74020 & 368 & 20.20 & 20.1$\pm^{1.1}_{1.2}$ & 5.6 & 0.7 & 11.7 & 30.76 & 30.77$\pm^{0.07}_{0.06}$ \\\hline
386; 416 & 75572 & 401 & 20.50 & 20.4$\pm^{1.5}_{1.6}$ & 7.2 & 0.7 & 15.3 & 30.75 & 30.75$\pm^{0.09}_{0.08}$ \\\hline
\end{tabular}
\tablefoot{Column~1: radii ($r_1,r_2$) that constrict each annular mask. Column~2: number of pixels within the mask. Column~3: radius associated with the weighted average SBF of each mask. Column~4: weighted average of the SBF in counts. Column~5: mean fitted SBF over 50 simulations in counts. Column~6: 90\% percentile of the relative error. Column~7: 90\% percentile of the relative standard deviation of the fitting. Column~8: relative 90\% width of the $\bar{N}_{\mathrm{fit}}$ distribution. Column~9: magnitude of the weighted average SBF. Column~10: mean fitted SBF over 50 simulations in magnitudes.}
\label{table:tableGradient}
\end{table*}

The results presented in Fig.~\ref{fig:DoubleAxisGradient75000px} and Table~\ref{table:tableGradient} serve as a test for the detection of SBF gradients. With $\langle \bar{N}_{\mathrm{fit}} \rangle$ we recover the profile correctly, with relative errors lower than $\varepsilon_{\mathrm{rel},90\%}<7.5\%$ and relative standard deviations lower than $\sigma_{\mathrm{fit},90\%}<1\%$.
The relative 90\% width of the $\bar{N}_{\mathrm{fit}}$ distribution shows a range of $9.8\%<\Delta_{90\%}<15.9\%$ precision. 
All three parameters are similar among the different masks, as they approximately share the same number of pixels. 
The mean value of the 50 solutions of the fitting follows its corresponding weighted average SBF, from $\langle \bar{N}_{\mathrm{fit}} \rangle=15.8$~counts in the inner mask up to $\langle \bar{N}_{\mathrm{fit}} \rangle=20.4$~counts in the external mask. 
Additionally, notice how the position of the radius associated with the weighted average SBF ($r_{\langle\mathrm{SBF} \rangle}$) tends to move towards the region with the larger number of pixels. 
On this account, the variation of $\langle \bar{N}_{\mathrm{fit}} \rangle$ through the consecutive masks could be detectable, however, if we consider the uncertainty associated with the $\Delta_{90\%}$ width of every annulus, only the results between the first mask and the rest of masks do not overlap. Any other mask comparison, besides the first annulus, has sufficiently large uncertainties for the results not to be distinguishable. 
For a clearer understanding of the behaviour, we presented in Fig.~\ref{fig:DoubleAxisGradient75000px} those two particular simulated galaxies (green and yellow stars) at each mask. 
In both cases the fitting solutions are found arbitrarily above and below the radial profile of the SBF gradient. This indicates how tracing the gradient with a sole observation could lead to a correct representation of the actual SBF profile. However, it also could lead to misinterpretations if, for instance, most of the mask results fall below the real SBF profile. Among our simulations, we have found cases that both underestimate and overestimate the SBF slope.

In actual observations we do not have 50 fittings from which we extract a mean value, hence, a sole simulation will trace the gradient of the galaxy but with varying points around the mean. So, even working under ideal conditions (as in this work) presents difficulties when comparing the SBF from radius to radius. At least, we can calculate the precision of the results in each annuli, as we advanced in Sect.~\ref{sec:Results} and we explain  in Sect.~\ref{sec:codePresentation}, also illustrated with Fig.~\ref{fig:flowchart}. To do so, we propose constructing a simulated set of Monte Carlo galaxies, as described in Sect.~\ref{sec:mockGalaxyCreation}, but considering as input the observed SBF of the different annuli, $\bar{N}_{\mathrm{obs}}$. Subsequently, we can determine $\sigma_{\mathrm{fit},90\%}$ and $\Delta_{90\%}$ with respect to the fitted $\bar{N}_{\mathrm{obs}}$, following the procedure explained in Sect.~\ref{sec:codePresentation} and illustrated with Fig.~\ref{fig:flowchart}.  
In any case, steeper SBF gradients and better observations would ease constraining the radial profile of the fluctuation.

Additionally, we replicate the fitting performed in Fig.~\ref{fig:PS} (between $r_1=80$~px and $r_2=332$~px), which would be considered as the global SBF of the galaxy, but keeping the SBF gradient of this section, instead of a constant SBF. We have that $\bar{N}(r_1=80)=15.46$ and $\bar{N}(r_2=332)=19.84$~counts. 
Within this mask, the weighted average of the SBF (from Eq.~(\ref{eq:SBFweightedAverage})) and the fitted SBF return the same value with $\langle \bar{N}_{\mathrm{fit}} \rangle=18.49\pm^{0.44}_{0.39}(=\bar{N}_{\langle\mathrm{SBF} \rangle})$~counts and a relative error of $\varepsilon_{\mathrm{rel},90\%}=2.29\%$. In magnitudes, this is $\bar{m}_{\mathrm{fit}}=30.86\pm^{0.023}_{0.026}(=\bar{m}_{\langle\mathrm{SBF} \rangle})$~mag. This shows that measuring the SBF of a galaxy as a whole may not reflect the variability of the fluctuation in different regions, which could lead to larger scatter in the calibration. Thus, it is necessary to take into account the potential non-homogeneity of the SBF distribution when constraining stellar populations. 
When studying the SBF profile with different annular masks (as in Fig.~\ref{fig:DoubleAxisGradient75000px}), we can validate the partial SBFs using the area covered by each one and comparing with the SBF of the galaxy as a whole, using Eq.~(\ref{eq:SBFweightedAverage}). We remind the reader that this also occurs in observations through the integrated light of the galaxy region.


\section{Summary}
\label{sec:summary}

In this paper, we have investigated the capabilities and limitations of the surface brightness fluctuation retrieval. To do so we have created a mock galaxy based on the data of NGC 4649 (VCC 1978) from \citet{cantiello2018next}. Then, we present the SBF derivation in detail and our fitting procedure. Creating a mock galaxy allows us to study a measurement under ideal laboratory conditions and determine the fidelity of the fitting in relation to the introduced SBF. 
When performing the fitting we consider the mean galaxy model, the sky background and the PSF as known quantities. Consequently, the errors studied reflect the variability of a system with inherent stochasticity due to the stellar population luminosity fluctuations and the instrumental noise. 
We define two parameters to evaluate the measurement: the relative error ($\varepsilon_{\mathrm{rel}}$), as a measure of the accuracy between the real SBF ($\bar{N}_{\mathrm{real}}$) of the mock galaxy and the fitted SBF we obtain ($\bar{N}_{\mathrm{fit}}$), and the relative standard deviation obtained from the least square fitting ($\sigma_{\mathrm{fit}}$). We assume as a valid SBF measurement those where $\varepsilon_{\mathrm{rel}}<10\%$ and $\sigma_{\mathrm{fit}}<10\%$. Additionally, we require all non-masked pixels of the galaxy to have values larger than 10 times the input SBF, to assure a Gaussian probability distribution of the integrated light among pixels \citep{tonry1988new,CL06} and be able to apply the traditional SBF derivation procedure.

We evaluate the SBF estimation under a wide range of conditions. To do so, we create a set of 50 mock galaxies each time a parameter is varied and obtain the distribution of SBF measurements for each set. From this distribution of results we calculate the 90\% percentile of the relative error and relative standard deviation ($\varepsilon_{\mathrm{rel,90\%}}$ and $\sigma_{\mathrm{fit,90\%}}$). Additionally, we calculate the relative 90\% width of the $\bar{N}_{\mathrm{fit}}$ distribution ($\Delta_{90\%}$), to measure the precision of the fitting results. 

We study the SBF fitting for different mask sizes, using all possible combinations of annular rings defined by an internal and external radius. The results show how the relative error ($\varepsilon_{\mathrm{rel},90\%}$), the relative standard deviation of the fitting ($\sigma_{\mathrm{fit},90\%}$) and the relative 90\% width of the $\bar{N}_{\mathrm{fit}}$ distribution ($\Delta_{90\%}$) are tightly related to the number of pixels within the mask ($n_{\mathrm{mask\;pix}}$). Therefore, the larger the number of pixels within the mask, the better the reliability of the measurement.
The precision ($\Delta_{90\%}$) is the most restrictive parameter, followed by the accuracy ($\varepsilon_{\mathrm{rel},90\%}$) and then by the relative standard deviation ($\sigma_{\mathrm{fit},90\%}$).
We focus our study on the relative error, as it is representative of the accuracy between the real SBF value ($\bar{N}_{\mathrm{real}}$) and the fitted ($\bar{N}_{\mathrm{fit}}$). 

We vary the brightness of the galaxy and its fluctuation contribution. We find that the best SBF measurements results are found for both high brightness and high fluctuation. When the fluctuation contribution is too low, the relative error does not fulfil the 10\% uncertainty threshold condition. When the brightness of the galaxy is too low we find that the pixels of the galaxy have lower count number than 10 times the input SBF. 
Under a traditional SBF extraction procedure this last condition is a first order requirement to obtain a physically reliable representation of the stellar population through the SBF. Otherwise, the estimate of the mean galaxy image can be biased and, so the SBF measurement \citep{cervino2004physical,CL06,cervino2008surface,cervino2013}. 

We also varied other parameters in search of ideal conditions for the SBF measurement. We find that the relative error shrinks with a small PSF up to the size of the pixel. A low sky value and a long exposure time improve the results. 
For the experiments performed in this work, varying the Sersic index shows no difference in the 90\% percentile of relative error ($\varepsilon_{\mathrm{rel},90\%}$). Varying the size of the galaxy, that is, the effective radius does not significantly change the result either, but does limit the maximum number of pixels in which we can perform the measurement. 

Aside from varying the parameters of the galaxy, we study the reliability of the SBF derivation. We have reviewed its mathematical development and verified that any approximations taken do not drastically influence the measurement of the SBF. Additionally, before the SBF fitting, we suggest modelling the instrumental noise in order to improve the accuracy of the SBF as the sole unknown parameter (see Table~\ref{table:relianceSBF_Can}). Of course, the modelling of the instrumental noise must adapt to the particular nuances of any observation (such as considering Poisson noise or any other).
Applying Eq.~(\ref{eq:ReadoutComp}) for modelling the noise requires proper approximations of the sky, the mean model and the PSF. Moreover, a poor estimation of these three parameters leads to biased SBF measurements. 
Also, we find that including the azimuthal margins found after creating the radial profile provides more conservative results, but improves the returned precision.

Finally, we study the possibility of SBF gradient detection by including a variable SBF in our mock galaxy. 
We propose an expression for measuring a region with a varying SBF, which corresponds to a pixel-weighted average of all the SBFs present in the masked region. In an actual observation the SBF value in each pixel is unknown, but a similar concept applies: an averaged fluctuation calculated with the different SBFs weighted by the area they occupy. 
If the galaxy fluctuation changes through different regions, measuring the SBF of the whole galaxy returns its weighted average. Consequently, it is crucial to take into account the heterogeneity of the SBF distribution when constraining stellar populations. This reaffirms previous knowledge, as the variability of the stellar population is already considered when studying the integrated colour of an observation. 
With our mock galaxies we attempt to recover the introduced SBF profile by performing the fitting in concentric annuli, using different masks. 
When looking for an SBF gradient, we suggest keeping the same number of pixels in every mask used, to ensure the consistency in the results. 
In our experiments, we can trace the introduced SBF profile using the mean fitted value of each annulus returned from our 50 simulations, allowing us to study the viability and uncertainty of an SBF gradient through the modelling. However, in an actual observation, there are no 50 versions of the observed galaxy from which to retrieve a mean value, so the result is limited to the sole observation performed in each annulus and its associated dispersion does not ensure a reliable gradient is recovered. The solutions will follow the real profile with varying points around the real value. 
In this case, at the very least we can calculate the precision of the results, as we explain in Sect.~\ref{sec:codePresentation}, illustrated in Fig.~\ref{fig:flowchart}.
In any case, working under ideal conditions is already challenging when comparing SBF from one radius to another. Better observations and a more pronounced SBF gradient will help to reveal the underlying fluctuation radial profile.


\section{SBF uncertainty estimator: A code application} 
\label{sec:codePresentation} 

We provide a simple code that estimates the uncertainty of an SBF retrieval, (\texttt{SBF\_uncertainty\_estimator}\footnote{\url{https://github.com/Pablo-IAC/SBF_uncertainty_estimator} \label{fn:code}}), due to the variability of a system that is intrinsically stochastic. It implements the procedure described in this document.

\begin{figure*}
\centering
\includegraphics[width=\textwidth]{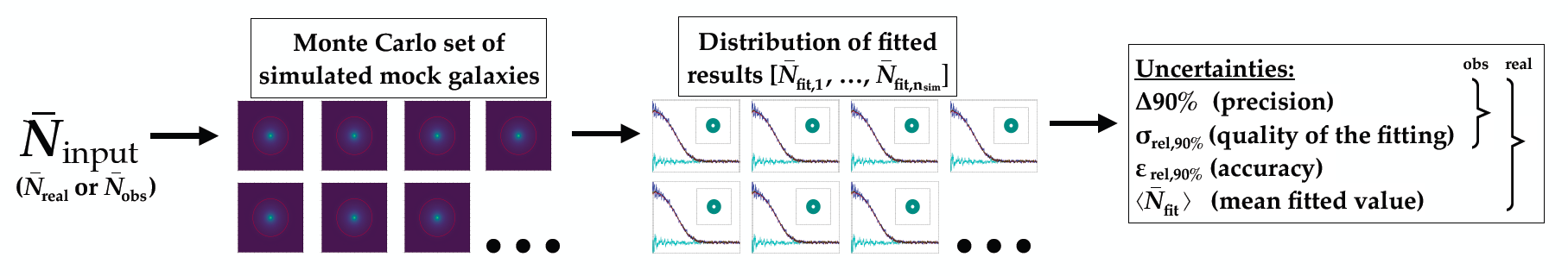}
\caption{Flowchart of the methodology followed throughout this work and applied in the code provided\footref{fn:code}.}
\label{fig:flowchart}
\end{figure*}

The aim of this code is twofold: (i) preliminary study the suitability of a forthcoming SBF measurement or (ii) estimate the uncertainty of an SBF observation. We explain below the steps that our code performs and the nuances in each case, (i) or (ii). Between both cases, the meaning of the introduced fluctuation ($\bar{N}_{\mathrm{input}}$) has different connotations. 
The flowchart shown in Fig.~\ref{fig:flowchart} also illustrates the procedure:

\begin{enumerate}

    \item Create a Monte Carlo set of simulated mock galaxies following the methodology described in this work.
        \begin{enumerate}[(i)]
        \item Introduce representative parameters of the target galaxy, including an initial estimate of its SBF, which can be taken from model predictions. Note that although the actual SBF of the galaxy is unknown, this estimate is
        needed to predict the quality of the SBF derivation. In the current work this value corresponds to $\bar{N}_{\mathrm{real}}$.
        \item Introduce the fitted SBF value ($\bar{N}_{\mathrm{obs}}$) obtained from the observation, along with the rest of parameters associated with the galaxy.
        \end{enumerate}
    \item Fit the SBF for each one of the mock galaxies of the simulated set, therefore retrieving a distribution of fitted SBFs from the simulations $[\bar{N}_{\mathrm{fit},1},...,\bar{N}_{\mathrm{fit,n_{sim}}}]$. Also, we obtain the mean value of the fitted SBFs, $\langle \bar{N}_{\mathrm{fit}} \rangle$.
        \begin{enumerate}[(i)]
        \item This distribution is conceptually equivalent to those presented throughout this work.
        \item This distribution would be a “second generation” of fitted mock SBFs with respect to the original observed SBF.
        \end{enumerate}

    \item Calculate the uncertainties associated with the variability of a stochastic system ($\varepsilon_{\mathrm{rel},90\%},\sigma_{\mathrm{fit},90\%},\Delta_{90\%}$) from the distribution of fitted mock SBFs. Check for low flux pixels ($N(\mathrm{Gal_{mock\;mask}}) - N(\mathrm{Sky_{mask}})\ge 10\cdot \bar{N}_{\mathrm{input}}$) where the modelling might not be correct (see Eq.~(\ref{eq:criteria}) and Sect.~\ref{sec:biasedSBF}) and warn the user if necessary.
    
        \begin{enumerate}[(i)]
        \item With the obtained uncertainties we can predict if our object of study will be a reliable source for the SBF study. The code performs an ideal experiment, so, if the result is untrustworthy according to these simulations, then the situation in a real observation could be more problematic.
        \item The quality of the fitting ($\sigma_{\mathrm{fit},90\%}$) and the precision ($\Delta_{90\%}$) will be useful for assigning such error bars to the observed SBF. We recall that precision is a more restrictive parameter. Given that the input SBF is $\bar{N}_{\mathrm{obs}}$ and the real SBF is unknown, the relative error does not measure the accuracy of the true SBF value of the galaxy. It is a metric for comparing the observed SBF with those derived from the 'second generation' Monte Carlo simulations.
        \end{enumerate}

\end{enumerate}

As in this work, the code assumes a Sersic profile, a fluctuation of the stellar population luminosity as a random Gaussian distribution, a flat sky, a 2-D Gaussian PSF and a random Poisson distribution for the instrumental noise. No GC nor background sources are considered. The code inputs are: the number of counts of the galaxy at the effective radius ($N_{\mathrm{Reff}}$), the number of counts associated with the SBF ($\bar{N}_{\mathrm{input}}$, either $\bar{N}_{\mathrm{real}}$ or $\bar{N}_{\mathrm{obs}}$), the sky counts ($N_{\mathrm{Sky}}$), the size of the point spread function in pixels ($\sigma_{\mathrm{PSF}}$), the size of the image in pixels ($n_{\mathrm{pix}}$), the effective radius in pixels ($R_{\mathrm{eff}}$), the Sersic index ($n$) and the applied annular mask ($r_{\mathrm{1}}$,$r_{\mathrm{2}}$) in pixels too. The code also asks for the range of frequencies ($k_{\mathrm{fit,i}}$,$k_{\mathrm{fit,f}}$) where the fitting will be performed. We recommend checking the azimuthally averaged power spectrum image to ensure that the experiment has been appropriately set and that the selected fitting range covers the frequencies of interest.
Finally, the code asks for the number of Monte Carlo simulations to be done, where the randomness comes from the stellar population luminosity fluctuation and the instrumental noise. The input parameters of our reference galaxy (Sect.~\ref{sec:galData}) can be found in Table~\ref{table:refgal}. The particular values of the reference galaxy shown in this that table, together with an annular mask of ($r_1=80;r_2=332$)~px and a range of fitting frequencies of ($k_{\mathrm{fit,i}}=75,k_{\mathrm{fit,f}}=400$)~px$^{-1}$, should replicate Fig.~\ref{fig:PS} as a validation test.
The code returns estimations for the uncertainties: the accuracy (the 90\% percentile of the relative error, $\varepsilon_{\mathrm{rel},90\%}$), the fitting quality (the 90\% percentile of the relative standard deviation of the fitting, $\sigma_{\mathrm{fit},90\%}$ and the precision (the relative 90\% width of the $\bar{N}_{\mathrm{fit}}$ distribution, $\Delta_{90\%}$). Additionally, it provides the mean value of the distribution of code fitted SBFs, $\langle \bar{N}_{\mathrm{fit}} \rangle$, as well as a warning if our third criterion from Eq.~(\ref{eq:criteria}) is not fulfilled, indicating a possible bias in the derived SBF.

We acknowledge that the programme is a very rough estimator of the SBF uncertainty. Nevertheless, as an ideal laboratory, it aims at providing a conservative benchmark for assessing the uncertainty, whilst real observations will likely yield less favourable results.
We encourage users to adapt and modify the code in a way that accomplishes their science cases.
 
\section{Recommendations for the SBF measurement}
\label{sec:conclusion}

Having summarised the work, we now propose a set of suggestions for future SBF calculations: 

\begin{itemize}

\item The recommended conditions for a proper SBF retrieval are as follows. 
As we find a strong reliance on the number of pixels within the mask, when applying a mask, the largest possible number of non-masked pixels should be kept (as long as they are adequate pixels for the measurement). 
Aside from the typical SBF measurements in old and metal-rich elliptical galaxies, we recommend selecting bright systems (avoiding faint sources) with high fluctuations (associated, for instance, with young or metal-poor populations, or both, \citealt{vazdekis2020surface}). 
One should prepare an observation with a small PSF, a low sky background, and a long exposure time. 
Having the same number of pixels within the mask, the size and light profile of the galaxy do not appear to affect the result drastically, although the size of the galaxy limits the number of pixels with enough light to study.

\item Before performing an observation, we can foresee the potential reliability of its SBF measurement, considering representative parameters of the target galaxy and applying the modelling described in this work. As explained in Sect.~\ref{sec:codePresentation}, case~(i), this provides a preliminary estimation of the uncertainty due to the stochasticity of the system under ideal conditions. If this initial test is faulty, the observation retrieves even worse errors.

\item When fitting the SBF, we strongly recommend modelling the instrumental noise ($P_1$ in the traditional notation of the SBF literature), in order to leave the SBF as a sole parameter to fit.
It can be done using a similar procedure to the one of \citet{mei2005acs} or as we proposed in Eq.~(\ref{eq:ReadoutComp}): one can begin by modelling the sky, the mean galaxy, and the PSF for the observation; then, the sky can be added to the mean model convolved with the PSF (in this work referred as $\mathrm{Gal_{mean,fluc,sky}}$); next, the Poissonian randomisation can be introduced to this composite image; lastly, the necessary noise image can be obtained by subtracting the randomised $\mathrm{Gal_{mean,fluc,sky}}$ from its original self. We note that to obtain the required noise image it is necessary to adapt this approach to the particularities of each observation.  
Additionally, a trustworthy estimation of the mean model, the PSF, and, especially, the sky are vital for the SBF derivation. An offset in these parameters leads to a bias in the SBF and high relative errors. In this situation the relative standard deviation of the fitting or the precision are not drastically influenced by an offset. Therefore, in an observation, we can find adequate results for these two uncertainties, while being unaware of how close the fitted SBF is to the real SBF value, that is, the accuracy and a possible bias.
This effect is particularly sensitive to small offsets in the sky. A way of preventing this issue is carefully checking the image of the azimuthally averaged power spectrum used for the fitting, in which any inconsistency should be noticeable and provides an indication about any possible offset.

\item When performing the SBF fitting in Fourier space, it is necessary to select a pair of limits ($k_{\mathrm{fit,i}}$, $k_{\mathrm{fit,f}}$) where the fitting is performed. Among different observations, changes in the size of the analysed image, the scale plate, or in the PSF require this range of frequencies to be adapted. At the moment, there is no universally applicable way of choosing the boundaries of the fitting, but one should manually avoid the lowest frequencies, which have larger uncertainties, and avoid the highest frequencies since the power spectrum profile becomes constant and only provides redundant information about the instrumental noise.

\item In a traditional SBF derivation procedure, pixels with a flux similar or lower than ten times the SBF ($N(\mathrm{Gal_{mock\;mask}}) - N(\mathrm{Sky_{mask}})\ge 10\cdot \bar{N}$) do not ensure a correct modelling when retrieving the fluctuation. After estimating the SBF, we suggest checking if all the pixels used in the measurement fulfil this condition. 
Deriving SBFs from faint sources, such as dwarf galaxies or low brightness regions of the galaxy, might be unreliable. This is because the integrated light among pixels does not follow a Gaussian distribution and the estimate of the mean galaxy image can be biased, and hence the SBF measurement as well \citep{CL06,cervino2008surface}. In every pixel of the reference 'mean image' (retrieved from the observation as an average of nearby pixels), its value must be a proper representation of the stellar population through the mean of the luminosity distribution.

\item Besides the traditional estimation of the SBF uncertainty on an observation, we also suggest computing the error presented in the current work. Specifically, we assessed the inherent stochastic nature of the system due to fluctuations in the luminosity of the stellar population and instrumental noise. To our knowledge, this aspect is not estimated in previous literature. 
In this case, the actual $\bar{N}_{\mathrm{real}}$ is unknown, and therefore we cannot evaluate the accuracy ($\varepsilon_{\mathrm{rel},90\%}$). However, we can estimate the relative standard deviation of the fitting and the precision of the measurement. Following what is detailed in this paper, we can simulate a Monte Carlo set of mock galaxies (taking $\bar{N}_{\mathrm{input}} = \bar{N}_{\mathrm{obs}}$), and then perform the fitting to calculate $\sigma_{\mathrm{fit},90\%}$ and $\Delta_{90\%}$ with respect to the observationally measured $\bar{N}_{\mathrm{obs}}$. This is illustratively explained in the flowchart of Fig.~\ref{fig:flowchart} and its description is provided in Sect.~\ref{sec:codePresentation}, case~(ii). In an observation, the intrinsic uncertainty due to the undetermined PSF, the sky estimate, the calibration, or any other source of variability can be complemented with our proposed $\sigma_{\mathrm{fit},90\%}$ and $\Delta_{90\%}$. 
We remind the reader that the relative standard deviation of the fitting ($\sigma_{\mathrm{fit},90\%}$) is a less restrictive parameter than the precision of the fitting ($\Delta_{90\%}$), so we recommend selecting $\Delta_{90\%}$ as an uncertainty indicator. 

\end{itemize}

In conclusion, retrieving reliable SBFs requires careful treatment of the measurement and the uncertainty. In order to exploit the potential of the SBFs fully, new studies with high-quality data and proper observational conditions are badly needed. \\


\begin{acknowledgements}
Special thanks to Javier Sánchez Sierras, Isaac Alonso Asensio, Martín Manuel Gómez Míguez and Ignacio Trujillo Cabrera for their invaluable advise and useful discussions. 
AV and PRB acknowledge support from grant PID2019-107427GB-C32 from the Spanish Ministry of Science, Innovation and Universities MCIU. Also, AV acknowledges support from grant PID2021-123313NA-I00 from the Spanish Ministry of Science, Innovation and Universities MCIU.
This work has also been supported through the IAC project TRACES, which is partially supported through the state budget and the regional budget of the Consejer{\'{\i}}a de Econom{\'{\i}}a, Industria, Comercio y Conocimiento of the Canary Islands Autonomous Community.
MC has been funded by the project PID2019-107408GB-C41 and PID2022-136598NB-C33 by the Spanish Ministry of Science and Innovation/State Agency of Research MCIN/AEI/10.13039/501100011033 and by “ERDF A way of making Europe”.
This research has made use of the SIMBAD data base, operated at CDS, Strasbourg, France. 
This research used the facilities of the Canadian Astronomy Data Centre operated by the National Research Council of Canada with the support of the Canadian Space Agency.
Based on observations obtained with MegaPrime/MegaCam, a joint project of CFHT and CEA/IRFU, at the Canada-France-Hawaii Telescope (CFHT) which is operated by the National Research Council (NRC) of Canada, the Institut National des Science de l'Univers of the Centre National de la Recherche Scientifique (CNRS) of France, and the University of Hawaii. This work is based in part on data products produced at Terapix available at the Canadian Astronomy Data Centre as part of the Canada-France-Hawaii Telescope Legacy Survey, a collaborative project of NRC and CNRS.
This paper made use of the IAC Supercomputing facility HTCondor (http://research.cs.wisc.edu/htcondor/), partly financed by the Ministry of Economy and Competitiveness with FEDER funds, code IACA13-3E-2493. 
We acknowledge the help from the members of the SIE team of IAC.

\end{acknowledgements}

%
%

\bibliographystyle{aa} 
\bibliography{SBF}  


\begin{appendix} 
      
\onecolumn

\section{SBF derivation considering a mask}
\label{sec:appendixApplyMask}

In this appendix we show the derivation of the SBF considering a mask, which is a generalisation of the simpler case without any mask (see Sect.~\ref{sec:SBFdef}).
We start from the definition of the mock galaxy, as in Eq.~(\ref{eq:obsGal}). Then, multiplying Eq.~(\ref{eq:obsGal}) with a mask image we obtain Eq.~(\ref{eq:obsGalMask}), which reads: $\;\mathrm{Gal_{mock}}\cdot \mathrm{Mask} = \left( \left( \mathrm{Gal_{mean}}+\mathrm{Gal_{fluc}}+\mathrm{Sky} \right)\otimes  \mathrm{PSF} + R \right)\cdot \mathrm{Mask}$, where $\mathrm{Sky}\otimes \mathrm{PSF} = \mathrm{Sky}$, as $\mathrm{Sky}$ is a constant image. In order to have all the images with the same size, the PSF image is resized by convolving it with a $n_{\mathrm{pix}} \times n_{\mathrm{pix}}$ image, where all the pixels are zero value but the central one, with a value of one. 

Next, we rearrange the terms in the previous equation so we can isolate the fluctuation and noise terms: 

\begin{equation}
        \left( \mathrm{Gal_{mock}} - \mathrm{Sky} - (\mathrm{Gal_{mean}}\otimes  \mathrm{PSF}) \right) \cdot \mathrm{Mask}= (\mathrm{Gal_{fluc}}\otimes  \mathrm{PSF})\cdot \mathrm{Mask} + R\cdot \mathrm{Mask}. 
        \label{eq:flucVSrest_Mask}
\end{equation}

\noindent Now, we divide by the square root of the mean model, $\sqrt{\mathrm{Gal_{mean}}\otimes  \mathrm{PSF}}$. 
Note that the SBF is obtained from the study of
the distribution of the square of the fluctuations around the mean ($\mu$) divided by the mean, that is, $(x-\mu)^2/\mu$. So we build an image of $(x-\mu)/\sqrt{\mu}$ which will be squared when the power spectrum is applied:

\begin{equation}
        \frac{\left( \mathrm{Gal_{mock}} - \mathrm{Sky} - (\mathrm{Gal_{mean}}\otimes \mathrm{PSF}) \right)\cdot \mathrm{Mask}}{\sqrt{\mathrm{Gal_{mean}}\otimes  \mathrm{PSF}}}=
        \frac{(\mathrm{Gal_{fluc}}\otimes  \mathrm{PSF}) \cdot \mathrm{Mask} + R\cdot \mathrm{Mask}}{\sqrt{\mathrm{Gal_{mean}}\otimes  \mathrm{PSF}}}.
        \label{eq:sqrtSBF_Mask}
\end{equation}

\noindent We acknowledge that the SBF definition (the ratio of the variance and the mean images) does not directly mention the PSF. However, for consistency, the definition implies that the term in the denominator must correspond to the one in the numerator, $\mathrm{Gal_{mean}}\otimes  \mathrm{PSF}$, instead of just dividing by $\mathrm{Gal_{mean}}$.

We denote as $\mathrm{Gal_{mock\;fluc\;mask}}$ the first term of Eq.~(\ref{eq:sqrtSBF_Mask}). Then, in order to disentangle the convolved PSF and the fluctuation contribution ($\mathrm{Gal_{fluc}}$) we shift to the Fourier space. We do so by applying the power spectrum $(PS(f)=|\mathfrak{F}(f)|^2)$:

\begin{equation}
\left|\mathfrak{F}\left( \mathrm{Gal_{mock\;fluc\;mask}} \right)\right|^2 = 
\left|\mathfrak{F}\left( \frac{(\mathrm{Gal_{fluc}}\otimes  \mathrm{PSF})\cdot \mathrm{Mask}+R\cdot \mathrm{Mask}}{\sqrt{\mathrm{Gal_{mean}}\otimes  \mathrm{PSF}}} \right)\right|^2.
\label{eq:PS_obsSBFraw_Mask}
\end{equation}

\noindent Developing the previous equation by expanding the right term as the square of complex numbers (see step between Eqs.~\ref{eq:PS_obsSBFraw} and \ref{eq:PS_obsSBFlong}), we find:  

\begin{equation}
    \begin{split}
    \left|\mathfrak{F}\left( \mathrm{Gal_{mock\;fluc\;mask}} \right)\right|^2
   & = \left|\mathfrak{F}\left( \frac{(\mathrm{Gal_{fluc}}\otimes  \mathrm{PSF})\cdot \mathrm{Mask}}{\sqrt{\mathrm{Gal_{mean}}\otimes  \mathrm{PSF}}} \right)\right|^2 + 
    \left|\mathfrak{F}\left( \frac{R\cdot \mathrm{Mask} }{\sqrt{\mathrm{Gal_{mean}}\otimes  \mathrm{PSF}}} \right)\right|^2 \\
   & + \mathfrak{F}\left( \frac{(\mathrm{Gal_{fluc}}\otimes  \mathrm{PSF})\cdot \mathrm{Mask}}{\sqrt{\mathrm{Gal_{mean}}\otimes  \mathrm{PSF}}} \right)^\dag \cdot      \mathfrak{F}\left( \frac{R\cdot \mathrm{Mask}}{\sqrt{\mathrm{Gal_{mean}}\otimes  \mathrm{PSF}}} \right) +  
     \mathfrak{F}\left( \frac{(\mathrm{Gal_{fluc}}\otimes  \mathrm{PSF})\cdot \mathrm{Mask}}{\sqrt{\mathrm{Gal_{mean}}\otimes  \mathrm{PSF}}} \right)\cdot     \mathfrak{F}\left( \frac{R\cdot \mathrm{Mask}}{\sqrt{\mathrm{Gal_{mean}}\otimes  \mathrm{PSF}}} \right)^\dag.
    \end{split}
    \label{eq:PS_obsSBFlong_Mask}
\end{equation}

\noindent The convolution theorem, 
$\mathfrak{F}(f\otimes  g)=\mathfrak{F}(f)\cdot \mathfrak{F}(g)$ 
and 
$\mathfrak{F}(f\cdot g)=\mathfrak{F}(f)\otimes  \mathfrak{F}(g)$, 
allows $\mathrm{Gal_{fluc}}$, the PSF contribution, and the mask to be separated. 
Thus, $\mathfrak{F}\left((f\otimes g)\cdot m\right)$ transforms into 
$\left(\mathfrak{F}(f)\cdot \mathfrak{F}(g)\right. )\otimes \mathfrak{F}(m)$. So, we can rewrite Eq.~(\ref{eq:PS_obsSBFlong_Mask}) as:

\begin{equation}
    \begin{split}
    \left|\mathfrak{F}\left( \mathrm{Gal_{mock\;fluc\;mask}}\right)\right|^2 & = 
    \left( \left|\mathfrak{F}\left( \frac{\mathrm{Gal_{fluc}}}{\sqrt{\mathrm{Gal_{mean}}\otimes  \mathrm{PSF}}} \right)\right|^2 \cdot \left|\mathfrak{F}\left( \mathrm{PSF} \right)\right|^2 \right) \otimes  \left|\mathfrak{F}\left( \mathrm{Mask} \right)\right|^2 
    + \left|\mathfrak{F}\left( \frac{R\cdot \mathrm{Mask}}{\sqrt{\mathrm{Gal_{mean}}\otimes  \mathrm{PSF}}} \right)\right|^2 \\
    & + \left( \mathfrak{F}\left( \frac{\mathrm{Gal_{fluc}}}{\sqrt{\mathrm{Gal_{mean}}\otimes  \mathrm{PSF}}} \right)^\dag \cdot \mathfrak{F}\left( \mathrm{PSF} \right)^\dag \right) \otimes  
    \mathfrak{F}\left( \mathrm{Mask} \right)^\dag \cdot \mathfrak{F}\left( \frac{R\cdot \mathrm{Mask} }{\sqrt{\mathrm{Gal_{mean}}\otimes  \mathrm{PSF}}} \right) \\
    & + \left( \mathfrak{F}\left( \frac{\mathrm{Gal_{fluc}}}{\sqrt{\mathrm{Gal_{mean}}\otimes  \mathrm{PSF}}} \right) \cdot \mathfrak{F}\left( \mathrm{PSF} \right) \right) \otimes 
    \mathfrak{F}\left( \mathrm{Mask}\right)\cdot \mathfrak{F}\left( \frac{R\cdot \mathrm{Mask} }{\sqrt{\mathrm{Gal_{mean}}\otimes  \mathrm{PSF}}} \right)^\dag.
    \end{split}
    \label{eq:applyingTheConvTh}
\end{equation}

\noindent For the first right term of Eq.~(\ref{eq:applyingTheConvTh}) we must calculate first the product $(|\mathfrak{F}(\mathrm{Gal_{fluc}}/\sqrt{\mathrm{Gal_{mean}})\otimes \mathrm{PSF}})|^2\cdot |\mathfrak{F}(\mathrm{PSF})|^2)$, and then convolve the result with $|\mathfrak{F}(\mathrm{Mask})|^2$.
If the fluctuation term is constant, it can exit the operation as a multiplicative factor. In Sect.~\ref{sec:SBFdef} we show that $(|\mathfrak{F}(\mathrm{Gal_{fluc}}/\sqrt{\mathrm{Gal_{mean}})\otimes \mathrm{PSF}})|^2$ can be represented by a constant number, at least in the simulations of this work. Therefore, $\left|\mathfrak{F}\left( \mathrm{PSF} \right)\right|^2 \otimes \left|\mathfrak{F}\left( \mathrm{Mask} \right)\right|^2$ can be operated first, as traditionally stated in the literature. 
The same idea applies to the Fourier transforms in the crossed term. 
Eventually, we rewrite the power spectrum as '$PS$' and an azimuthal average is applied to Eq.~(\ref{eq:applyingTheConvTh}) (note that the azimuthal average of sums is equal to the sum of azimuthal averages). Hence, radial profiles are obtained, denoted with the subindex 'r':

\begin{equation}
\begin{split}
PS&(\mathrm{Gal_{mock\;fluc\;mask}})_{\mathrm{r}} = \bar{N}\cdot \left( PS(\mathrm{PSF})\otimes  PS(\mathrm{Mask}) \right)_\mathrm{r} + 
PS\left(\frac{R\cdot\mathrm{Mask}}{\sqrt{\mathrm{Gal_{mean}}\otimes  \mathrm{PSF}}}\right)_{\mathrm{r}} \\
& + \sqrt{\bar{N}^\dag} \cdot \left(\mathfrak{F}\left( \mathrm{PSF} \right)^\dag  \otimes  \mathfrak{F}\left( \mathrm{Mask} \right)^\dag  \cdot \mathfrak{F}\left( \frac{R\cdot \mathrm{Mask} }{\sqrt{\mathrm{Gal_{mean}}\otimes  \mathrm{PSF}}} \right) \right)_{\mathrm{r}} + 
\sqrt{\bar{N}} \cdot \left(\mathfrak{F}\left( \mathrm{PSF} \right)\otimes \mathfrak{F}\left( \mathrm{Mask} \right)\cdot \mathfrak{F}\left( \frac{R\cdot \mathrm{Mask} }{\sqrt{\mathrm{Gal_{mean}}\otimes  \mathrm{PSF}}} \right)^\dag \right)_{\mathrm{r}}.
\end{split}
\label{eq:PS_conTerminoCruzadoLong_radMask}
\end{equation}

\noindent After applying the azimuthal average, the SBF value ($\bar{N}$) appears as a constant, coming from $PS(\mathrm{Gal_{fluc}}/\sqrt{\mathrm{Gal_{mean}}\otimes \mathrm{PSF}})_\mathrm{r}$. If we assume the fluctuation term to have only the real component ($\bar{N}^\dag=\bar{N}$), 

\begin{equation}
\begin{split}
PS&(\mathrm{Gal_{mock\;fluc\;mask}})_{\mathrm{r}}
 = \bar{N}\cdot \left( PS(\mathrm{PSF})\otimes  PS(\mathrm{Mask}) \right)_\mathrm{r} + 
 PS\left(\frac{R\cdot\mathrm{Mask}}{\sqrt{\mathrm{Gal_{mean}}\otimes  \mathrm{PSF}}}\right)_{\mathrm{r}} \\ 
 & + \sqrt{\bar{N}} \cdot \left( \mathfrak{F}\left( \mathrm{PSF} \right)^\dag  \otimes  \mathfrak{F}\left( \mathrm{Mask} \right)^\dag  \cdot \mathfrak{F}\left( \frac{R\cdot \mathrm{Mask} }{\sqrt{\mathrm{Gal_{mean}}\otimes  \mathrm{PSF}}} \right)\right)_{\mathrm{r}} + 
 \sqrt{\bar{N}} \cdot \left(\mathfrak{F}\left( \mathrm{PSF} \right)\otimes \mathfrak{F}\left( \mathrm{Mask} \right)\cdot \mathfrak{F}\left( \frac{R\cdot \mathrm{Mask}}{\sqrt{\mathrm{Gal_{mean}}\otimes  \mathrm{PSF}}} \right)^\dag  \right)_{\mathrm{r}}.
\end{split}
\label{eq:PS_conTerminoCruzado_radMask}
\end{equation}

\noindent Here, we have computed the real and imaginary part of the crossed term, finding a null imaginary component (see Fig.~\ref{fig:CT_components}). We could fit the SBF value directly from Eq.~(\ref{eq:PS_conTerminoCruzado_radMask}), however numerical calculations show that the influence of the real part of the crossed term is negligible (see Table~\ref{table:relianceSBF_Can}). 
Thus, we find the expression most commonly used in the literature for the SBF fitting ($P(k) = P_0\cdot  E(k) + P_1$), which matches Eq.~(\ref{eq:PS_obsSBFshort_rad_Mask}):

\begin{equation}
PS(\mathrm{Gal_{mock\;fluc\;mask}})_{\mathrm{r}} \approx \bar{N}\cdot \left(PS(\mathrm{PSF})\otimes  PS(\mathrm{Mask})\right)_{\mathrm{r}} + 
PS\left(\frac{R\cdot \mathrm{Mask}}{\sqrt{\mathrm{Gal_{mean}}\otimes  \mathrm{PSF}}} \right)_{\mathrm{r}}.
\label{eq:PS_obsSBFshort_rad_Mask_repeated}
\end{equation}

\newpage

\section{Notation}
\label{sec:notation}
This section contains an exhaustive summary of the notation used in the current work. 
\begin{table*}[h]
\centering
\caption{Glossary.} 
\renewcommand{\arraystretch}{1.2}
\begin{tabular}{|l|l|}
\hline
$L$ & Integrated light of the galaxy. \\ \hline
$m$ & Apparent magnitude of the galaxy. \\ \hline
$\bar{m}$ & SBF magnitude of the galaxy. \\ \hline
$R_{\mathrm{eff}}$ & Effective radius of the galaxy. \\ \hline
$I_{\mathrm{eff}}$ & Intensity per unit area at the effective radius of the galaxy. \\ \hline
$n$  & Sersic index. \\ \hline
$b_n$ & Indicator of the Sersic profile shape (2$n$ - 0.324). \\ \hline
$\mu_{\mathrm{eff}}$ & Surface brightness at the effective radius. \\ \hline 
$N_{\mathrm{Reff}}$ & Number of counts at the effective radius. \\ \hline
PHOT\_C0 & Nominal camera zero point (CFHT/MegaCam). \\ \hline 
$t_{\mathrm{exp}}$ & Exposure time in seconds. \\ \hline
$N_{\mathrm{Sky}}$ & Sky background counts. \\ \hline
$n_{\mathrm{pix}}$ & Size of the image in number of pixels. \\ \hline
$\sigma_{\mathrm{PSF}}$ & Standard deviation of the PSF, a 2-D Gaussian in our work. \\ \hline
$n_{\mathrm{mask\;pix}}$ & Number of pixels within a mask. \\ \hline
$N(\mathrm{Gal_{mean}})$ & Count number from any pixel of $\mathrm{Gal_{mean}}$. \\ \hline
$\bar{N}$ & Number of counts associated with the SBF. If present, a subscript indicates the source of the $\bar{N}$ value. \\ \hline
$\sigma_{\mathrm{fluc}}$ & Standard deviation of the fluctuation of the stellar population. \\ \hline
$\bar{N}_{\langle\mathrm{SBF} \rangle}$ & Pixel-weighted average of the variable SBFs found in the studied region of the galaxy. \\ \hline
$\mathrm{Rand_{Gauss}}$ & Randomisation of a given value following a normal (Gaussian) distribution. \\ \hline
$\mathrm{Rand_{Poi}}$ & Randomisation of a given value following a Poisson distribution. \\ \hline
$\mathrm{Gal_{mean}}$ & Reference image of the galaxy (as the following, size $n_{\mathrm{pix}}\times n_{\mathrm{pix}}$), here calculated with a Sersic profile. \\ \hline
$\mathrm{Gal_{fluc}}$ & Image with the fluctuation due to the stellar population.\\ \hline
$\mathrm{Gal_{mean,fluc}}$ & Reference image of the galaxy including the fluctuation of the stellar population. \\ \hline
$\mathrm{Gal_{mean,fluc,sky}}$ & Image of the galaxy including the fluctuation of the stellar population, plus the sky background. \\ \hline
$\mathrm{Gal_{mean,fluc,sky,PSF}}$ & Image of the galaxy including the fluctuation of the stellar population, plus the sky background, \\ & convolved with the PSF. \\ \hline
$\mathrm{Gal_{mock}}$ & Image of the mock galaxy, built as the mean reference galaxy with the fluctuation, plus the \\ & sky background, convolved with the PSF, plus a Poisson randomisation as instrumental noise.\\ \hline
$\mathrm{Gal_{mock\;fluc\;mask}}$ & Normalised image of the fluctuation due to the stellar population and the instrumental noise. \\ \hline
$\mathrm{Sky}$ & Image of the sky, with constant value $N_{\mathrm{Sky}}$. \\ \hline
$\mathrm{PSF}$ & Image of the point spread function. \\ \hline
$R$ & Image of the instrumental noise. \\ \hline
$R_{\mathrm{approx}}$ & Image of the modelled instrumental noise. \\ \hline
norm (superindex) & Normalised, i.e. divided by $\sqrt{\mathrm{Gal_{mean}}\otimes \mathrm{PSF}}$. \\ \hline
Mask & Image of the mask as zero-one (false-true) pixels. \\ \hline
$PS$ & Power spectrum image. \\ \hline
r (subindex) & Azimuthally averaged. \\ \hline
$k$ & Frequency in the Fourier space. \\ \hline
[$k_{\mathrm{fit,i}}$, $k_{\mathrm{fit,i}}$] & Range of frequencies within which the fitting is performed. \\ \hline
$\varepsilon_{\mathrm{rel}}$ & Relative error of the measurement, calculated as $|\bar{N}_{\mathrm{real}}-\bar{N}_{\mathrm{fit}}|/\bar{N}_{\mathrm{real}}$.  \\ \hline
$\sigma_{\mathrm{fit}}$ & Relative standard deviation of the fitting. \\ \hline
$\varepsilon_{\mathrm{rel},90\%}$ & Higher 90\% percentile of the distribution of $\varepsilon_{\mathrm{rel}}$ values. \\ \hline
$\sigma_{\mathrm{fit},90\%}$ & Higher 90\% percentile of the distribution of $\sigma_{\mathrm{fit}}$ values. \\ \hline
$\Delta{90\%}$ & Relative 90\% width of the $\bar{N}_{\mathrm{fit}}$ distribution. \\ \hline
\end{tabular}
\end{table*}
\end{appendix}

\end{document}